\documentstyle[epsfig]{article}

\newcommand{\lra}{\mbox{$\leftrightarrow$}}

\newcommand{\bone}{\mbox{\bf 1}}
\newcommand{\tr}{\mbox{tr$\,$}}
\newcommand{\str}{\mbox{str$\,$}}

\newcommand{\e}{\mbox{$\epsilon$}}
\newcommand{\half}{\mbox{$1\over2$}}
\newcommand{\third}{\mbox{$1\over3$}}
\newcommand{\quarter}{\mbox{$1\over4$}}
\newcommand{\eigth}{\mbox{$1\over8$}}
\newcommand{\ihalf}{\mbox{$i\over2$}}
\newcommand{\NN}{\mbox{$\cal N$}}
\newcommand{\OO}{\mbox{$\cal O$}}
\newcommand{\PP}{\mbox{$\cal P$}}
\newcommand{\RR}{\mbox{$\cal R$}}
\newcommand{\FF}{\mbox{$\cal F$}}
\newcommand{\KK}{\mbox{$\cal K$}}
\newcommand{\LL}{\mbox{$\cal L$}}
\newcommand{\YY}{\mbox{$\cal Y$}}

\begin{document}

\begin{titlepage}
\begin{center}

{\hbox to\hsize{
hep-th/0109064
\hfill 
UCLA/01/TEP/17 
}}
{\hbox to\hsize{
\hfill 
September 2001
}}

\vskip 1in
 
{\Large \bf 
Quarter BPS Operators 
in \NN=4 SYM 
}

\vskip 0.45in

\renewcommand{\thefootnote}{\fnsymbol{footnote}}
{\bf 
Anton V. Ryzhov%
\footnote{
	ryzhovav@physics.ucla.edu
	}
}
\setcounter{footnote}{0}

\vskip 0.15in

{\em Department of Physics and Astronomy, \\
University of California, Los Angeles, \\ 
LA, CA 90095-1547 \\ }

\vskip 0.1in

\end{center}

\vskip .5in

\begin{abstract}
Chiral primary operators annihilated
by a quarter of the supercharges are constructed 
in the four dimensional $\NN=4$ Super-Yang-Mills theory 
with gauge group $SU(N)$. 
These \quarter-BPS operators 
share many non-renormalization properties with 
the previously studied \half-BPS operators. 
However, they are much more involved, 
which renders their construction nontrivial 
in the fully interacting theory. 
In this paper we calculate $\OO(g^2)$ two-point functions 
of 
local, polynomial, scalar composite operators 
within a given representation 
of the $SU(4)$ $R$-symmetry group. 
By studying these two-point functions, 
we identify the eigenstates of the dilatation operator, 
which turn out to be complicated mixtures of single and
multiple trace operators. 

Given the elaborate combinatorics of this problem, 
we concentrate on two special cases. 
First, we present explicit computations for
\quarter-BPS operators with scaling dimension $\Delta \le 7$. 
In this case, the discussion applies to 
arbitrary $N$ of the gauge group. 
Second, we carry out 
a leading plus subleading large $N$ analysis 
for the particular class of operators 
built out of double and single trace operators only. 
The large $N$ construction addresses 
\quarter-BPS operators of general dimension.

\end{abstract}

\end{titlepage}

\newpage

\section{Introduction}

During the past years, 
there has been a renewed interest in the study of chiral operators 
in the \NN=4 supersymmetric 
Yang-Mills theory in four dimensions. 
Forming short representations 
of the global $SU(2,2|4)$ superconformal symmetry group, 
chiral operators have tightly constrained quantum 
numbers. 
In particular, the scaling dimension of a chiral operator 
is not renormalized.%
\footnote{
\label{footnote:non-renormalized nonBPS}
	The possibility that certain non-chiral operators 
	may have vanishing anomalous dimension
	was raised in \cite{AF}. 
	}

Chiral primary operators have been classified in \cite{DP}, \cite{AFSZ}. 
They can be \half-BPS, \quarter-BPS, and \eigth-BPS. 
The \half-BPS operators provide 
the simplest example of chiral primaries. 
These are 
scalar composite operators in the $[0,q,0]$ representations 
of the $R$-symmetry group $SU(4) \sim SO(6)$; 
their scaling dimension is $\Delta = q$, see \cite{AFSZ}. 
\half-BPS operators are annihilated by eight out of the sixteen Poincar\'e 
supercharges 
of the theory. 
Similarly, \quarter-BPS primaries belong to 
$[p,q,p]$ representations of the $R$-symmetry group, 
are annihilated by four supercharges, and have protected 
scaling dimension of $\Delta = 2p+q$. 
Finally, \eigth-BPS primaries 
live in $[p,q,p+2k]$ of $SU(4)$, 
are killed by only two supercharges, 
and their $\Delta = 3k+2p+q$. 
Quantum numbers of the descendant operators 
are related to those of their primaries by the 
\NN=4 superconformal algebra.

\half-BPS operators have been much studied. 
Using the conjectured AdS/CFT correspondence \cite{AdS-CFT}, 
it was shown, 
that for gauge groups 
$SU(N)$ with $N$ large, two and three point functions 
of \half-BPS chiral primaries are the same at weak 
and strong coupling \cite{LMRS}.%
\footnote{
	Higher $n$-point functions also 
	agree with supergravity predictions in the large $N$ limit 
	\cite{DHKMV}. 
	}
It was then verified that these correlators get no 
$\OO(g^2)$ corrections on the SYM side, for arbitrary $N$ \cite{DFS}. 
Chiral descendant operators share these 
nonrenormalization properties with their parent primaries \cite{DFS}. 
Order $g^4$ and  instanton contributions to 
two and three point functions 
of \half-BPS chiral primaries 
turn out to vanish as well 
\cite{HSW}, \cite{BKRS}, \cite{AF}, \cite{ZANON}. 
Nonrenormalization of these correlators 
was further established on general grounds 
in \cite{Intriligator}, \cite{HSW2}. 
Besides $SU(N)$ theories and single trace chiral primaries, 
multiple trace operators with the same $SU(2,2|4)$ quantum numbers, 
as well as 
arbitrary gauge groups 
were considered \cite{Skiba}. 
In these cases, two and three point functions 
were also found to receive no $\OO(g^2)$ corrections.

It is natural to ask whether other chiral operators, 
for example \quarter-BPS primaries, 
have protected correlators. 
Here the situation is much less straightforward 
than for $[0,q,0]$ operators. 
In fact, except for 
the simplest operator 
found in \cite{BKRS}, 
no other \quarter-BPS chiral primaries 
were written down%
\footnote{
	\quarter-BPS operators have been studied indirectly 
	through OPEs of \half-BPS chiral primaries, see 
	\cite{BKRS}, \cite{AF}, \cite{ADS:4-pt}, \cite{ADS:4-pt:superspace}.
	} 
in the fully interacting theory. 
The main difficulty is that unlike in the free theory, 
where a kinematical (group theoretical) treatment 
of \cite{AFSZ} is sufficient, for nonzero coupling 
the problem of determining primary operators 
becomes a dynamical question.%
\footnote{
	I would like to thank Sergio Ferrara 
	for bringing this to my attention. 
	}

Apart from the 
double trace scalar composite operators 
in the $[p,q,p]$ of the $R$-symmetry (flavor) group $SU(4)$ 
(the free theory chiral primaries from the classification of \cite{AFSZ}), 
there are other single and multiple trace scalar composite operators 
with the same $SU(4)$ quantum numbers 
and the same Born level scaling dimension. 
Unlike in the \half-BPS case where this 
phenomenon occurs \cite{Skiba}, 
scalar composites in the $[p,q,p]$ generally do not have 
a well defined scaling dimension. 
Therefore, one should first find their linear combinations 
which are eigenstates of the dilatation operator, 
which we call pure operators. 
To this end, we calculate two point functions 
of 
local, gauge invariant, polynomial, scalar composite 
operators in a given $[p,q,p]$ representation; 
diagonalize the dilatation operator within 
each representation of $SU(4)$; 
and find that 
some of the pure operators receive no $\OO(g^2)$ corrections to 
their scaling dimension or normalization. 
These operators have the right $SU(4)$ quantum numbers 
and protected $\Delta = 2p+q$, 
and are the only candidates for being 
the \quarter-BPS chiral primaries 
from the classification of \cite{AFSZ}.

Calculating the symmetry factors for Feynman diagrams 
is a formidable combinatorial problem for general 
representation $[p,q,p]$ of $SU(4)$, 
and general $N$ of the gauge group $SU(N)$. 
So to keep the formulae manageable, in this paper 
we concentrate on two special cases. 
For low dimensional operators ($2p+q < 8$), 
we perform explicit computations for arbitrary $N$; 
in particular, we recover the simplest 
\quarter-BPS operator 
studied previously in \cite{BKRS}. 
Alternatively, we give a leading plus subleading 
large $N$ argument (valid for general $[p,q,p]$ representations) 
for a class of 
\quarter-BPS chiral primaries, 
which are linear combinations 
of double- and single-trace scalar composite operators.

The paper is arranged as follows. 
First we review some aspects of 
$SU(2,2|4)$ group theory, and 
describe the scalar composite operators 
we will be dealing with. 
Then we set the stage for $\OO(g^2)$ 
calculations of two-point functions, 
and outline the main ingredients 
of these calculations. 
After that, we explicitly compute the simplest 
sets of correlators. 
In the course of these computations, 
it turns out that only one 
type of Feynman diagrams contributes 
to the correlators at order $g^2$, 
and we provide a simple explanation of this fact.%
\footnote{
	The argument we give applies more generally. 
	In particular, it provides 
	an alternative interpretation of the work in 
	\cite{DFS} and \cite{Skiba}. 
	}
We present the full calculation for these two point functions. 
For higher $\Delta$, calculations were done using {\it Mathematica} 
and only the results are shown. 
Several new features come into play, and we 
describe them as we go along. 
Finally, 
we switch gears and do a large $N$ analysis 
of \quarter-BPS operators with arbitrary scaling dimension.

In this paper, we 
properly identify the \quarter-BPS primaries in the 
fully interacting theory. 
Analysis of two point functions 
is the first step in a systematic study of 
nonrenormalization properties of chiral operators. 
Three-point correlators of \quarter-BPS primaries 
will be studied in 
\cite{DR}.

\section{$SU(2,2|4)$ group theory}
\label{section:superconformal group}

Four dimensional  \NN=4 superconformal Yang-Mills theory 
has been studied extensively 
for a long time, and we begin by reviewing 
some well known facts. 

\NN=4 SYM can be formulated in several (equivalent) ways; 
see Appendix  \ref{n=4 susy:section} for some of the descriptions. 
None of them shows all the features of the 
theory explicitly. For example, working with 
six scalars $\phi^I = \phi^I_a t^a$ 
(where 
$a=1, ... , N^2-1$ runs over the gauge group $SU(N)$, 
and $\phi^I_a(x)$ are real scalar fields), 
and grouping the fermions 
as $\lambda^i_a$, $i=1,...,4$, 
makes the full $SU(4)$ $R$-symmetry group manifest, but hides all the 
supersymmetries. On the other hand, formulating 
the theory in terms of \NN=1 superfields shows 
some of the supersymmetry, but the Lagrangian 
looks invariant just under the $SU(3) \times U(1)$
subgroup of the full $SU(4)$. 
In practice, the more supersymmetries we use, the
simpler it is to perform actual calculations.%
\footnote{
	E.g., the order $g^4$ calculations 
	in \cite{BKRS} were done in the \NN=2 
	harmonic superspace formalism. 
	}
For our purposes 
it suffices to use the component fields of the \NN=1 superfield 
formulation of the theory, 
with the (Euclidean signature) Lagrangian 
\cite{DFS} 
\begin	{eqnarray}
\label	{lagrangian:for Feynman rules}
\LL &\!\!=\!\!& \tr \left\{ 
\quarter F_{\mu\nu} F^{\mu\nu} 
+ \half \bar \lambda \gamma^\mu D_\mu \lambda 
+ \overline{D_\mu z_j} D^\mu z_j 
+ \half \bar \psi^j \gamma^\mu D_\mu \psi^j 
\right\} 
\nonumber\\ 
&&
+ i \sqrt{2} g f^{abc} \left( 
\bar \lambda_a \bar z_b^j L \psi_c^j - 
\bar \psi_a^j R z_b^j \lambda_c 
\right) 
- \half Y f^{abc} \e_{ijk} \left( 
\bar \psi_a^i z_b^j L \psi_c^k - 
\bar \psi_a^i R \bar z_b^j \psi_c^k 
\right) 
\nonumber\\ 
&&
- \half g^2 (f^{abc} \bar z_b^j z_c^j ) (f^{ade} \bar z_d^k z_e^k ) 
+ \quarter Y^2 f^{abc} f^{ade} \e_{ijk} \e_{ilm} 
z_b^j z_c^k \bar z_d^l \bar z_e^m 
\end	{eqnarray}
($L$ and $R$ are chirality projectors). 
The theory defined by (\ref{lagrangian:for Feynman rules}) 
has \NN=1 supersymmetry. 
We use separate coupling constants $g$ and $Y$ 
to distinguish the terms coming from the gauge and 
superpotential sectors. 
When $Y = g \sqrt{2}$, SUSY is enhanced to \NN=4.

Since the manifest symmetry group is now $SU(3) \times U(1)$, 
we first project onto it the representations of 
the full $SU(4)$. 
This can be done by mapping the quantum numbers as 
\begin{equation}
\label{projection:def}
[p,q,r] \mapsto [p,q]^{ - {1\over2} (p+2q+3r)}
\end{equation}
Under this projection, the fermions in the theory 
are mapped as:
$\lambda_{1,2,3} \mapsto \psi_{1,2,3} \in [1,0]^{-{1\over2}}$, 
$\lambda_4 \mapsto \lambda = [0,0]^{{3\over2}}$, so 
${\bf 4} = [1,0,0] \rightarrow [1,0]^{-{1\over2}} \oplus [0,0]^{{3\over2}}$. 
Similarly the scalars are projected 
as 
\begin{equation}
{\bf 6} ~=~ 
[0,1,0] ~~\rightarrow~~ 
[1,0]^1 \oplus [0,1]^{-1} ~=~ \{ z_j \} \oplus \{ \bar z^k \} 
\end{equation}
Put more simply, 
this amounts to 
rewriting the real scalars $\phi^I$, and fermions $\lambda^i$ 
as 
$\phi^i = {1\over\sqrt{2}}(z_i + \bar z_i)$, 
$\phi^{i+3} = {1\over i \sqrt{2}}(z_i - \bar z_i)$, 
and 
$\lambda^i = \psi_i$, $\lambda^4 = \lambda$. 
Index $i=1,2,3$ labels the {\bf 3} or ${\bf \bar 3}$
of the $SU(3)$ factor of the manifest symmetry group 
of (\ref{lagrangian:for Feynman rules}).

The $R$-symmetry group 
of the theory is $SU(4) \sim SO(6)$, which 
is a part of the larger superconformal 
$SU(2,2|4)$. 
Unitary representations of \NN=4 SYM 
were classified in \cite{DP}. 
As in any conformal theory, 
operators 
are classified by their scaling dimension $\Delta$. 
Each multiplet 
of $SU(2,2|4)$ 
contains an operator of lowest 
dimension, which is called a primary operator. 
The action of generators of the conformal group%
\footnote{
	See for example \cite{MS}, 
	or the big review \cite{AGMOO}, p. 32. 
	}
on a primary operator $\Phi(x)$ is given by
\begin{eqnarray}
\label{begin:conformal algebra}
\left[ P_\mu , \Phi(x) \right] &=& i \partial_\mu \Phi(x) \\
\left[ M_{\mu\nu} , \Phi(x) \right] &=& 
\left[ i (x_\mu \partial_\nu - x_\nu \partial_\mu) 
+ \Sigma_{\mu\nu} \right] \Phi(x) \\
\left[ D , \Phi(x) \right] &=& 
i \left( - \Delta + x^\mu \partial_\mu \right) \Phi(x) \\
\left[ K_\mu , \Phi(x) \right] &=& 
\left[ i (x^2 \partial_\mu - 2 x_\mu x^\nu \partial_\nu + 2 x_\mu \Delta) 
- 2 x^\nu \Sigma_{\mu\nu} \right] \Phi(x) 
\label{end:conformal algebra}
\end{eqnarray}
Notice that 
$\left[ M_{\mu\nu} , \Phi(0) \right] = \Sigma_{\mu\nu} \Phi(0)$, 
$\left[ D , \Phi(0) \right] = - i \Delta \Phi(0)$, and 
$\left[ K_\mu , \Phi(0) \right] = 0$. 
Together with the 16 
Poincar\'e supersymmetry generators $Q$ (and $\bar Q$), 
and 16 special conformal fermionic generators $S$ (and $\bar S$),
these close in a superconformal algebra of $SU(2,2|4)$. 
The additional (anti)commutation relations are schematically 
given by 
\begin{eqnarray}
\label{begin:superconformal algebra}
\left[ D , Q \right] &=& - \ihalf Q , \quad 
\left[ D , S \right] ~=~ + \ihalf S , \quad 
\left[ K , Q \right] ~\sim~ S \quad 
\left[ P , S \right] ~\sim~ Q , \\ 
\left[ Q , S \right] &\sim& M + D + R , \quad 
\left[ S , S \right] ~\sim~ K , \quad 
\left[ Q_i , Q_j \right] ~\sim~ P \, \delta_{ij} ~ (i,j=1,...,4) \quad~ 
\label{end:superconformal algebra}
\end{eqnarray}
where $R$ stands for the quantum numbers of the $R$-symmetry group $SU(4)$. 
The Lagrangian of the theory, 
as well as the action of supersymmetry generators 
on the elementary fields, 
are listed in Appendix \ref{n=4 susy:section}.

Primary operators of the superconformal group 
which are annihilated by at least some of the $Q$-s 
are called chiral primaries. 
Descendants of chiral primaries are then 
chiral operators, in the \NN=4 sense. 
Chirality is a property of the whole $SU(2,2|4)$ 
multiplet; 
just being annihilated by say 8 Poincar\'e SUSY generators
doesn't make an operator \half-BPS. 
Since the supercharges anticommute, 
we can take a 
non-chiral operator and act on it with 
some of the $Q$-s. 
The resulting (non-chiral!) operator 
will be annihilated by the same $Q$-s. 

For a chiral primary field $\Phi$ 
annihilated by a Poincar\'e supercharge $Q$, 
we can write 
$\left[ Q , \Phi(x) \right] =0$ and $\left[ K , \Phi(0) \right] = 0$, 
and so 
$\left[ S , \Phi(0) \right] \sim \left[ [K,Q] , \Phi(0) \right] = 0$ 
as well. 
Hence we can express the conformal dimension $\Delta$ of $\Phi$ 
entirely in terms of its spin $\Sigma$ 
and $SU(4)$ quantum numbers 
$R$ 
\begin{equation}
0 = \left[ [Q,S] , \Phi(0) \right] \sim 
\left[ M+D+R , \Phi(0) \right] 
= \left( \Sigma - i \Delta + R \right) \Phi(0)
\end{equation}
by the superconformal algebra 
(\ref{begin:conformal algebra}-\ref{end:superconformal algebra}). 
Quantum numbers of descendants 
are related to those of their parent primaries 
by 
(\ref{begin:conformal algebra}-\ref{end:superconformal algebra}) 
as well. 
In particular, 
$\Delta$ of any chiral operator 
can not receive quantum corrections.

\section{Gauge invariant scalar composite operators}
\label{section:operators}

A kinematic (group theoretic) classification of 
BPS operators was given in \cite{AFSZ}. 
Chiral primaries%
\footnote{
	When referring to ``primary'' fields, 
	we often have in mind the entire 
	$SU(4)$ multiplet to which 
	the actual primary belongs. 
	This slight abuse of notation is common in the literature. 
	} 
are Lorentz scalars, 
which are made by taking local 
gauge invariant polynomial combinations of the 
$\phi^I(x)$, $I=1, ... , 6$. 
They fall into one of the three families \cite{DP}. 
The simplest one consists of \half-BPS operators. These 
chiral primaries are annihilated by half of the $Q$-s, 
and live in short multiplets with spins ranging 
from zero to 2. 
\half-BPS chiral primaries 
are totally symmetric traceless rank $q$ tensors 
of the flavor $SO(6)$. 
$SU(4)$ labels of these representations are $[0,q,0]$ 
with the corresponding 
$SO(6)$ Young tableau%
\footnote{
	See for example \cite{Hamermesh} for a general 
	discussion on constructing irreducible tensors 
	of $SO(n)$. 
	}
$\mbox{
\setlength{\unitlength}{0.7em}
\begin{picture}(3.5,1)
\put(-.8,0){\framebox (1,1){}}
\put(0.2,0){\framebox (2,1){\scriptsize $...$}}
\put(2.2,0){\framebox (1,1){}}
\end{picture}}$, 
one row of length $q$. 
Operators with the highest $SU(4)$ weight in the $[0,q,0]$ 
have the form 
$\tr (\phi^1)^q$, modulo the $SO(6)$ traces.%
\footnote{
	For example, the highest weight state in the [2,0,2] 
	is $\tr (\phi^1)^2 - {1\over6} \sum_{I=1}^6 \tr \phi^I \phi^I$. 
	Operators in this representation are usually referred to 
	as ``$\tr X^2$'' in the literature, and are special since 
	their descendants include the $SU(4)$ flavor currents 
	and the stress tensor. 
	}
Because the color group is $SU(N)$ rather than $U(N)$, 
$\tr \phi^I = 0$ 
so $q \ge 2$. 
Conformal dimension of a \half-BPS chiral primary is related to 
its flavor quantum numbers as $\Delta = q$.

\quarter-BPS operators form 
the next simplest family of chiral operators 
in the classification of \cite{AFSZ}. 
Their multiplets have spins from zero to 3. 
The primaries 
belong to $[p,q,p]$ representations, 
and are annihilated by four out of sixteen 
Poincar\'e supercharges. 
There is a restriction $p \ge 2$: 
for $p=0$ the operators are \half-BPS; 
and in the case $p=1$, they vanish 
after we take the $SU(N)$ traces. 
The highest weight state 
of $[p,q,p]$ 
corresponds to the 
\begin{equation}
\label{pqp:representation} 
\mbox{
\setlength{\unitlength}{1em}
\begin{picture}(7.5,2)
\put(0,1){\framebox (1,1){\scriptsize $1$}}
\put(1,1){\framebox (2,1){\scriptsize $...$}}
\put(3,1){\framebox (1,1){\scriptsize $1$}}
\put(4,1){\framebox (1,1){\scriptsize $1$}}
\put(5,1){\framebox (1,1){\scriptsize $...$}}
\put(6,1){\framebox (1,1){\scriptsize $1$}}
\put(0,0){\framebox (1,1){\scriptsize $2$}}
\put(1,0){\framebox (2,1){\scriptsize $...$}}
\put(3,0){\framebox (1,1){\scriptsize $2$}}
\put(1.8,-.7){\scriptsize $p$}
\put(5.4,.3){\scriptsize $q$}
\end{picture}}
\end{equation}
$SO(6)$ Young tableau. 
In the free theory, \quarter-BPS primaries 
corresponding to (\ref{pqp:representation}) 
are of the form 
$\tr (\phi^1)^{p+q} \, \tr (\phi^2)^{p}$ 
(modulo $(\phi^1,\phi^2)$ antisymmetrizations, 
and subtraction of the $SO(6)$ traces). 
However, there are many other ways to 
partition a given Young tableau, 
and each may result in 
a different operator after we take the $SU(N)$ traces. 
A priori, we do not know if any of them 
are pure (i.e. eigenstates of the dilatation operator $D$), 
or are mixtures of operators with different 
scaling dimensions. 
So these operators should be regarded 
just as a basis of gauge invariant, local, 
polynomial, scalar composite operators in the $[p,q,p]$ 
of $SU(4)$. 
By taking linear combinations of these, we will 
construct eigenstates of $D$ in general, 
and \quarter-BPS primaries in particular.


For completeness, let us mention the \eigth-BPS operators, 
which form the last family of chiral operators in the classification 
of \cite{AFSZ}. \eigth-BPS multiplets are also short, with 
spins from zero to 7/2, and the chiral primaries 
are of the form 
$\tr (\phi^1)^{p+k+q} \, \tr (\phi^2)^{p+k} \, \tr (\phi^3)^{k}$ 
(modulo $(\phi^1,\phi^2,\phi^3)$ antisymmetrizations, 
and minus the $SO(6)$ traces), 
in the free theory. 
As before, there is a $k \ge 2$ restriction on the 
quantum numbers: $k \ge 1$ so the operators 
are annihilated by exactly two Poincar\'e supercharges;
while operators with $k=1$ necessarily contain commutators 
after we take the $SU(N)$ traces, as $\tr \phi^I = 0$. 
\eigth-BPS chiral primaries have $SU(4)$ labels $[p,q,p+2k]$, and 
their scaling dimensions have protected values of 
$\Delta = 3 k + 2 p + q$. 
Although 
these operators are also interesting, 
we will not study them in this paper.

When calculating $n$-point functions, it suffices to consider 
one (nonzero) correlator for a given choice of representations; 
all others will be related to it by 
$SU(4)$ Clebsch-Gordon coefficients 
(by the Wigner-Eckart theorem). 
Therefore, we are free to take the most 
convenient representatives of the full $SU(4)$ 
representations, or of the smaller $SU(3) \times U(1)$ 
bits into which a given representation of $SU(4)$ breaks down.%
\footnote{
	All correlators in the 
	resulting $SU(3) \times U(1)$ representations 
	will have identical spatial dependence, 
	since they come from the same $SU(4)$ representation. 
	}
The combinatorics of the problem 
simplifies if we consider operators 
of the form $[(z_1)^{p+q} (z_2)^p]$ 
and their conjugates, 
which is what we will do in this paper.

Finally, suppose we have disentangled the mixtures 
of $[p,q,p]$ scalar composite operators 
annihilated by a quarter of the Poincar\'e supercharges,
into linear combinations of operators 
with definite scaling dimension. 
Furthermore, assume we found an operator $\YY$ 
whose scaling dimension is protected. 
Since $\YY$ is a pure operator annihilated by four 
Poincar\'e supercharges, it can be either 
a \quarter-BPS primary; 
or a level two descendant of a \eigth-BPS primary, 
but this case is excluded%
\footnote{
	If $\YY$ came from a \eigth-BPS primary, 
	the parent primary would in the $[p',q',p'+2 k]$ 
	representation of $SU(4)$, with $k \ge 2$. 
	On the other hand, to make the scaling dimension 
	and $SU(4)$ Dynkin labels 
	work out right, the only allowed choice is 
	$[p,q,p+2]$, or $k=1$. 
	}
by group theory; 
or a level four descendant of a non-chiral primary. 
If $\YY$ were non-chiral, 
its primary would be 
a scalar composite operator of the form $[z^{2p+q-3} \bar z]$;
and in all examples that we studied in this paper, such 
operators do receive $\OO(g^2)$ corrections to their 
scaling dimension.%
\footnote{
	But see footnote \ref{footnote:non-renormalized nonBPS}. 
	}
We conclude that 
a scalar composite operator in the $[p,q,p]$, 
which is annihilated by a quarter of the supercharges 
and has a protected scaling dimension $\Delta = 2p+q$, 
is a \quarter-BPS chiral primary.

\section{Contributing diagrams}
\label{section:contributing diagrams}

The two point functions we will be calculating in this 
paper are 
of the 
form 
\begin{equation}
\label{eq:general two-point}
\langle 
\left[ {z_1}^{(p+q)} {z_2}^p \right] \!(x) 
\,
\left[ {\bar z_1}^{(p+q)} {\bar z_2}^p \right] \!(y) \rangle 
\end{equation}
where $[...]$ stands for gauge invariant combinations. 
The free field part of such a correlator 
is given by a power of the free
scalar propagator $[G(x,y)]^{(2p+q)}$, times a combinatorial factor. 
At order $g^2$, there are corrections to the scalar propagator 
coming from a fermion loop and a gauge boson semi-loop. Also, 
blocks involving four scalars get contributions from a single 
gauge boson exchange, and from the four-scalar vertex. Gauge fixing
and ghost terms in the Lagrangian do not contribute 
to (\ref{eq:general two-point}) at $\OO(g^2)$. 

\begin{figure}
{\begin{center}
\epsfig{width=4.75in, file=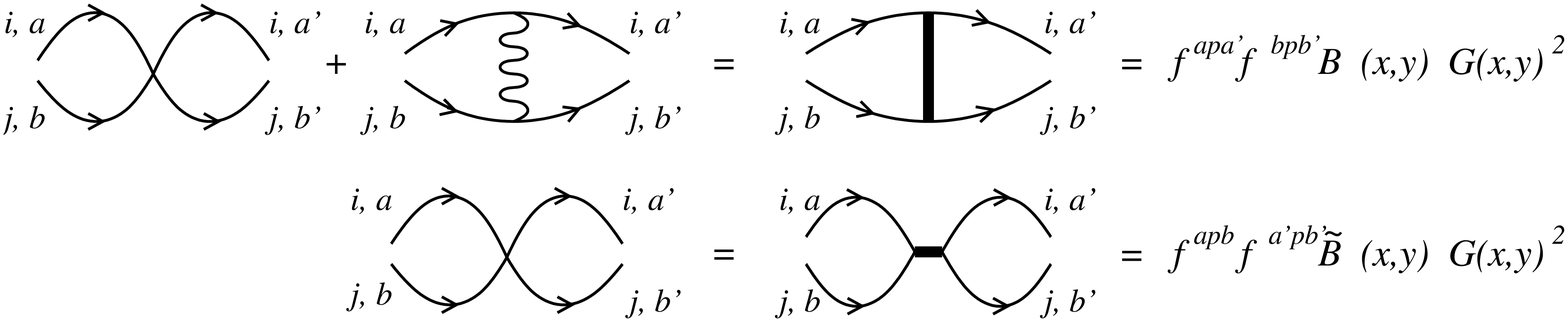, angle=0}
\vskip 0.15in
\epsfig{width=4.6in, file=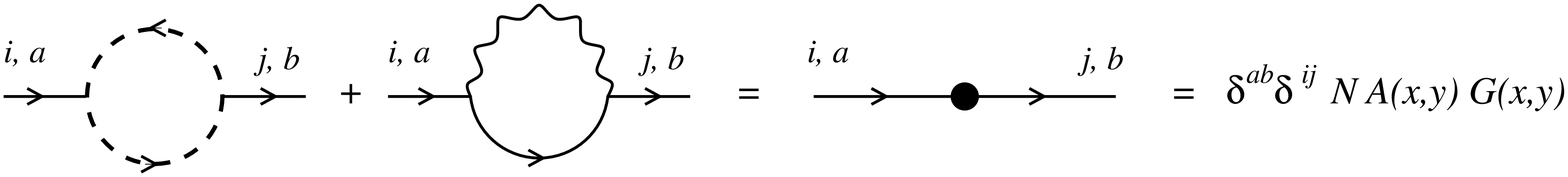, angle=0}\quad
\end{center}}
\vskip -0.2in
\caption{%
Structures contributing to two-point functions 
at order $g^2$. 
Thick lines correspond to exchanges of the gauge boson 
and 
auxiliary fields $F_i$ or $D$
in the \NN=1 formulation. 
\label {fig:four-scalar and propagator}
}%
\end {figure}

\begin{figure}[t!]
{\begin{center}
\epsfig{width=0.7in, file=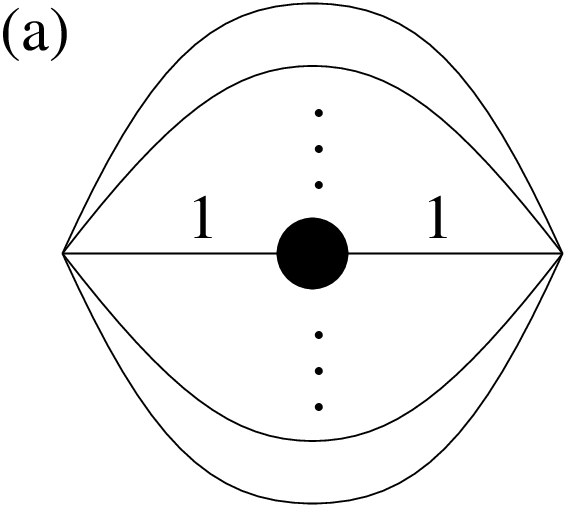, angle=-90} 
\epsfig{width=0.7in, file=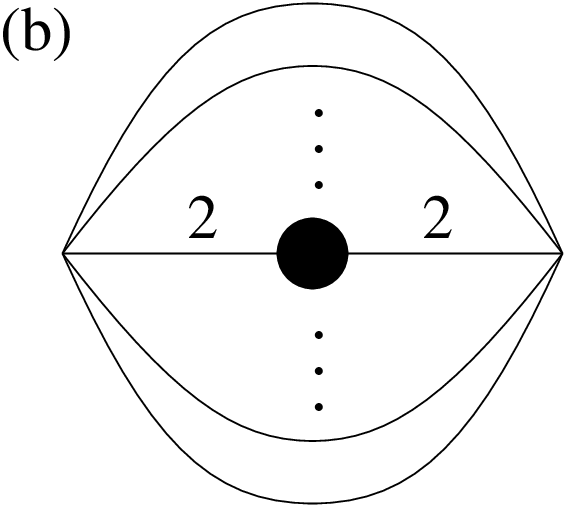, angle=-90} 
\epsfig{width=0.7in, file=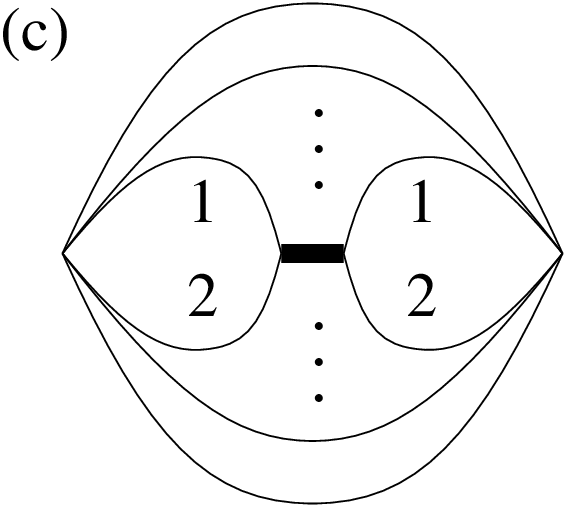, angle=-90}
\epsfig{width=0.7in, file=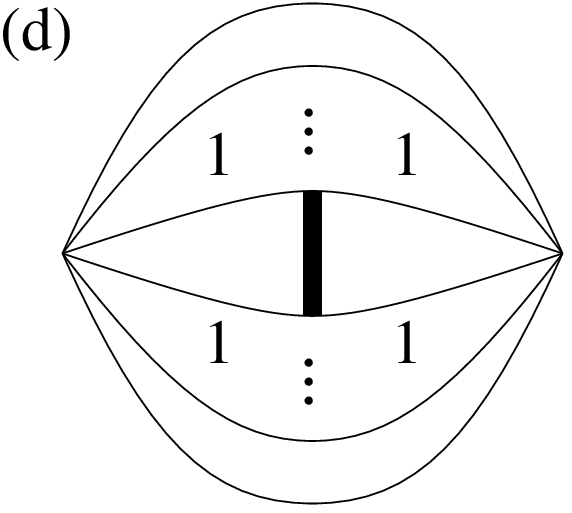, angle=-90} 
\epsfig{width=0.7in, file=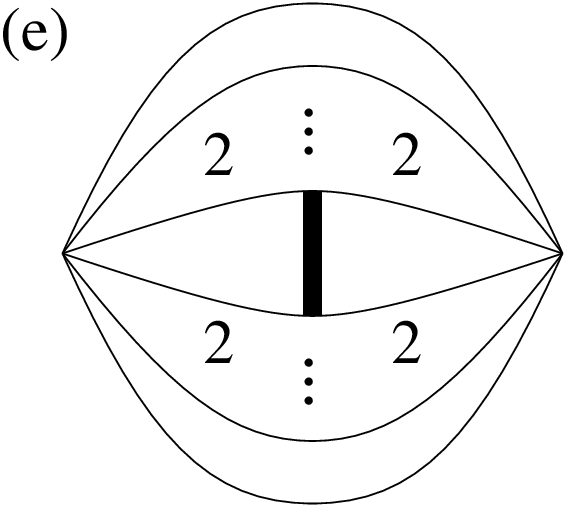, angle=-90} 
\epsfig{width=0.7in, file=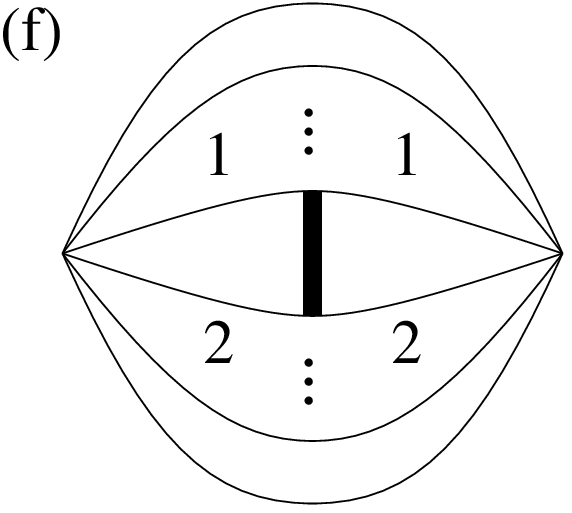, angle=-90} 
\end{center}}
\vskip -0.2in
\caption{%
Diagrams contributing to two-point functions of scalars at order $g^2$. 
\label {fig:contributing diagrams}
}%
\end {figure}

From the Lagrangian (\ref{lagrangian:for Feynman rules}) 
we can read off the structures
for the four-scalar blocks, and the leading
correction to the propagator. 
These are shown in Figure \ref{fig:four-scalar and propagator}, 
where they are categorized according to their 
gauge group (color) index structure 
(we will use the same notation as in \cite{DFS}). 
The scalar propagator remains diagonal in both color and 
flavor indices at order $g^2$. 
Notice that the corrections proportional to $\tilde B$ 
are antisymmetric in $i$ and $j$, hence they are absent when the 
scalars in the four legs have the same flavor. 
Thus we will have to compute contributions of six types%
\footnote{
	If all scalars were of the same flavor, say 1 
	(as is the case for \half-BPS operators considered 
	in \cite{DFS}and \cite{Skiba}), we would 
	only have to consider diagrams of types (a) and (d). 
	}
(see Figure \ref{fig:contributing diagrams}). 

Most Feynman diagrams we come across are easier 
to evaluate in position space, where they factorize 
into products of free propagators and the 
blocks shown in Figure \ref{fig:four-scalar and propagator}, 
and everything except for the combinatorial 
factors out front is almost trivial. 
In momentum space, on the other hand, even the simplest 
$\OO(g^2)$ graphs contain divergent subdiagrams.

The functions $A$ and $B$ will be discussed in detail 
in Section \ref{section:gauge-dependent}.
Coordinate dependence of 
$\tilde B$ is parametrically determined %
by conformal invariance, 
$\tilde B(x,0) = \tilde a \log (x^2 \mu^2) + \tilde b$. 
The coefficients $\tilde a$ and $\tilde b$ can be found  
using, for example, differential regularization \cite{FJMV}, 
or a simpler equivalent prescription:
replace 
$1/x^2 \to 1/(x^2 + \e^2)$ for 
propagators inside 
integrals ($\e \sim \mu^{-1}$ is related to the renormalization scale). 
With this, 
\begin{eqnarray}
\tilde B (x,0) &=& 
\label{b-tilde}
- \quarter Y^2 
\int { (d^4 z) \left[ 4 \pi^2 x^2 \right]^2 \over 
\left[ 4 \pi^2 ((z-x)^2 + \e^2)\right]^2 \left[ 4 \pi^2 (z^2 + \e^2)\right]^2}
\nonumber\\&=& 
- Y^2 
{1 \over 32 \pi^2 } 
\left[ \log (x^2/\e^2) - 1\right] 
\end{eqnarray}
(for \NN=4 SUSY, $Y^2 = 2 g^2$); 
The same result is obtained in dimensional regularization.

\section{The simplest cases} 
\label{section:simplest}

We begin by considering scalar composite operators 
in representations $[p,q,p]$ of the color $SU(4)$, 
which have $2p+q = $ 4 and 5. 

The case of $\Delta = 4 + \OO(g^2)$ has been 
studied before. 
For example, the authors of \cite {BKRS}
argued that there are two operators%
\footnote{
	For the notation and definitions, see Section \ref{2-0-2}. 
	}
$\OO^{[2,0,2]}_1$ and $\OO^{[2,0,2]}_2$, 
which are made of four scalars and 
annihilated by four supercharges. 
$\OO^{[2,0,2]}_1$ is a descendant 
of the Konishi scalar $\left( \sum_{I=1}^6 \tr \phi^I \phi^I \right)$
and therefore is pure 
(i.e. is an eigenstate of the dilatation operator), 
since the Konishi scalar is pure. 
The other operator, $\OO^{[2,0,2]}_2$, 
contains 
a piece proportional to $\OO^{[2,0,2]}_1$, 
but the rest is
a chiral primary. The method in \cite{BKRS} was to analyze four-point
correlators of certain \half-BPS operators, and to look at 
the possible operators in exchange channels. 
They found that there is a \quarter-BPS operator exchanged
by demonstrating that there is a pole 
corresponding to an operator of 
scaling dimension $\Delta = 4$. 
They determined this operator to be 
$\YY_{[2,0,2]} = \OO^{[2,0,2]}_2 - {4\over N} \OO^{[2,0,2]}_1$. 

Unfortunately, this method does not generalize to 
chiral primaries with scaling dimension $\Delta \ge 6$, 
as we shall see in Section \ref{6 and higher}. 
So instead we explicitly compute two-point functions
of scalar composite operators of a 
given scaling dimension, and find the ones
which do not get corrected. 
This allows us to fix the normalization 
of \quarter-BPS operators as well.

\subsection{Scalar composites with weight [2,0,2]}
\label{2-0-2}

The simplest operators annihilated by 
four out of sixteen Poincar\'e supercharges 
correspond to the highest weight state of the 
{\bf 84} = [2,0,2] of $SU(4)$. 
The $SO(6)$ 
Young tableau for representation is 
$
\mbox{
\setlength{\unitlength}{.5em}
\begin{picture}(1.7,1.5)
\put(-.8,0.5){\framebox (1,1){}}
\put(0.2,0.5){\framebox (1,1){}}
\put(-.8,-.5){\framebox (1,1){}}
\put(0.2,-.5){\framebox (1,1){}}
\end{picture}}
$. 
An $SO(6)$ irreducible tensor $T$ 
with this symmetry is made from the corresponding 
$Gl(6)$ irreducible tensor $T^0$ by subtracting 
all possible $SO(6)$ traces: 
\begin{eqnarray}
\label{202:representation} 
T_{\mbox{\setlength{\unitlength}{.7em}
\begin{picture}(2,2.2)
\put(-.3,1){\framebox (1,1){\scriptsize $a$}}
\put(0.7,1){\framebox (1,1){\scriptsize $b$}}
\put(-.3,0){\framebox (1,1){\scriptsize $c$}}
\put(0.7,0){\framebox (1,1){\scriptsize $d$}}
\end{picture}}}
&=& 
T^0_{\mbox{\setlength{\unitlength}{.7em}
\begin{picture}(2,2.2)
\put(-.3,1){\framebox (1,1){\scriptsize $a$}}
\put(0.7,1){\framebox (1,1){\scriptsize $b$}}
\put(-.3,0){\framebox (1,1){\scriptsize $c$}}
\put(0.7,0){\framebox (1,1){\scriptsize $d$}}
\end{picture}}}
- {1\over4} \Big( 
T^0_{\mbox{\setlength{\unitlength}{.7em}
\begin{picture}(2,2.2)
\put(-.3,1){\framebox (1,1){\scriptsize $\bullet$}}
\put(0.7,1){\framebox (1,1){\scriptsize $\bullet$}}
\put(-.3,0){\framebox (1,1){\scriptsize $a$}}
\put(0.7,0){\framebox (1,1){\scriptsize $b$}}
\put(-.3,1){\framebox (2,1){\scriptsize $-$}}
\end{picture}}} \delta_{cd} + 
T^0_{\mbox{\setlength{\unitlength}{.7em}
\begin{picture}(2,2.2)
\put(-.3,1){\framebox (1,1){\scriptsize $\bullet$}}
\put(0.7,1){\framebox (1,1){\scriptsize $\bullet$}}
\put(-.3,0){\framebox (1,1){\scriptsize $c$}}
\put(0.7,0){\framebox (1,1){\scriptsize $d$}}
\put(-.3,1){\framebox (2,1){\scriptsize $-$}}
\end{picture}}} \delta_{ab} + 
T^0_{\mbox{\setlength{\unitlength}{.7em}
\begin{picture}(2,2.2)
\put(-.3,1){\framebox (1,1){\scriptsize $\bullet$}}
\put(0.7,1){\framebox (1,1){\scriptsize $\bullet$}}
\put(-.3,0){\framebox (1,1){\scriptsize $a$}}
\put(0.7,0){\framebox (1,1){\scriptsize $d$}}
\put(-.3,1){\framebox (2,1){\scriptsize $-$}}
\end{picture}}} 
\delta_{bc} + T^0_{\mbox{\setlength{\unitlength}{.7em}
\begin{picture}(2,2.2)
\put(-.3,1){\framebox (1,1){\scriptsize $\bullet$}}
\put(0.7,1){\framebox (1,1){\scriptsize $\bullet$}}
\put(-.3,0){\framebox (1,1){\scriptsize $b$}}
\put(0.7,0){\framebox (1,1){\scriptsize $c$}}
\put(-.3,1){\framebox (2,1){\scriptsize $-$}}
\end{picture}}} \delta_{ad} 
\Big)
\nonumber\\ 
&& \hspace{2.5em} 
+ {1\over20} 
\left( 
\delta_{ab} \delta_{cd} - \delta_{ad} \delta_{bc} 
\right) T^0_{\mbox{\setlength{\unitlength}{.7em}
\begin{picture}(2,2.2)
\put(-.3,1){\framebox (1,1){\scriptsize $\bullet$}}
\put(0.7,1){\framebox (1,1){\scriptsize $\bullet$}}
\put(-.3,0){\framebox (1,1){\scriptsize $\bullet$}}
\put(0.7,0){\framebox (1,1){\scriptsize $\bullet$}}
\put(-.3,1){\framebox (2,1){\scriptsize $-$}}
\put(-.3,0){\framebox (2,1){\scriptsize $-$}}
\end{picture}}} 
\end{eqnarray}
where 
$
\mbox{
\setlength{\unitlength}{.7em}
\begin{picture}(1.7,1)
\put(-.8,0){\framebox (1,1){\scriptsize $\bullet$}}
\put(0.2,0){\framebox (1,1){\scriptsize $\bullet$}}
\put(-.8,0){\framebox (2,1){\scriptsize $-$}}
\end{picture}}
= \sum_{a=1}^6 \!\!\!\! 
\mbox{
\setlength{\unitlength}{.7em}
\begin{picture}(2.5,1)
\put(0,0){\framebox (1,1){\scriptsize $a$}}
\put(1,0){\framebox (1,1){\scriptsize $a$}}
\end{picture}}
$. 
Recall that the $SU(4) \to SU(3)\times U(1)$ projection 
(\ref{projection:def}) is realized on the elementary fields 
as $\phi_a = {1\over\sqrt{2}}(z_a + \bar z_a)$, 
$\phi_{a+3} = {1\over i \sqrt{2}}(z_a - \bar z_a)$, $a=1,2,3$. 
Under this ``3+1 split,'' 
the highest weight state of [2,0,2] becomes 
\begin{eqnarray}
\label{202:highest weight} 
T_{\mbox{\setlength{\unitlength}{.7em}
\begin{picture}(2,2.2)
\put(-.3,1){\framebox (1,1){\scriptsize $1$}}
\put(0.7,1){\framebox (1,1){\scriptsize $1$}}
\put(-.3,0){\framebox (1,1){\scriptsize $2$}}
\put(0.7,0){\framebox (1,1){\scriptsize $2$}}
\end{picture}}}
&=&
{1\over4} \Big(
T_{\mbox{\setlength{\unitlength}{.7em}
\begin{picture}(2,2.2)
\put(-.3,1){\framebox (1,1){\tiny $\bf 1$}}
\put(0.7,1){\framebox (1,1){\tiny $\bf 1$}}
\put(-.3,0){\framebox (1,1){\tiny $\bf 2$}}
\put(0.7,0){\framebox (1,1){\tiny $\bf 2$}}
\end{picture}}}
+ 2 
T_{\mbox{\setlength{\unitlength}{.7em}
\begin{picture}(2,2.2)
\put(-.3,1){\framebox (1,1){\tiny $\bf \bar 1$}}
\put(0.7,1){\framebox (1,1){\tiny $\bf 1$}}
\put(-.3,0){\framebox (1,1){\tiny $\bf 2$}}
\put(0.7,0){\framebox (1,1){\tiny $\bf 2$}}
\end{picture}}}
+ 2 
T_{\mbox{\setlength{\unitlength}{.7em}
\begin{picture}(2,2.2)
\put(-.3,1){\framebox (1,1){\tiny $\bf 1$}}
\put(0.7,1){\framebox (1,1){\tiny $\bf 1$}}
\put(-.3,0){\framebox (1,1){\tiny $\bf \bar 2$}}
\put(0.7,0){\framebox (1,1){\tiny $\bf 2$}}
\end{picture}}}
+ 2 
T_{\mbox{\setlength{\unitlength}{.7em}
\begin{picture}(2,2.2)
\put(-.3,1){\framebox (1,1){\tiny $\bf \bar 1$}}
\put(0.7,1){\framebox (1,1){\tiny $\bf 1$}}
\put(-.3,0){\framebox (1,1){\tiny $\bf \bar 2$}}
\put(0.7,0){\framebox (1,1){\tiny $\bf 2$}}
\end{picture}}}
+ 
T_{\mbox{\setlength{\unitlength}{.7em}
\begin{picture}(2,2.2)
\put(-.3,1){\framebox (1,1){\tiny $\bf \bar 1$}}
\put(0.7,1){\framebox (1,1){\tiny $\bf \bar 1$}}
\put(-.3,0){\framebox (1,1){\tiny $\bf 2$}}
\put(0.7,0){\framebox (1,1){\tiny $\bf 2$}}
\end{picture}}}
\Big) + \mbox{c.c.} 
\nonumber\\ 
&=& 
{1\over4} 
T^0_{\mbox{\setlength{\unitlength}{.7em}
\begin{picture}(2,2.2)
\put(-.3,1){\framebox (1,1){\tiny $\bf 1$}}
\put(0.7,1){\framebox (1,1){\tiny $\bf 1$}}
\put(-.3,0){\framebox (1,1){\tiny $\bf 2$}}
\put(0.7,0){\framebox (1,1){\tiny $\bf 2$}}
\end{picture}}}
+ \mbox{terms with lower $U(1)$ charge}
\end{eqnarray}
where in the left hand side, $1 = \phi^1$; 
and in the right hand side, 
${\bf 1} = z_1$, ${\bf \bar 1} = \bar z_1$, etc. 
We see that after this projection, we don't 
have to worry about subtracting the $SO(6)$ 
traces, if we are only interested in the 
highest $U(1)$ charge operators. 
Henceforth, we will consider operators made by 
applying the Young (anti)symmetrizer corresponding 
to this tableau, to the string of $z_i$-s.%
\footnote{
	Computing two point functions of operators 
	in representations of $SU(3)\times U(1)$ 
	is as good as computing two point functions of the 
	original $SU(4)$ irreducible operators, 
	see Section \ref{section:superconformal group}. 
	In the remainder of the paper, 
	we will neglect the $SO(6)$ traces 
	without a comment. 
	}

We can construct one single trace and one double trace 
operators with the highest [2,0,2] weight. 
When projected onto $SU(3)\times U(1)$, they become 
\begin{eqnarray}
\OO^{[2,0,2]}_1 
&=& \tr z_1 z_1 z_2 z_2 - \tr z_1 z_2 z_1 z_2 
= 
- \half \tr [z_1 , z_2] [z_1 , z_2] 
\\ 
\OO^{[2,0,2]}_2 
&=& 2 \left( \tr z_1 z_1 ~ \tr z_2 z_2 - 
\tr z_1 z_2 ~ \tr z_1 z_2 \right) 
\end{eqnarray}
$\OO^{[2,0,2]}_1$ 
is a descendant; by consecutively applying 
four SUSY transformations,%
\footnote{
	Supersymmetry transformations are listed in 
	Appendix \ref{n=4 susy:section}, see 
	equations (\ref{SUSY-su(3) form:begin}-\ref{SUSY-su(3) form:end}). 
	}
it can be obtained from the Konishi scalar, 
$\KK_0 = \tr z_j \bar z^j$. 
More explicitly, acting with the SUSY generator 
$\bar Q_{\bar \zeta}$ 
gives 
\begin{equation}
\delta_{\bar \zeta} z_j = 0, 
\quad 
\delta_{\bar \zeta} \bar z^j = \sqrt{2} \bar\zeta \bar\psi^j, 
\quad 
\delta_{\bar \zeta} \bar\psi^j = i \e^{jkl} [z_k,z_l] \bar\zeta, 
\end{equation}
and with 
$Q_{\zeta_3}$, 
\begin{equation}
\delta_{\zeta_3} z_j = -\sqrt{2} \zeta_3 \lambda \delta_{3j}, 
\quad 
\delta_{\zeta_3} \lambda = 2 i [z_1,z_2] \zeta_3, 
\end{equation}
thus 
\begin{eqnarray}
(\bar Q_{\bar \zeta})^2 \KK_0 &=& 
\bar Q_{\bar \zeta} \tr z_j \sqrt{2} \bar\zeta \bar\psi^j = 
6 i \sqrt{2} (\bar\zeta \bar\zeta) \tr [z_1,z_2] z_3 
,
\nonumber\\
(Q_{\zeta_3})^2 (\bar Q_{\bar \zeta})^2 \KK_0 &=& 
- 12 i (\bar\zeta \bar\zeta) Q_{\zeta_3} \tr [z_1,z_2] \zeta_3 \lambda = 
24 (\bar\zeta \bar\zeta) (\zeta_3 \zeta_3) \tr [z_1,z_2]^2 
\nonumber\\
&=&
- 48 (\bar\zeta \bar\zeta) (\zeta_3 \zeta_3) \OO^{[2,0,2]}_1 
\label{konishi:descendant}
\end{eqnarray}
The four generators which annihilate $\OO^{[2,0,2]}_1$, 
are the ones we acted with to obtain it from $\KK_0$. 
However, since $\KK_0$ is a non-chiral primary, 
its descendant $\OO^{[2,0,2]}_1$ 
is not \quarter-BPS, 
despite being annihilated by 
a quarter of SUSY generators.

The free field results for two point functions of 
$\OO^{[2,0,2]}_1$ and $\OO^{[2,0,2]}_2$ are 
\begin{eqnarray}
\label{2:0:2}
\left( \matrix{
\langle \OO_1 \bar{\OO}_1 \rangle & 
\langle \OO_1 \bar{\OO}_2 \rangle \cr 
\langle \OO_2 \bar{\OO}_1 \rangle & 
\langle \OO_2 \bar{\OO}_2 \rangle  
} \right)_{\rm free} 
&=&
{3 (N^2 -1) G^4 \over 16} 
\left( \matrix{
N^2 & 4 N \cr 
4 N & 8 (N^2 - 2)  
} \right)
\end{eqnarray}
while the leading corrections are found to be 
\begin{eqnarray}
\label{2:0:2:corrections}
\left( \matrix{
\langle \OO_1 \bar{\OO}_1 \rangle & 
\langle \OO_1 \bar{\OO}_2 \rangle \cr 
\langle \OO_2 \bar{\OO}_1 \rangle & 
\langle \OO_2 \bar{\OO}_2 \rangle  
} \right)_{g^2} 
&=&
{9 (N^2 - 1) G^4 (\tilde B N) \over 16}
\left( \matrix{
N^2 & 4 N \cr 
4 N & 16 
} \right) 
; 
\end{eqnarray}
here 
$\langle \OO_i \bar{\OO}_j \rangle \equiv 
\langle \OO^{[2,0,2]}_i (x) \bar{\OO}^{[2,0,2]}_j (y) \rangle$, 
and $G \equiv G(x,y)$, $\tilde B \equiv \tilde B(x,y)$. 
Some helpful formulae we used for deriving 
(\ref{2:0:2}-\ref{2:0:2:corrections}) 
are collected in Appendix \ref{suN-identities}.

By looking at (\ref{2:0:2:corrections}), we conclude that 
neither 
$\OO^{[2,0,2]}_1$ nor $\OO^{[2,0,2]}_2$ are chiral. 
However, there is a linear combination of these two operators 
which has protected two point functions at order $g^2$. 
The operator 
\begin{eqnarray}
\YY_{[2,0,2]}(x) &\equiv& \OO^{[2,0,2]}_2(x) - 
{4\over N} \OO^{[2,0,2]}_1(x) 
\label{2:0:2:protected}
\end{eqnarray}
satisfies 
$\langle \YY \bar {\YY} \rangle = 
\langle \YY \bar {\OO}_1 \rangle = 0$, 
so $\YY_{[2,0,2]}$ is orthogonal to the 
descendant of the Konishi operator $\OO^{[2,0,2]}_1$, and 
has protected dimension $\Delta_{\tiny \YY} = 4$ at order $g^2$. 
Computationally, 
this cancellation is rather intricate: 
all $\langle \OO_i \bar{\OO}_j \rangle$ 
have very different large $N$ behavior.

We can also calculate the two point 
function of the Konishi scalar with itself: 
\begin{eqnarray}
\langle \KK_0(x) \bar{\KK}_0(y) \rangle &=& 
3 (N^2-1) [G(x,y)]^2 
\left\{
1 + 3 \tilde B(x,y) N + \OO(g^4) 
\right\}
\end{eqnarray}
so 
$
\langle \OO^{[2,0,2]}_1(x) \bar{\OO}^{[2,0,2]}_1(y) \rangle = 
\mbox{$1\over16$} N^2 [G(x,y)]^2
\langle \KK_0(x) \bar{\KK}_0(y) \rangle + \OO(g^4)$, which 
is just the free theory result. 
In particular, $\KK_0$ and its descendant $\OO^{[2,0,2]}_1$ 
have the same normalization and 
their scaling dimensions differ by 2, as they must: 
with $\tilde B(x,0)$ given by (\ref{b-tilde}), 
the scaling dimension of $\OO^{[2,0,2]}_1$ is 
$\Delta_1 = 4 + {3 g^2 N \over 16 \pi^2} + \OO(g^4)$, 
and that of $\KK_0$ is 
$\Delta_{\tiny \KK} = 2 + {3 g^2 N \over 16 \pi^2} + \OO(g^4)$, 
in agreement with 
\cite{BKRS}.
The mixture 
$\OO^{[2,0,2]}_2 = \YY_{[2,0,2]} + {4\over N} \OO^{[2,0,2]}_1$
has the two-point function with itself which 
breaks down into two pieces, 
\begin{eqnarray}
\langle \OO^{[2,0,2]}_2(x) \bar{\OO}^{[2,0,2]}_2(y) \rangle 
&=& 
\langle \YY_{[2,0,2]}(x) \bar{\YY}_{[2,0,2]}(y) \rangle +
{16\over N^2} \, 
\langle \OO^{[2,0,2]}_1(x) \bar{\OO}^{[2,0,2]}_1(y) \rangle  
\nonumber\\
&=&
{C_{\tiny \YY} \over (x-y)^{2 \Delta_{\tiny \YY}}} + 
{16\over N^2} {C_1 \over (x-y)^{2 \Delta_1}} 
\nonumber
\end{eqnarray}
Logarithmic corrections to 
$\langle \OO^{[2,0,2]}_2(x) \bar{\OO}^{[2,0,2]}_2(y) \rangle$
are due entirely to the second term 
${16\over N^2} \langle \OO^{[2,0,2]}_1(x) \bar{\OO}^{[2,0,2]}_1(y) \rangle$.

$\OO(g^2)$ corrections to all two-point functions just considered 
are proportional to $\tilde B$. 
In other words, only the contributions due to diagrams of 
type (c) in Figure \ref{fig:contributing diagrams} 
survive, and all other corrections cancel.
As we will show in 
Section \ref{section:gauge-dependent}, 
this is a general phenomenon. 
Also, in the large $N$ limit 
the fraction 
$\langle \OO_1 \bar{\OO}_2 \rangle / \langle \OO_1 \bar{\OO}_1 \rangle$ 
is suppressed; 
it vanishes in the limit $N\rightarrow\infty$, $g^2 N$ fixed.%
\footnote{
	The large $N$ limit will be analyzed in more 
	detail in Section \ref{section:large N}. 
	}

One can show that conformal dimension of $\YY_{[2,0,2]}$ is protected 
perturbatively (to order $g^4$) 
and nonperturbatively (for any instanton number) 
as well, see \cite{BKRS}. 
Non-renormalization of scaling dimension of $\YY$ hints at its 
BPS property. 
Following the authors of \cite{BKRS}, 
we suggest that it is 
indeed a \quarter-BPS chiral primary  operator.

\subsection{Scalar composites with weight [2,1,2]}
\label{2-1-2}

The story is similar in the case of the {\bf 300} = [2,1,2] of $SU(4)$.
The only scalar composite operators corresponding to the highest 
weight of this representation are  
(after the projection onto $SU(3)\times U(1)$, 
as discussed in Sections \ref{section:superconformal group} and \ref{2-0-2}) 
\begin{eqnarray}
\OO^{[2,1,2]}_1 &=& \tr z_1 z_1 z_1 z_2 z_2 - \tr z_1 z_1 z_2 z_1 z_2 
= 
- \half \tr [z_1 , z_2] [z_1^2 , z_2] 
\\ 
\OO^{[2,1,2]}_2 &=& \tr z_1 z_1 z_1 ~ \tr z_2 z_2 - 
2 \tr z_1 z_2 ~ \tr z_1 z_2 z_1 + \tr z_1 z_2 z_2 ~ \tr z_1 z_1 
\end{eqnarray}
$\OO^{[2,1,2]}_1$ is a descendant; 
$(Q_{\zeta_3})^2 (\bar Q_{\bar \zeta})^2 
\str z_1 z_i \bar z^i 
\propto \OO^{[2,1,2]}_1$
just like in Equation (\ref{konishi:descendant}).
Born level and order $g^2$ two point functions are 
\begin{eqnarray}
\label{2:1:2-begin}
\left( \matrix{
\langle \OO_1 \bar{\OO}_1 \rangle & 
\langle \OO_1 \bar{\OO}_2 \rangle \cr 
\langle \OO_2 \bar{\OO}_1 \rangle & 
\langle \OO_2 \bar{\OO}_2 \rangle 
} \right) 
&=&
{(N^2 -1) (N^2 -4) G^5 \over 16 N} 
\left\{
\left( \matrix{
N^2 & 6 N \cr 
6 N & 6 (N^2 -3)  
} \right)
\right. \nonumber\\ && \left. \quad\quad\quad\quad\quad\quad
+ ~
4 \tilde B N  
\left( \matrix{
N^2 & 6 N \cr 
6 N & 36 
} \right)
+ \OO(g^4) 
\right\}
\quad\quad
\label{2:1:2-end}
\end{eqnarray}
where 
$\langle \OO_i \bar{\OO}_j \rangle \equiv 
\langle \OO^{[2,1,2]}_i (x) \bar{\OO}^{[2,1,2]}_j (y) \rangle$, 
and $G \equiv G(x,y)$, $\tilde B \equiv \tilde B(x,y)$ as before. 
(Note that corrections proportional to $A$ and $B$ cancel again.) 
There is a linear combination of $\OO^{[2,1,2]}_1$ and $\OO^{[2,1,2]}_2$
\begin{eqnarray}
\label{2:1:2:o-hat}
\YY_{[2,1,2]}(x) &\equiv& \OO^{[2,1,2]}_2(x) - {6\over N} \OO^{[2,1,2]}_1(x) 
\end{eqnarray}
whose two point functions with arbitrary operators 
do not receive perturbative order $g^2$ corrections. 
Again, it seems reasonable to conclude that $\YY_{[2,1,2]}$ is a 
\quarter-BPS operator, 
as it is annihilated by four out of sixteen supercharges, 
has a protected scaling dimension $\Delta_{\tiny \YY} = 5$ 
(at order $g^2$), 
and contains no descendant pieces, 
$\langle \YY_{[2,1,2]}(x) \bar{\OO}^{[2,1,2]}_1(y) \rangle = 0$.

\section{A Gauge Invariance Argument}
\label{section:gauge-dependent}

In Sections \ref{2-0-2} and \ref{2-1-2}, we explicitly 
calculated the $\OO(g^2)$ corrections to 
two point functions of scalar composite operators. 
We found that corrections proportional to $A$ and $B$ cancel, 
i.e. gauge group combinatorics 
demands that diagrams containing a gauge boson exchange 
do not arise in the correlator. 
Here we give a general derivation of this fact, 
which boils down to gauge invariance of the operators in question, 
and gauge dependence of $A$ and $B$. 

The two point functions we have been considering are of the form 
\begin{equation}
\label	{two-point:D-term}
\langle 
\left[ {z}^m \right] \!(x) 
\,
\left[ {\bar z}^{m} \right] \!(y) \rangle 
\end{equation}
where $\left[ {z}^m \right] \!(x)$ is some gauge-invariant 
homogeneous polynomial (of degree $m$) in the $z_i^a$-s. 
Diagrams involving a gauge boson exchange 
which contribute to the two-point functions of the form 
(\ref{two-point:D-term}), are proportional to either $A(x,y)$ 
or $B(x,y)$, 
see 
Figure \ref{fig:four-scalar and propagator}.
By using nonrenormalization of 
the two point function 
$\langle \tr z_1 z_2 (x) \; \tr \bar z_1 \bar z_2 (y) \rangle$, 
one can immediately see \cite{DFS} that 
$B(x,y) = - 2 A(x,y)$, so these contributions add up to 
\begin{eqnarray}
\label	{two-point:D-term:gauge-inv}
\langle 
\left[ {z}^m \right](x) 
\left[ {\bar z}^{m} \right](y) \rangle_{(A+B)} 
&=& c_g A(x,y) [G(x,y)]^m 
\end{eqnarray}
where $c_g$ is some combinatorial coefficient.

Conformal invariance restricts 
$A(x,y) = a \log x^2 \mu^2 + b$. 
The constants $a$ and $b$ turn out to be gauge dependent. 
The gauge fixing parameter $\xi$ enters the expression 
for the scalar propagator as%
\footnote{
	By changing 
	$\xi$, 
	we can vary both the pole piece and the $\OO(\e^0)$ term
	(but we can't make them both zero simultaneously; 
	the $\xi$-independent part is proportional
	to a different integral). 
	Compare this with \cite{Kovacs}, where 
	order $g^2$ corrections to the scalar propagators 
	were found to vanish in super-Feynman gauge 
	of \NN=1 formulation of the theory.
	}
\begin 	{eqnarray}
\mbox{
\setlength{\unitlength}{0.7em}
\begin{picture}(4,1)
\put(0,0.4){\framebox (4,0){\Large $\bullet$}}
\end{picture}}
&=& 
\xi g^2 \mu^{4 - d} 
{1 \over p^4}
\int 
{ (d^d k) \over (2\pi)^d} 
{
[(2 p + k) \cdot k]^2
\over
k^4 (p+k)^2
}
+ (\mbox{$\xi$-independent}) 
\nonumber\\&=& 
\xi g^2 
\half \pi^{2+\e} 
{1 \over (p^2)^{1-\e} \mu^{2 \e}} 
\left[
{1\over\e} + \gamma + \OO(\e)
\right]
+ (\mbox{$\xi$-independent}) \quad\quad
\end 	{eqnarray}
in momentum space 
(in dimensional regularization \cite{Collins}; 
$\e = {d\over2}-2$ and 
$\gamma$ 
is Euler's gamma constant), 
so in position space 
\begin 	{eqnarray}
A(x,0) &=& 
\half \pi^2 g^2 \xi 
\left[ 
\log x^2 \mu^2 + \log 4\pi - \gamma 
\right]
+ (\mbox{$\xi$-independent}) 
\nonumber\\
&\equiv& a \log x^2 \mu^2 + b 
\end 	{eqnarray}
after factoring out the free propagator $G(x,0)$. 
Both 
$a$ and $b$ have pieces linear in the 
gauge fixing parameter $\xi$. 

Since a correlator of gauge invariant 
operators 
must be gauge independent, 
the combinatorial coefficient multiplying 
$A(x,y)$ in equation (\ref{two-point:D-term:gauge-inv}) 
must vanish; 
we necessarily have $c_g=0$. 
This is a general phenomenon, 
illustrated by an explicit calculation of 
Sections \ref{2-0-2} and \ref{2-1-2}: 
gauge dependent contributions 
are proportional to $2 A + B = 0$. 

In the $\OO(g^2)$ calculations of correlators of 
\half-BPS operators \cite{DFS} and \cite{Skiba}, 
there were no other contributions to 
two-point functions except for 
the ones proportional to $A$ and $B$. 
Thus, gauge invariance together with \NN=4 SUSY 
(which is needed to make $2 A + B = 0$) guarantees 
that the correlators of \cite{DFS} and \cite{Skiba} 
receive no order $g^2$ corrections.


\section{Operators of dimension 6 and higher}
\label{6 and higher}

At this point, we would like to consider 
operators made of $2p+q \ge 6$ scalar fields. 
According to the classification of \cite{AFSZ}, 
these $[p,q,p]$ operators are the candidates for 
\quarter-BPS chiral primaries. 
However, a new complication arises 
compared to the cases of $2p+q \le 5$ studied in 
Sections \ref{2-0-2} and \ref{2-1-2}. 
Now, there are many ways in which we can make 
gauge invariant combinations of fields, and hence 
many scalar composites have to be taken into account. 
Apart from single and double trace operators 
we have seen so far, 
operators made of three or more traces also
have to be considered. 

This phenomenon has a counterpart in the context 
of \half-BPS operators, see \cite{Skiba}. 
The crucial difference is 
that in our case, 
none of the scalar composites are pure, 
and only some special mixtures have a well 
defined scaling dimension. 
In general, the ``naive'' 
scalar composite operators 
will have nonvanishing two point functions 
with each other, whenever this is allowed by group theory. 
Neither is it easy to extract the descendant operators. 
Unlike in the simplest cases of Section \ref{section:simplest}, 
operators containing commutators are not pure, but 
contain pieces which are descendants of different operators. 

Thus the prescription of Sections \ref{2-0-2} and \ref{2-1-2} 
(find the pure non-BPS primaries, list their descendants, 
then subtract these pieces from the candidate 
\quarter-BPS operator) no longer goes through; we need to 
do something else. 
So instead we calculate the 
two point functions of highest weight $[p,q,p]$ scalar composites 
$\OO^{[p,q,p]}_i$, 
and arrange them as%
\footnote{
	The operators we are working with are 
	after the projection onto $SU(3)\times U(1)$, 
	as discussed in Sections \ref{section:superconformal group} 
	and \ref{2-0-2}.
	The $\OO^{[p,q,p]}_i$ are made of only $z$-s and no $\bar z$-s. 
	}
\begin	{equation}
\label{eq:oo-correlator-general}
\langle \OO^{[p,q,p]}_i(x) \bar{\OO}^{[p,q,p]}_j(y) \rangle
\equiv 
[G(x,y)]^{(2p+q)} 
\left[ {\bf F}_{ij} + \tilde B(x,y) N \, {\bf G}_{ij} + \OO(g^4) \right] 
\end	{equation}
with ${\bf F}$ the matrix of combinatorial factors at free level, 
and ${\bf G}$, of order $g^2$ correction combinatorial factors. 
Note that there can be no corrections proportional to $A$ or $B$, 
as was argued in Section \ref{section:gauge-dependent}. 
Both ${\bf F}$ and ${\bf G}$ are matrices of pure numbers; 
they are still functions of $N$, but coordinate and $g^2$ 
dependence are all absorbed in 
$\tilde B$ and $[G(x,y)]^{(2p+q)}$. 
Now the problem becomes one of linear algebra: 
starting with a basis of 
$\OO^{[p,q,p]}_i$, we want to find their linear combinations 
$\YY^{[p,q,p]}_j$ that are pure operators. 
The $\YY^{[p,q,p]}_j$ have a well defined renormalized scaling dimension 
$\Delta_j = \Delta^0 + \Delta^1_j + \OO(g^4)$; 
$\Delta^0 = 2 p + q$ for all $\OO^{[p,q,p]}_i$ 
and hence for all $\YY^{[p,q,p]}_j$. 
Such operators can be chosen orthogonal 
at Born level, and so 
\begin	{equation}
\langle \YY^{[p,q,p]}_i \bar{\YY}^{[p,q,p]}_j \rangle 
= {C^{[p,q,p];0}_i \delta_{ij} \over x^{2 \Delta^0}} 
\left[ 
1 + \beta_i - \Delta^1_i \log \mu^2 x^2 + \OO(g^4)
\right] 
\end	{equation}
to order $g^2$. 
Coefficients $\Delta^1_j \sim \beta_j \sim g^2$ 
correspond to corrections of $\YY^{[p,q,p]}_j$'s scaling dimension 
and its normalization; $\beta_j$ 
depends on the 
renormalization scale $\mu$. 
To distinguish the pure operators which do receive 
corrections to their scaling dimension, we will denote them 
by $\tilde{\YY}$, and reserve the notation $\YY$ 
for the ones that 
have $\OO(g^2)$ protected two point functions.

This is a standard problem, 
analogous to finding the normal modes of small oscillations 
of a mechanical system 
(see \cite{Goldstein}, for example). 
We have to diagonalize 
a symmetric matrix ${\bf G}$ of corrections 
with respect to the symmetric positive definite matrix 
${\bf F}$ of free correlators. 
In other words, we need to find the eigenvalues of 
matrix ${\bf F}^{-1} {\bf G}$. 
If some of them 
vanish, the corresponding eigenvectors 
are operators whose two point functions 
(with themselves as well as with other operators) 
do not get order $g^2$ corrections. 
We conjecture that 
these are in fact the \quarter-BPS operators we are after.

\subsection{Scalar composites with weight [2,2,2]}
\label{section:2:2:2}

Operators with $\Delta^0 = 6$ are the lowest dimension 
operators which illustrate the new phenomenon. 
Corresponding to the highest weight of the {\bf 729} = [2,2,2] 
of $SU(4)$, 
there are five linearly independent%
\footnote{
	For $N \le 4$, the number of independent 
	gauge invariant operators is smaller.
	}
operators: 
\begin	{eqnarray}
\noalign{\noindent
one single trace operator
}
\OO^{[2,2,2]}_1 &\equiv& 
\tr z_1 z_1 z_1 z_1 z_2 z_2 
- \mbox{$2\over3$} \, \tr z_1 z_1 z_1 z_2 z_1 z_2 
- \mbox{$1\over3$} \, \tr z_1 z_1 z_2 z_1 z_1 z_2 
\quad\quad
\\
\noalign{\noindent
three double trace operators
}
\OO^{[2,2,2]}_2 &\equiv& 
\tr z_1 z_1 z_1 z_1 ~\tr z_2 z_2 
- 2 \, \tr z_1 z_1 z_1 z_2 ~\tr z_1 z_2 
\nonumber\\ && \quad\quad\quad\quad
+ \mbox{$1\over3$} 
\left( 
2 \, \tr z_1 z_1 z_2 z_2 + \tr z_1 z_2 z_1 z_2 
\right) ~\tr z_1 z_1 
\\ 
\OO^{[2,2,2]}_3 &\equiv& 
\tr z_1 z_1 z_1  ~\tr z_1 z_2 z_2 
- \tr z_1 z_1 z_2 ~\tr z_1 z_1 z_2 
\\ 
\OO^{[2,2,2]}_4 &\equiv& 
\half \tr z_1 z_1 z_1 z_1 ~\tr z_2 z_2 
- \tr z_1 z_1 z_1 z_2 ~\tr z_1 z_2 
\nonumber\\ && \quad\quad\quad\quad
+ \half
\left( 
4 \, \tr z_1 z_1 z_2 z_2 - 3 \, \tr z_1 z_2 z_1 z_2 
\right) ~\tr z_1 z_1 
\\
\noalign{\noindent
and one triple trace operator 
}
\OO^{[2,2,2]}_5 &\equiv& 
\tr z_1 z_1 
\left( 
\tr z_1 z_1 ~\tr z_2 z_2 - 
\tr z_1 z_2 ~\tr z_1 z_2
\right) 
\end	{eqnarray}
The 
$\OO^{[2,2,2]}_i$ are constructed by applying the proper
Young operator to all possible gauge invariant combinations 
of $(z_1)^4 (z_2)^2$. 
The Young operator corresponds to the 
tableau 
$
\mbox{
\setlength{\unitlength}{0.5em}
\begin{picture}(4,1.5)
\put(-.5,-.5){\framebox (2,2){}}
\put(-.5,-.5){\framebox (1,2){}}
\put(-.5,.5){\framebox (4,1){}}
\put(-.5,.5){\framebox (3,1){}}
\end{picture}}$,
while gauge 
invariant combinations amount to 
grouping the $z_i$-s into traces.%
\footnote{
	We will give a more explicit discussion about how 
	the fields are grouped into traces, 
	when we talk about [2,3,2] operators in 
	Section \ref{section:2:3:2} 
	(equations 
	\ref{2-3-2:partitions:begin}-\ref{2-3-2:partitions:end}). 
	}
%
None 
of the $\OO^{[2,2,2]}_{1, ... , 5}$ 
have a well defined scaling dimension. 

So we calculate explicitly 
the $\half \cdot 5 \cdot (5+1) = 15$ 
two point functions of [2,2,2] operators,  
and arrange them as in Equation (\ref{eq:oo-correlator-general}). 
We calculate the matrix ${\bf F}$ of free correlator 
combinatorial factors; and the matrix 
${\bf G}$, of order $g^2$ correction combinatorial factors 
(the expressions are not transparent so we 
list them in Appendix \ref{app:2:2:2}). 
It turns out that two eigenvalues of ${\bf F}^{-1} {\bf G}$ 
vanish, signaling \quarter-BPS operators; the other three 
eigenvalues satisfy a cubic and so can be easily computed for 
general $N$. 
The two linear combinations of operators which satisfy 
$\langle \YY^{[2,2,2]}_j \bar{\OO}^{[2,2,2]}_i \rangle = 0$ for all $i$, are 
\begin	{eqnarray}
\label	{eq:2-2-2:y1}
\YY^{[2,2,2]}_1 = 
-{\frac{8 N}{\left( N^2 - 4 \right) }} \OO^{[2,2,2]}_1 + 
\OO^{[2,2,2]}_2 + 
{\frac{8}{3 \left( N^2 - 4 \right) }} 
\left( 2 \OO^{[2,2,2]}_3 + \OO^{[2,2,2]}_4 \right) 
\end	{eqnarray}
and the one orthogonal to it 
(in the sense that 
$\langle \YY^{[2,2,2]}_1(x) \bar{\YY}^{[2,2,2]}_2(y) \rangle = 0$) 
\begin	{eqnarray}
\label	{eq:2-2-2:y2}
\YY^{[2,2,2]}_2 &\!\!\!\!=\!\!\!\!& 
{144 \left( N^2 - 4 \right) \left( {N^2} -2 \right) 
\over 3 {N^6} - 47 {N^4} + 248 {N^2} -192 } \OO^{[2,2,2]}_1 
-
{3 N \left( {N^2} -7 \right) \left( 3 {N^2} + 8 \right) 
\over 3 {N^6} - 47 {N^4} + 248 {N^2} -192 } \OO^{[2,2,2]}_2 
\hspace{-3em}
\nonumber\\ &&
-
{2 N \left( 3 {N^4} - 23 {N^2} + 104 \right) 
\over 3 {N^6} - 47 {N^4} + 248 {N^2} -192 } \left( 2 \OO^{[2,2,2]}_3 + \OO^{[2,2,2]}_4 \right)
+ 
\OO^{[2,2,2]}_5 
\end	{eqnarray}
These are the only candidates for \quarter-BPS operators in 
representation [2,2,2]. 
We see that $\YY^{[2,2,2]}_1$ is constructed out 
of double and single trace operators, while 
$\YY^{[2,2,2]}_2$ is made up of everything that is available. 
Both 
(\ref{eq:2-2-2:y1}) and (\ref{eq:2-2-2:y2})
are exact in $N$ and are not large $N$ approximations.

The remaining three pure operators involve mixtures of all 
the triple, double, and single trace operators $\OO^{[2,2,2]}_{1, ... , 5}$. 
The coefficients are nondescript irrational (unlike for 
$\YY^{[2,2,2]}_1$ and $\YY^{[2,2,2]}_2$) functions of $N$. 
In the large $N$ limit, 
another operator appears which has a vanishing correction 
to its two point functions with all $\OO_i$-s. 

\subsection{Scalar composites with weight [3,1,3]}
\label{section:3:1:3}

Two $[p,q,p]$ representations have $2p+q=7$. 
These are [3,1,3] = ${\bf 960}$ and [2,3,2] = ${\bf 1470}$. 
In the first case, the 
scalar composite operators are 
\begin	{eqnarray}
\noalign{\noindent
one single trace 
}
\OO^{[3,1,3]}_1 &\equiv& 
\third \tr z_1 z_1 z_1 z_1 z_2 z_2 z_2 
-\half \tr z_1 z_1 z_1 z_2 z_1 z_2 z_2 
-\half \tr z_1 z_1 z_1 z_2 z_2 z_1 z_2 
\hspace{-2em}\nonumber\\&&
+ 
\third \tr z_1 z_1 z_2 z_1 z_2 z_1 z_2 + 
\third \tr z_1 z_1 z_2 z_2 z_1 z_1 z_2 
\quad\quad
\\
\noalign{\noindent
two double trace operators
}
\OO^{[3,1,3]}_2 &\equiv& 
\tr z_1 z_1 z_1 z_1 ~\tr z_2 z_2 z_2 
- 3 \, \tr z_1 z_1 z_1 z_2 ~\tr z_1 z_2 z_2 
\\ && 
+ 
\left( 
2 \, \tr z_1 z_1 z_2 z_2 + \tr z_1 z_2 z_1 z_2 
\right) ~\tr z_1 z_1 z_2 
- \tr z_1 z_2 z_2 z_2 ~\tr z_1 z_1 z_1 
\nonumber\\ 
\OO^{[3,1,3]}_3 &\equiv& 
- \left( 
\tr z_1 z_1 z_2 z_2 z_2 - \tr z_1 z_2 z_1 z_2 z_2 
\right) ~\tr z_1 z_1  
\nonumber\\ && 
+ 
\left( 
\tr z_1 z_1 z_1 z_2 z_2 - \tr z_1 z_1 z_2 z_1 z_2 
\right) ~\tr z_1 z_2 
\\
\noalign{\noindent
and one triple trace operator 
}
\OO^{[3,1,3]}_4 &\equiv& 
\tr z_1 z_2 
\left( 
2 \, \tr z_1 z_2 ~\tr z_1 z_2 z_2 - 
\tr z_2 z_2 ~\tr z_1 z_1z_1 
- 3 \tr z_1 z_1 ~\tr z_1 z_2z_2 
\right) 
\hspace{-2em}\nonumber\\&&
+
\tr z_1 z_1 
\left( 
\tr z_2 z_2 ~\tr z_1 z_1 z_2 + 
\tr z_1 z_1 ~\tr z_2 z_2z_2
\right) 
\end	{eqnarray}
	The operators are obtained in the same way as before:
	the Young operator corresponds to the diagram
	$
	\mbox{
	\setlength{\unitlength}{0.5em}
	\begin{picture}(4,1.5)
	\put(-.5,-.5){\framebox (3,2){}}
	\put(-.5,-.5){\framebox (2,2){}}
	\put(-.5,-.5){\framebox (1,2){}}
	\put(-.5,.5){\framebox (4,1){}}
	\end{picture}}$
	of $SO(6)$,
	and the partitions are 7=7, 4+3, 5+2, and 2+2+3. 
	Operators resulting from other partitions turn out 
	to be linear combinations of the $\OO^{[3,1,3]}_i$ above. 
	We see that even the number of linearly independent 
	scalar composites for a given Dynkin label is a 
	complicated function of the scaling dimension $\Delta$. 
Like in the [2,2,2] case, 
none of the $\OO^{[3,1,3]}_i$ are pure. 
$\OO^{[3,1,3]}_1$ and $\OO^{[3,1,3]}_3$ contain commutators, 
and are likely to be linear combinations of descendants of 
non-BPS primaries. 

The matrix ${\bf F}^{-1} {\bf G}$ is now $4 \times 4$; it has 
two zero eigenvalues, while the other two satisfy a quadratic 
equation. 
The formulae are more tractable than in the [2,2,2] case,
so we present some of the details here. 
We find 
\begin	{equation} \!
\small
{\frac{768}{5 {N^3} \left( {N^2} -1\right) \left( {N^2} -4 \right) }}
\, {\bf F} 
= 
\left( 
\matrix{ {\frac{{N^2} + 3}{{N^2}}} & {\frac{12}{N}} & -{\frac{12}{N}} & {\frac{72}
    {{N^2}}} \cr {\frac{12}{N}} & {\frac{36 \left( {N^4} - 8 {N^2} + 18 \right) }{{N^4}}} & 
    -{\frac{108}{{N^2}}} & {\frac{72 \left( 2 {N^2} -9\right) }{{N^3}}} \cr {\frac{-12}
    {N}} & -{\frac{108}{{N^2}}} & {\frac{36 \left( {N^2} + 6 \right) }{5 {N^2}}} & -{\frac{72}
    {N}} \cr {\frac{72}{{N^2}}} & {\frac{72 \left( 2 {N^2} -9 \right) }{{N^3}}} & -{\frac{72}
    {N}} & {\frac{72 \left( {N^2} -3 \right) }{{N^2}}} }
\right)
\end	{equation}
for the matrix of free combinatorial factors, and 
\begin{eqnarray}
\small
{\frac{128}{25 {N^3} \left( {N^2} -1 \right) \left( {N^2} -4 \right) }}
\, {\bf G} 
\hspace{-5em} && \nonumber\\ 
&=& 
\small
\left( 
\matrix{ {\frac{{N^2} +7 }{{N^2}}} & {\frac{12 \left( {N^2} +3 \right) }{{N^3}}} & -{\frac{72 
      \left( {N^2}+1 \right) }{5 {N^3}}} & {\frac{96}{{N^2}}} \cr {\frac{12 
      \left( {N^2}+3 \right) }{{N^3}}} & {\frac{144}{{N^2}}} & -{\frac{144}{{N^2}}} & {\frac{864}
    {{N^3}}} \cr -{\frac{72 \left( {N^2}+1 \right) }{5 {N^3}}} & -{\frac{144}{{N^2}}} & {
     \frac{144 \left( {N^2}+16 \right) }{25 {N^2}}} & 
-{\frac{288 \left( {N^2} +6\right) }
    {5 {N^3}}} \cr {\frac{96}{{N^2}}} & {\frac{864}{{N^3}}} & 
-{\frac{288 \left( {N^2} + 6 \right) }{5 {N^3}}} & {\frac{576}{{N^2}}} }
\right) 
\nonumber\\
\end	{eqnarray}
for the matrix of corrections proportional to $\tilde B(x,y) N$.

The vectors killed by ${\bf F}^{-1} {\bf G}$ 
work out to be 
\begin	{eqnarray}
\label	{eq:3-1-3:y1}
\YY^{[3,1,3]}_1 &=& 
- {\frac{12 N}{{N^2}-2}} \, \OO^{[3,1,3]}_1 + \OO^{[3,1,3]}_2 
- {\frac{5}{{N^2}-2}} \, \OO^{[3,1,3]}_3
\\
\YY^{[3,1,3]}_2 &=& 
{\frac{96}{{N^2}-4}} \OO^{[3,1,3]}_1 
- {\frac{4 N}{{N^2}-4}} \OO^{[3,1,3]}_2 + 
{\frac{10 N}{{N^2}-4}} \OO^{[3,1,3]}_3 + \OO^{[3,1,3]}_4
\quad
\end{eqnarray}
They correspond to 
zero eigenvalues of ${\bf F}^{-1} {\bf G}$, 
and so are the candidates for \quarter-BPS 
primaries in the [3,1,3].%
\footnote{
	We chose $\YY^{[3,1,3]}_2$ to be orthogonal 
	to $\YY^{[3,1,3]}_1$, 
	in the sense that 
	$\langle \YY^{[3,1,3]}_2 \bar{\YY}^{[3,1,3]}_1\rangle = 0$. 
	}
The remaining eigenvectors of ${\bf F}^{-1} {\bf G}$ are 
\begin{eqnarray}
\tilde{\YY}^{[3,1,3]}_3 &=& 
\OO^{[3,1,3]}_1 
- {\frac{10}{3 \left( N + {\sqrt{ N^2 + 160 }} \right) }} \OO^{[3,1,3]}_3 
\\
\tilde{\YY}^{[3,1,3]}_4 &=& 
\OO^{[3,1,3]}_3 
- {\frac{3 \left( N - {\sqrt{ N^2 + 160 }} \right) }{10}} \OO^{[3,1,3]}_1 
\label	{eq:3-1-3:y4}
\end	{eqnarray}
corresponding to the eigenvalues 
$27 + {\frac{3 {\sqrt{160 + {N^2}}}}{N}}$ 
for $\tilde{\YY}^{[3,1,3]}_3$, 
and $27 - {\frac{3 {\sqrt{160 + {N^2}}}}{N}}$ 
for $\tilde{\YY}^{[3,1,3]}_4$. 
Expressions 
(\ref{eq:3-1-3:y1}-\ref{eq:3-1-3:y4})
are exact in $N$, 
and are not just large $N$ approximations. 
As expected, the descendants 
$\tilde{\YY}$ 
are mixtures of operators involving commutators. 

	Note that both the $g^2$ corrections to the scaling 
	dimension of $\tilde{\YY}$ 
	and their expansion 
	coefficients involve radicals (so there is really no way to ``guess'' 
	the pure primaries such operators came from). 
	Also, radiative corrections to all $\Delta$-s are non-negative, 
	since at free field level the $\OO^{[3,1,3]}_i$ are annihilated by 
	a quarter of the supercharges, and hence saturate the BPS 
	bound. 


\subsection{Scalar composites with weight [2,3,2]}
\label{section:2:3:2}

The other $[p,q,p]$ of $SU(4)$ with $2p+q=7$ is the 
[2,3,2] = ${\bf 1470}$. 
Here we have seven linearly independent operators 
corresponding to the highest weight state: 
\begin	{eqnarray}
\noalign{\noindent
one single trace operator 
}
\label{2-3-2:operators:begin}
\OO^{[2,3,2]}_1 &\equiv& 
2 \, \tr z_1 z_1 z_1 z_1 z_1 z_2 z_2 
- \tr z_1 z_1 z_1 z_1 z_2 z_1 z_2 
- \tr z_1 z_1 z_1 z_2 z_1 z_1 z_2 
\quad\quad
\\
\noalign{\noindent
four double trace operators
}
\OO^{[2,3,2]}_2 &\equiv& 
2 \, \tr z_1 z_1 z_1 z_1 z_1 ~\tr z_2 z_2 
- 4 \, \tr z_1 z_1 z_1 z_1 z_2 ~\tr z_1 z_2 
\\ && 
+ 
\left( 
\tr z_1 z_1 z_1 z_2 z_2 + \tr z_1 z_1 z_2 z_1 z_2 
\right) ~\tr z_1 z_1 
\nonumber\\ 
\OO^{[2,3,2]}_3 &\equiv& 
\tr z_1 z_1 z_1 z_1 z_1 ~\tr z_2 z_2 
- 2 \, \tr z_1 z_1 z_1 z_1 z_2 ~\tr z_1 z_2 
\\ && 
+ 
\left( 
8 \, \tr z_1 z_1 z_1 z_2 z_2 - 7 \, \tr z_1 z_1 z_2 z_1 z_2 
\right) ~\tr z_1 z_1 
\nonumber\\ 
\OO^{[2,3,2]}_4 &\equiv& 
3 \, \tr z_1 z_1 z_1 z_1 ~\tr z_1 z_2 z_2 
- 6 \, \tr z_1 z_1 z_1 z_2 ~\tr z_1 z_1 z_2 
\nonumber\\ && 
+ 
\left( 
2 \, \tr z_1 z_1 z_2 z_2 + \tr z_1 z_2 z_1 z_2 
\right) ~\tr z_1 z_1 z_1 
\\ 
\OO^{[2,3,2]}_5 &\equiv& 
3 \, \tr z_1 z_1 z_1 z_1 ~\tr z_1 z_2 z_2 
- 6 \, \tr z_1 z_1 z_1 z_2 ~\tr z_1 z_1 z_2 
\nonumber\\ && 
+ 
\left( 
7 \, \tr z_1 z_1 z_2 z_2 - 4 \, \tr z_1 z_2 z_1 z_2 
\right) ~\tr z_1 z_1 z_1 
\\
\noalign{\noindent
and two triple trace operators 
}
\OO^{[2,3,2]}_6 &\equiv& 
-8 \, \tr z_1 z_2 \tr z_1 z_2 \tr z_1 z_1 z_1 
-6 \, \tr z_1 z_1 \tr z_1 z_2 \tr z_1 z_1 z_2 
\hspace{-2em}\nonumber\\&&
+
\tr z_1 z_1 
\left( 
11 \, \tr z_2 z_2 ~\tr z_1 z_1 z_1 + 
 3 \, \tr z_1 z_1 ~\tr z_1 z_2 z_2 
\right) 
\\
\OO^{[2,3,2]}_7 &\equiv& 
7 \, \tr z_1 z_2 \tr z_1 z_2 \tr z_1 z_1 z_1 
-6 \, \tr z_1 z_1 \tr z_1 z_2 \tr z_1 z_1 z_2 
\hspace{-2em}\nonumber\\&&
+
\tr z_1 z_1 
\left( 
-4 \, \tr z_2 z_2 ~\tr z_1 z_1 z_1 + 
 3 \, \tr z_1 z_1 ~\tr z_1 z_2 z_2 
\right) 
\label{2-3-2:operators:end}
\end	{eqnarray}
More explicitly, the operators 
listed in (\ref{2-3-2:operators:begin}-\ref{2-3-2:operators:end})
are constructed as 
\begin{eqnarray}
\label{2-3-2:partitions:begin}
7=7: 
&&
\OO^{[2,3,2]}_1 \sim 
\left(\mbox{
\setlength{\unitlength}{0.7em}
\begin{picture}(6,1.7)
\put(0,0.5){\framebox (1,1){}}
\put(1,0.5){\framebox (1,1){}}
\put(2,0.5){\framebox (1,1){}}
\put(3,0.5){\framebox (1,1){}}
\put(4,0.5){\framebox (1,1){}}
\put(0,-0.5){\framebox (1,1){}}
\put(1,-0.5){\framebox (1,1){}}
\end{picture}}\right)
;
\\
7=5+2: 
&&
\OO^{[2,3,2]}_2 \sim 
\left(\mbox{
\setlength{\unitlength}{0.7em}
\begin{picture}(6,1.7)
\put(0,0.7){\framebox (1,1){}}
\put(1,0.7){\framebox (1,1){}}
\put(2,0.7){\framebox (1,1){}}
\put(3,0.7){\framebox (1,1){}}
\put(4,0.7){\framebox (1,1){}}
\put(0,-0.8){\framebox (1,1){}}
\put(1,-0.8){\framebox (1,1){}}
\end{picture}}\right)
,
\quad
\OO^{[2,3,2]}_3 \sim 
\left(\mbox{
\setlength{\unitlength}{0.7em}
\begin{picture}(6,1.7)
\put(-.3,0.5){\framebox (1,1){}}
\put(0.7,0.5){\framebox (1,1){}}
\put(1.7,0.5){\framebox (1,1){}}
\put(3.3,0.5){\framebox (1,1){}}
\put(4.3,0.5){\framebox (1,1){}}
\put(-.3,-0.5){\framebox (1,1){}}
\put(0.7,-0.5){\framebox (1,1){}}
\end{picture}}\right)
;
\\
7=4+3: 
&&
\OO^{[2,3,2]}_4 \sim 
\left(\mbox{
\setlength{\unitlength}{0.7em}
\begin{picture}(6,1.7)
\put(-.3,0.2){\framebox (1,1){}}
\put(1.3,0.7){\framebox (1,1){}}
\put(2.3,0.7){\framebox (1,1){}}
\put(3.3,0.7){\framebox (1,1){}}
\put(4.3,0.7){\framebox (1,1){}}
\put(-.3,-0.8){\framebox (1,1){}}
\put(0.7,-0.8){\framebox (1,1){}}
\end{picture}}\right)
,
\quad
\OO^{[2,3,2]}_5 \sim 
\left(\mbox{
\setlength{\unitlength}{0.7em}
\begin{picture}(6,1.7)
\put(-.3,0.5){\framebox (1,1){}}
\put(0.7,0.5){\framebox (1,1){}}
\put(2.3,0.5){\framebox (1,1){}}
\put(3.3,0.5){\framebox (1,1){}}
\put(4.3,0.5){\framebox (1,1){}}
\put(-.3,-0.5){\framebox (1,1){}}
\put(0.7,-0.5){\framebox (1,1){}}
\end{picture}}\right)
;
\\
\hspace{-2em} 
7=3+2+2: 
&&
\OO^{[2,3,2]}_6 \sim 
\left(\mbox{
\setlength{\unitlength}{0.7em}
\begin{picture}(6,1.7)
\put(-.3,0.7){\framebox (1,1){}}
\put(0.7,0.7){\framebox (1,1){}}
\put(2.3,0.7){\framebox (1,1){}}
\put(3.3,0.7){\framebox (1,1){}}
\put(4.3,0.7){\framebox (1,1){}}
\put(-.3,-0.8){\framebox (1,1){}}
\put(0.7,-0.8){\framebox (1,1){}}
\end{picture}}\right)
,
\quad
\OO^{[2,3,2]}_7 \sim 
\left(\mbox{
\setlength{\unitlength}{0.7em}
\begin{picture}(6,1.7)
\put(-0.6,0.5){\framebox (1,1){}}
\put(1,0.5){\framebox (1,1){}}
\put(2.6,0.5){\framebox (1,1){}}
\put(3.6,0.5){\framebox (1,1){}}
\put(4.6,0.5){\framebox (1,1){}}
\put(-0.6,-0.5){\framebox (1,1){}}
\put(1,-0.5){\framebox (1,1){}}
\end{picture}}\right)
\label{2-3-2:partitions:end}
\quad\quad
\end{eqnarray}
where each continuous group of boxes stands for a single trace. 
Other partitions give rise 
to operators which are linear combinations of 
the ones shown in 
(\ref{2-3-2:partitions:begin}-\ref{2-3-2:partitions:end}). 
Now it matters not only how we partition the string 
of seven letters, but exactly which letters we put in 
the groups.
For example, $\OO^{[2,3,2]}_{4,5}$  correspond to the same partition 
7=4+3, but it is important which $z_i$ appear in 
which trace before we apply the Young operator.

Again, the $\OO^{[2,3,2]}_i$ have nonzero two point functions 
with each other, none are pure, and 
it is impossible to guess the primaries which 
the descendants come from. 
Matrices 
$\langle \OO^{[2,3,2]}_i \bar{\OO}^{[2,3,2]}_j \rangle$ 
of two point functions 
are listed in Appendix \ref{app:2:3:2}. 
The operators we predict to be \quarter-BPS are: 
a combination of double and single trace operators only 
\begin	{equation}
\label	{eq:3-2-3:y1}
\YY^{[2,3,2]}_1 = 
-{\frac{10 N}{N^2-7}} \, \OO^{[2,3,2]}_1 + \OO^{[2,3,2]}_2 + 
{\frac{2}{N^2-7}} \left( \OO^{[2,3,2]}_3 + \OO^{[2,3,2]}_4 + \OO^{[2,3,2]}_5 \right) 
\end{equation}
and two linear combinations involving all types of operators 
\begin	{equation}
\YY^{[2,3,2]}_2 = -20 \, \OO^{[2,3,2]}_1 
+ {\frac{2 \left( N^2+2 \right) }{N}} \, \OO^{[2,3,2]}_2 
-{\frac{2}{N}} \left( \OO^{[2,3,2]}_3 + \OO^{[2,3,2]}_4 \right) 
+ \OO^{[2,3,2]}_6 
\end{equation}
\begin	{equation}
\YY^{[2,3,2]}_3 = 10 \, \OO^{[2,3,2]}_1 
-{\frac{\left( N^2-4 \right) }{N}} \, \OO^{[2,3,2]}_2 
-{\frac{2}{N}} \left( \OO^{[2,3,2]}_3 + \OO^{[2,3,2]}_4 \right) 
+ \OO^{[2,3,2]}_7
\end{equation}
($\YY^{[2,3,2]}_{1,2,3}$ are not orthogonal. 
Although orthonormal linear combinations are easy to find, 
they look rather messy and we don't list them here.) 
Again we emphasize that 
these expressions are exact in $N$ and 
are not large $N$ approximations. 

The remaining four pure operators involve mixtures of all 
$\OO^{[2,3,2]}_{1, ... , 7}$. 
The coefficients are again irrational functions of $N$ 
(unlike for $\YY^{[2,3,2]}_{1,2,3}$), and so are the 
eigenvalues. 
In the large $N$ limit, 
one of them has a $g^2$ correction to its scaling dimension 
which is suppressed by $N^{-2}$ --- 
another operator 
becomes \quarter-BPS in the large $N$ limit. 

\section{Summary of $\Delta \le 7$ results}
\label{section:tables}

Having carried out these explicit calculations, 
let us bring together the results for $[p,q,p]$ highest 
weight, gauge invariant, local, polynomial 
scalar composite operators we have discussed so far.  
Table \ref{table:results} below lists 
the representations $[p,q,p]$ with $2p+q \le 7$;  
the operators $\OO^{[p,q,p]}_i$ constructed by taking traces 
in various combinations; 
and the resulting pure operators $\YY^{[p,q,p]}_j$. 
We omitted the $\YY^{[p,q,p]}_j$ with corrected 
scaling dimension, and listed  
only the \quarter-BPS chiral primaries. 
The notation and definitions are spelled out in 
Sections \ref{section:simplest} and \ref{6 and higher}.

\begin{table}
\begin{tabular}{ | r | l | l | }
\hline 
$\phantom{\Big|^A}\hspace{-3.5ex}$
$[p,q,p]$ \hspace{-1ex} & 
\hspace{7ex} $\OO^{[p,q,p]}_i$ & \hspace{17ex} $\YY^{[p,q,p]}_j$ \\
\hline 
\hspace{-1ex}
$[2,0,2]$ 
\hspace{-1ex}
& 
\hspace{-3.5ex}
{\begin{tabular}{l}
$\phantom{\Big|^B}\hspace{-2ex}$
$\OO^{[2,0,2]}_1 \sim 
\left(\hspace{-1ex}\mbox{
\setlength{\unitlength}{0.7em}
\begin{picture}(2,1.7)
\put(0,0.5){\framebox (1,1){}}
\put(1,0.5){\framebox (1,1){}}
\put(0,-0.5){\framebox (1,1){}}
\put(1,-0.5){\framebox (1,1){}}
\end{picture}} \; \right)_{\phantom{{}_A}\hspace{-1ex}} 
$ 
\\
$\phantom{\Big|^B}\hspace{-2ex}$
$\OO^{[2,0,2]}_2 \sim 
\left(\hspace{-1ex}\mbox{
\setlength{\unitlength}{0.7em}
\begin{picture}(2,1.7)
\put(0,0.6){\framebox (1,1){}}
\put(1,0.6){\framebox (1,1){}}
\put(0,-0.8){\framebox (1,1){}}
\put(1,-0.8){\framebox (1,1){}}
\end{picture}} \; \right)_{\phantom{{}_A}\hspace{-1ex}} 
$ 
\\
\end{tabular}}
& 
\hspace{-2.5ex}
{\begin{tabular}{l}
$\YY^{[2,0,2]}_1 = - {4\over N} \OO^{[2,0,2]}_1 + \OO^{[2,0,2]}_2$ \\
\end{tabular}}
\\
\hline 
\hspace{-1ex}
$[2,1,2]$ 
\hspace{-1ex}
& 
\hspace{-3.5ex}
{\begin{tabular}{l}
$\phantom{\Big|^B}\hspace{-2ex}$
$\OO^{[2,1,2]}_1 \sim 
\left(\hspace{-1ex}\mbox{
\setlength{\unitlength}{0.7em}
\begin{picture}(3,1.7)
\put(0,0.5){\framebox (1,1){}}
\put(1,0.5){\framebox (1,1){}}
\put(2,0.5){\framebox (1,1){}}
\put(0,-0.5){\framebox (1,1){}}
\put(1,-0.5){\framebox (1,1){}}
\end{picture}} \; \right)_{\phantom{{}_A}\hspace{-1ex}} 
$ 
\\
$\phantom{\Big|^B}\hspace{-2ex}$
$\OO^{[2,1,2]}_2 \sim 
\left(\hspace{-1ex}\mbox{
\setlength{\unitlength}{0.7em}
\begin{picture}(3,1.7)
\put(0,0.6){\framebox (1,1){}}
\put(1,0.6){\framebox (1,1){}}
\put(2,0.6){\framebox (1,1){}}
\put(0,-0.8){\framebox (1,1){}}
\put(1,-0.8){\framebox (1,1){}}
\end{picture}} \; \right)_{\phantom{{}_A}\hspace{-1ex}} 
$ 
\\
\end{tabular}}
& 
\hspace{-2.5ex}
{\begin{tabular}{l}
$\YY^{[2,1,2]}_1 = - {6\over N} \OO^{[2,1,2]}_1 + \OO^{[2,1,2]}_2$ \\
\end{tabular}}
\\
\hline 
\hspace{-1ex}
$[2,2,2]$ 
\hspace{-1ex}
& 
\hspace{-3.5ex}
{\begin{tabular}{l}
$\phantom{\Big|^B}\hspace{-2ex}$
$\OO^{[2,2,2]}_1 \sim 
\left(\hspace{-1ex}\mbox{
\setlength{\unitlength}{0.7em}
\begin{picture}(4,1.7)
\put(0,0.5){\framebox (1,1){}}
\put(1,0.5){\framebox (1,1){}}
\put(2,0.5){\framebox (1,1){}}
\put(3,0.5){\framebox (1,1){}}
\put(0,-0.5){\framebox (1,1){}}
\put(1,-0.5){\framebox (1,1){}}
\end{picture}} \; \right)_{\phantom{{}_A}\hspace{-1ex}} 
$ 
\\
$\phantom{\Big|^B}\hspace{-2ex}$
$\OO^{[2,2,2]}_2 \sim 
\left(\hspace{-1ex}\mbox{
\setlength{\unitlength}{0.7em}
\begin{picture}(4,1.7)
\put(0,0.6){\framebox (1,1){}}
\put(1,0.6){\framebox (1,1){}}
\put(2,0.6){\framebox (1,1){}}
\put(3,0.6){\framebox (1,1){}}
\put(0,-0.8){\framebox (1,1){}}
\put(1,-0.8){\framebox (1,1){}}
\end{picture}} \; \right)_{\phantom{{}_A}\hspace{-1ex}} 
$ 
\\
$\phantom{\Big|^B}\hspace{-2ex}$
$\OO^{[2,2,2]}_3 \sim 
\left(\hspace{-1ex}\mbox{
\setlength{\unitlength}{0.7em}
\begin{picture}(4.5,1.7)
\put(0,0.5){\framebox (1,1){}}
\put(1,0.5){\framebox (1,1){}}
\put(0,-0.5){\framebox (1,1){}}
\put(2.5,0.5){\framebox (1,1){}}
\put(3.5,0.5){\framebox (1,1){}}
\put(2.5,-0.5){\framebox (1,1){}}
\end{picture}} \; \right)_{\phantom{{}_A}\hspace{-1ex}} 
$ 
\\
$\phantom{\Big|^B}\hspace{-2ex}$
$\OO^{[2,2,2]}_4 \sim 
\left(\hspace{-1ex}\mbox{
\setlength{\unitlength}{0.7em}
\begin{picture}(4.5,1.7)
\put(0,0.5){\framebox (1,1){}}
\put(1,0.5){\framebox (1,1){}}
\put(2.5,0.5){\framebox (1,1){}}
\put(3.5,0.5){\framebox (1,1){}}
\put(0,-0.5){\framebox (1,1){}}
\put(1,-0.5){\framebox (1,1){}}
\end{picture}} \; \right)_{\phantom{{}_A}\hspace{-1ex}} 
$ 
\\
$\phantom{\Big|^B}\hspace{-2ex}$
$\OO^{[2,2,2]}_5 \sim 
\left(\hspace{-1ex}\mbox{
\setlength{\unitlength}{0.7em}
\begin{picture}(4.5,1.7)
\put(0,0.6){\framebox (1,1){}}
\put(1,0.6){\framebox (1,1){}}
\put(2.5,0.6){\framebox (1,1){}}
\put(3.5,0.6){\framebox (1,1){}}
\put(0,-0.8){\framebox (1,1){}}
\put(1,-0.8){\framebox (1,1){}}
\end{picture}} \; \right)_{\phantom{{}_A}\hspace{-1ex}} 
$ 
\\
\end{tabular}}
\hspace{-2.5ex}
& 
\hspace{-2.5ex}
{\begin{tabular}{l}
$\YY^{[2,2,2]}_1 = - {8 N \over N^2-4} \OO^{[2,2,2]}_1 + \OO^{[2,2,2]}_2$ \\ 
\hspace{10ex} 
$+ {8 \over 3 (N^2-4)} \left(2 \OO^{[2,2,2]}_3 + \OO^{[2,2,2]}_4 \right)$ \\
\\
$\YY^{[2,2,2]}_2 = - {8\over N} \OO^{[2,2,2]}_1 + \OO^{[2,2,2]}_2 
+ {4\over 3 N} \OO^{[2,2,2]}_5$ 
\\
\end{tabular}}
\hspace{-3ex}
\\
\hline 
\hspace{-1ex}
$[3,1,3]$ 
\hspace{-1ex}
& 
\hspace{-3.5ex}
{\begin{tabular}{l}
$\phantom{\Big|^B}\hspace{-2ex}$
$\OO^{[3,1,3]}_1 \sim 
\left(\hspace{-1ex}\mbox{
\setlength{\unitlength}{0.7em}
\begin{picture}(4,1.7)
\put(0,0.5){\framebox (1,1){}}
\put(1,0.5){\framebox (1,1){}}
\put(2,0.5){\framebox (1,1){}}
\put(3,0.5){\framebox (1,1){}}
\put(0,-0.5){\framebox (1,1){}}
\put(1,-0.5){\framebox (1,1){}}
\put(2,-0.5){\framebox (1,1){}}
\end{picture}} \; \right)_{\phantom{{}_A}\hspace{-1ex}} 
$ 
\\
$\phantom{\Big|^B}\hspace{-2ex}$
$\OO^{[3,1,3]}_2 \sim 
\left(\hspace{-1ex}\mbox{
\setlength{\unitlength}{0.7em}
\begin{picture}(4,1.7)
\put(0,0.6){\framebox (1,1){}}
\put(1,0.6){\framebox (1,1){}}
\put(2,0.6){\framebox (1,1){}}
\put(3,0.6){\framebox (1,1){}}
\put(0,-0.8){\framebox (1,1){}}
\put(1,-0.8){\framebox (1,1){}}
\put(2,-0.8){\framebox (1,1){}}
\end{picture}} \; \right)_{\phantom{{}_A}\hspace{-1ex}} 
$ 
\\
$\phantom{\Big|^B}\hspace{-2ex}$
$\OO^{[3,1,3]}_3 \sim 
\left(\hspace{-1ex}\mbox{
\setlength{\unitlength}{0.7em}
\begin{picture}(4.5,1.7)
\put(0,0.2){\framebox (1,1){}}
\put(1,0.2){\framebox (1,1){}}
\put(0,-0.8){\framebox (1,1){}}
\put(1,-0.8){\framebox (1,1){}}
\put(2,-0.8){\framebox (1,1){}}
\put(2.5,0.6){\framebox (1,1){}}
\put(3.5,0.6){\framebox (1,1){}}
\end{picture}} \; \right)_{\phantom{{}_A}\hspace{-1ex}} 
$ 
\\
$\phantom{\Big|^B}\hspace{-2ex}$
$\OO^{[3,1,3]}_4 \sim 
\left(\hspace{-1ex}\mbox{
\setlength{\unitlength}{0.7em}
\begin{picture}(4.5,1.7)
\put(0,0.6){\framebox (1,1){}}
\put(1,0.6){\framebox (1,1){}}
\put(2.5,0.6){\framebox (1,1){}}
\put(3.5,0.6){\framebox (1,1){}}
\put(0,-0.8){\framebox (1,1){}}
\put(1,-0.8){\framebox (1,1){}}
\put(2,-0.8){\framebox (1,1){}}
\end{picture}} \; \right)_{\phantom{{}_A}\hspace{-1ex}} 
$ 
\\
\end{tabular}}
\hspace{-2.5ex}
& 
\hspace{-2.5ex}
{\begin{tabular}{l}
$\YY^{[3,1,3]}_1 = - {12 N \over N^2-2} \OO^{[3,1,3]}_1 + \OO^{[3,1,3]}_2$ \\ 
\hspace{10ex} 
$- {5 \over N^2-2} \OO^{[3,1,3]}_3 $ \\
\\
$\YY^{[3,1,3]}_2 = - {24\over N} \OO^{[3,1,3]}_1 
+ \OO^{[3,1,3]}_2 
+ {1\over 2 N} \OO^{[3,1,3]}_4$ 
\\
\end{tabular}}
\hspace{-3.3ex}
\\
\hline 
\hspace{-1ex}
$[2,3,2]$ 
\hspace{-1ex}
& 
\hspace{-3.5ex}
{\begin{tabular}{l}
$\phantom{\Big|^B}\hspace{-2ex}$
$\OO^{[2,3,2]}_1 \sim 
\left(\hspace{-1ex}\mbox{
\setlength{\unitlength}{0.7em}
\begin{picture}(5,1.7)
\put(0,0.5){\framebox (1,1){}}
\put(1,0.5){\framebox (1,1){}}
\put(2,0.5){\framebox (1,1){}}
\put(3,0.5){\framebox (1,1){}}
\put(4,0.5){\framebox (1,1){}}
\put(0,-0.5){\framebox (1,1){}}
\put(1,-0.5){\framebox (1,1){}}
\end{picture}} \; \right)_{\phantom{{}_A}\hspace{-1ex}} 
$ 
\\
$\phantom{\Big|^B}\hspace{-2ex}$
$\OO^{[2,3,2]}_2 \sim 
\left(\hspace{-1ex}\mbox{
\setlength{\unitlength}{0.7em}
\begin{picture}(5,1.7)
\put(0,0.6){\framebox (1,1){}}
\put(1,0.6){\framebox (1,1){}}
\put(2,0.6){\framebox (1,1){}}
\put(3,0.6){\framebox (1,1){}}
\put(4,0.6){\framebox (1,1){}}
\put(0,-0.8){\framebox (1,1){}}
\put(1,-0.8){\framebox (1,1){}}
\end{picture}} \; \right)_{\phantom{{}_A}\hspace{-1ex}} 
$ 
\\
$\phantom{\Big|^B}\hspace{-2ex}$
$\OO^{[2,3,2]}_3 \sim 
\left(\hspace{-1ex}\mbox{
\setlength{\unitlength}{0.7em}
\begin{picture}(5.5,1.7)
\put(0,0.5){\framebox (1,1){}}
\put(1,0.5){\framebox (1,1){}}
\put(2,0.5){\framebox (1,1){}}
\put(0,-0.5){\framebox (1,1){}}
\put(1,-0.5){\framebox (1,1){}}
\put(3.5,0.5){\framebox (1,1){}}
\put(4.5,0.5){\framebox (1,1){}}
\end{picture}} \; \right)_{\phantom{{}_A}\hspace{-1ex}} 
$ 
\\
$\phantom{\Big|^B}\hspace{-2ex}$
$\OO^{[2,3,2]}_4 \sim 
\left(\hspace{-1ex}\mbox{
\setlength{\unitlength}{0.7em}
\begin{picture}(5.5,1.7)
\put(0,0.2){\framebox (1,1){}}
\put(1.5,.6){\framebox (1,1){}}
\put(2.5,0.6){\framebox (1,1){}}
\put(3.5,0.6){\framebox (1,1){}}
\put(4.5,0.6){\framebox (1,1){}}
\put(0,-0.8){\framebox (1,1){}}
\put(1,-0.8){\framebox (1,1){}}
\end{picture}} \; \right)_{\phantom{{}_A}\hspace{-1ex}} 
$ 
\\
$\phantom{\Big|^B}\hspace{-2ex}$
$\OO^{[2,3,2]}_5 \sim 
\left(\hspace{-1ex}\mbox{
\setlength{\unitlength}{0.7em}
\begin{picture}(5.5,1.7)
\put(0,0.5){\framebox (1,1){}}
\put(1,0.5){\framebox (1,1){}}
\put(2.5,0.5){\framebox (1,1){}}
\put(3.5,0.5){\framebox (1,1){}}
\put(4.5,0.5){\framebox (1,1){}}
\put(0,-0.5){\framebox (1,1){}}
\put(1,-0.5){\framebox (1,1){}}
\end{picture}} \; \right)_{\phantom{{}_A}\hspace{-1ex}} 
$ 
\\
$\phantom{\Big|^B}\hspace{-2ex}$
$\OO^{[2,3,2]}_6 \sim 
\left(\hspace{-1ex}\mbox{
\setlength{\unitlength}{0.7em}
\begin{picture}(5.5,1.7)
\put(0,0.6){\framebox (1,1){}}
\put(1,0.6){\framebox (1,1){}}
\put(2.5,0.6){\framebox (1,1){}}
\put(3.5,0.6){\framebox (1,1){}}
\put(4.5,0.6){\framebox (1,1){}}
\put(0,-0.8){\framebox (1,1){}}
\put(1,-0.8){\framebox (1,1){}}
\end{picture}} \; \right)_{\phantom{{}_A}\hspace{-1ex}} 
$ 
\\
$\phantom{\Big|^B}\hspace{-2ex}$
$\OO^{[2,3,2]}_7 \sim 
\left(\hspace{-1ex}\mbox{
\setlength{\unitlength}{0.7em}
\begin{picture}(6,1.7)
\put(0,0.5){\framebox (1,1){}}
\put(1.5,0.5){\framebox (1,1){}}
\put(3,0.5){\framebox (1,1){}}
\put(4,0.5){\framebox (1,1){}}
\put(5,0.5){\framebox (1,1){}}
\put(0,-0.5){\framebox (1,1){}}
\put(1.5,-0.5){\framebox (1,1){}}
\end{picture}} \; \right)_{\phantom{{}_A}\hspace{-1ex}} 
$ 
\\
\end{tabular}}
\hspace{-3.5ex}
& 
\hspace{-2.5ex}
{\begin{tabular}{l}
$\YY^{[2,3,2]}_1 = - {10 N \over N^2-7} \OO^{[2,3,2]}_1 + \OO^{[2,3,2]}_2$ \\ 
\hspace{6ex} 
$+ {2 \over N^2-7} \left(\OO^{[2,3,2]}_3 + \OO^{[2,3,2]}_4 
+ \OO^{[2,3,2]}_5 \right)$ \\
\\
$\YY^{[2,3,2]}_2 = - 20 \OO^{[2,3,2]}_1 
+ {2 N^2 + 4 \over N} \OO^{[2,3,2]}_2$ \\ 
\hspace{6ex}  
$- {2\over N} \left(\OO^{[2,3,2]}_3 + \OO^{[2,3,2]}_4 \right) + 
\OO^{[2,3,2]}_6$ \\
\\
$\YY^{[2,3,2]}_3 = 10 \OO^{[2,3,2]}_1 
- {N^2 - 4 \over N} \OO^{[2,3,2]}_2$ \\ 
\hspace{6ex}  
$- {2\over N} \left(\OO^{[2,3,2]}_3 + \OO^{[2,3,2]}_4 \right) + 
\OO^{[2,3,2]}_7$ 
\\
\end{tabular}}
\hspace{-3.3ex}
\\
\hline
\end{tabular}
\caption{Gauge invariant, local, polynomial, 
scalar composite operators in the $[p,q,p]$ 
representations of $SU(4)$, with $2p+q \le 7$. 
Each continuous string of boxes in the $\OO^{[p,q,p]}_i$ 
corresponds to a single trace. 
The $\YY^{[p,q,p]}_i$ listed are the \quarter-BPS 
chiral primaries; other pure operators are not shown. 
}
\label{table:results}
\end{table}


\section{Large $N$ analysis}
\label{section:large N}

As we have seen, computations get more and more 
cumbersome as one tries to find \quarter-BPS 
operators for bigger representations of the color group; 
even the number of operators one has to consider is a 
nontrivial function of the representation. 
Symmetry factors multiplying 
the Feynman graphs show no immediate pattern, 
and most of the results presented in 
Sections \ref{2-0-2}, \ref{2-1-2}, and \ref{6 and higher} 
had to be calculated using {\it Mathematica}.%
\footnote{
	The calculations took from 
	0.003 hours for the [2,0,2] representation 
	to
	23 hours for [3,1,3], 
	per single $\OO(g^2)$ two point function. 
	We used a Sparc 10 with 2048 M memory 
	and 440 MHz speed. 
	Born level calculations were considerably 
	(about 20 times) faster.  
	} 

The next best thing we can do is 
consider the large $N$ limit. Specifically, we shall 
concentrate on the leading behavior as $N \to \infty$,
plus the first $1/N$ correction. 

\subsection{Operators $\OO_{[p,q,p]}$ and $\KK_{[p,q,p]}$}
\label{section:oo and kk}

Let us take another look at the results of 
Section 
\ref{6 and higher}, 
where we managed to perform $\OO(g^2)$ analysis exactly 
in $N$ rather than in the large $N$ approximation. 
In all cases considered so far, there is a special 
\quarter-BPS chiral primary $\YY^{[2,3,2]}_1$
(equations \ref{eq:2-2-2:y1}, \ref{eq:3-1-3:y1}, and \ref{eq:3-2-3:y1}), 
which is made of only the double trace and single trace operators. 
At large $N$, this operator is a combination 
of only 
a particular double trace operator, 
and the single trace operator, whose contribution 
is $1/N$ suppressed. 
The goal of Section \ref{section:large N} 
is to show that this is in fact what happens for 
general $[p,q,p]$ representations. 
Here, we begin by defining 
these operators.

Recall that the $SO(6)$ Young tableau for the $[p,q,p]$ of $SU(4)$ 
consists of two rows 
(one of length $p+q$, and the other of length $p$). 
Among the possible partitions of the highest weight tableau, 
there are two special ones 
\begin{eqnarray}
\OO_{[p,q,p]} \sim 
\left(
\mbox{
\setlength{\unitlength}{1em}
\begin{picture}(7.5,1.6)
\put(0,.5){\framebox (1,1){\scriptsize $1$}}
\put(1,.5){\framebox (2,1){\scriptsize $...$}}
\put(3,.5){\framebox (1,1){\scriptsize $1$}}
\put(4,.5){\framebox (1,1){\scriptsize $1$}}
\put(5,.5){\framebox (1,1){\scriptsize $...$}}
\put(6,.5){\framebox (1,1){\scriptsize $1$}}
\put(0,-1){\framebox (1,1){\scriptsize $2$}}
\put(1,-1){\framebox (2,1){\scriptsize $...$}}
\put(3,-1){\framebox (1,1){\scriptsize $2$}}
\put(1.8,-1.7){\scriptsize $p$}
\put(5.4,-0.7){\scriptsize $q$}
\end{picture}}
\right) 
, \quad 
\KK_{[p,q,p]} \sim 
\left(
\mbox{
\setlength{\unitlength}{1em}
\begin{picture}(7.5,1.6)
\put(0,.2){\framebox (1,1){\scriptsize $1$}}
\put(1,.2){\framebox (2,1){\scriptsize $...$}}
\put(3,.2){\framebox (1,1){\scriptsize $1$}}
\put(4,.2){\framebox (1,1){\scriptsize $1$}}
\put(5,.2){\framebox (1,1){\scriptsize $...$}}
\put(6,.2){\framebox (1,1){\scriptsize $1$}}
\put(0,-.8){\framebox (1,1){\scriptsize $2$}}
\put(1,-.8){\framebox (2,1){\scriptsize $...$}}
\put(3,-.8){\framebox (1,1){\scriptsize $2$}}
\put(1.8,-1.5){\scriptsize $p$}
\put(5.4,-0.5){\scriptsize $q$}
\end{picture}}
\right) 
\end{eqnarray}
where each continuous group of boxes stands for 
a single trace, as before. 
Explicitly, the corresponding operators are 
\begin{eqnarray}
\label{def:OO}
\OO_{[p,q,p]} &=& 
\sum_{k=0}^p {(-1)^k p! \over k! (p-k)!} ~
\tr \left( {z_1}^{p+q-k} {z_2}^{k} \right)_s ~ 
\tr \left( {z_1}^{k} {z_2}^{p-k} \right)_s 
\\
\label{def:KK}
\KK_{[p,q,p]} &=& 
\sum_{k=0}^p {(-1)^k p! \over k! (p-k)!} ~
\tr \left[ \left( {z_1}^{p+q-k} {z_2}^{k} \right)_s 
\left( {z_1}^{k} {z_2}^{p-k} \right)_s \right] 
\end{eqnarray}
(after projecting onto $SU(4) \to SU(3) \times U(1)$ 
and keeping only 
the highest $U(1)$-charge pieces, 
as discussed in 
Sections \ref{section:superconformal group} and \ref{2-0-2}). 
Made of only $z_1$ and $z_2$, 
both types of operators 
are annihilated by four out of the sixteen 
Poincar\'e supersymmetry generators: 
using the SUSY 
transformations 
spelled out in Appendix \ref{n=4 susy:section}, 
we find 
$\bar Q_{\bar \zeta} z_j = 0$, 
$Q_{\zeta_3} z_j = - \sqrt{2} (\lambda \zeta_3) \delta_{3j}$, 
so 
\begin{eqnarray}
\bar Q_{\bar \zeta} \OO_{[p,q,p]} = 
Q_{\zeta_3} \OO_{[p,q,p]} = 
\bar Q_{\bar \zeta} \KK_{[p,q,p]} = 
Q_{\zeta_3} \KK_{[p,q,p]} = 
0 
. 
\end{eqnarray}

It is clear why $\KK_{[p,q,p]}$ is special: it is 
the only single trace $[p,q,p]$ operator 
which can be constructed out of these fields. 
On the other hand, $\OO_{[p,q,p]}$ is 
``the most natural'' double trace composite operator 
in this representation. 
We also recognize it as the 
free theory chiral primary 
from the classification of \cite{AFSZ}.

As we have seen in the previous Sections, neither
the single trace $\KK_{[p,q,p]}$ nor the double trace $\OO_{[p,q,p]}$ 
are eigenstates of the dilation operator, for general $N$. 
Below we calculate correlators 
$\langle \OO \bar{\OO} \rangle$, 
$\langle \OO \bar{\KK} \rangle$, 
$\langle \KK \bar{\OO} \rangle$, and 
$\langle \KK \bar{\KK} \rangle$, 
in the large $N$ 
limit, and determine the pure operators 
and their scaling dimension 
in this approximation.

\subsection{General correlators 
$\langle {\OO}_{[p,q,p]} \bar{\OO}_{[p,q,p]} \rangle$ 
to order $g^2$}
\label{section:OO}

Let us first consider the 
$\langle {\OO}_{[p,q,p]}(x) \bar{\OO}_{[p,q,p]}(y) \rangle$ 
correlators. 
The free contribution is just a power of the free scalar propagator 
$G(x,y) = [4\pi (x-y)^2]^{-1}$, times a combinatorial factor: 
\begin{eqnarray}
\label{def:RR}
\langle {\OO}_{[p,q,p]}(x) \bar{\OO}_{[p,q,p]}(y) \rangle |_{\rm free} 
&=&
\sum_{k,l=0}^p 
{(-1)^k p! \over k! (p-k)!} ~ {(-1)^l p! \over l! (p-l)!} ~
(\RR_{k,l}^{p+q,p}) |_{\rm free} 
\\&=& \nonumber 
[G(x,y)]^{(2p+q)} 
\sum_{k,l=0}^p 
{(-1)^k p! \over k! (p-k)!} ~ {(-1)^l p! \over l! (p-l)!} ~
\FF_{k,l}^{p+q,p}
\end{eqnarray}
where 
\begin{eqnarray}
\RR_{k,l}^{p+q,p} &=& 
\langle 
\left[ 
\str (z_1)^{(p+q-k)} (z_2)^{k} 
\right] (x) 
\left[ 
\str (z_1)^{k} (z_2)^{(p-k)} 
\right] (x) 
\nonumber\\&&~\,
\left[ 
\str (z_1)^{(p+q-l)} (z_2)^{l} 
\right] (y) 
\left[ 
\str (z_1)^{l} (z_2)^{(p-l)} 
\right] (y) 
\rangle 
\end{eqnarray}
and 
\begin{eqnarray}
\label{def:FF}
\FF_{k,l}^{p+q,p} &=& 
\sum_{\sigma , \rho } 
\left[ \str t^{a_{k+1}} ... t^{a_{p+q}} t^{b_1} ... t^{b_k} \right] 
\left[ \str t^{a_1} ... t^{a_k} t^{b_{k+1}} ... t^{b_p} \right] 
\nonumber \\ && \quad 
\left[ \str t^{a_{\sigma(l+1)}} ... t^{a_{\sigma(p+q)}} 
t^{b_{\rho(1)}} ... t^{b_{\rho(l)}} \right] 
\left[ \str t^{a_{\sigma(1)}} ... t^{a_{\sigma(l)}} 
t^{b_{\rho(l+1)}} ... t^{b_{\rho(p)}} \right] 
\nonumber\\ 
\end{eqnarray}
($\sigma$ and $\rho$ sample over groups of permutations 
$S_{p+q}$ and $S_p$ on $p+q$ and $p$ letters, respectively). 

Like in the \half-BPS case, 
the leading contribution%
\footnote{
	See Appendix \ref{suN-identities} for useful $SU(N)$ identities. 
	}
to 
$\FF_{k,l}^{p+q,p} \sim (N/2)^{(2p+q)}$ 
comes 
from terms in which generators appear in reverse 
order for $z$-s and $\bar z$-s. 
To estimate the large $N$ behavior 
we can use equation (\ref{merging traces}) to ``merge traces,'' 
$(\tr t^{d_1} ... t^{d_s} t^c) (\tr t^c t^{d_s} ... t^{d_1}) 
\sim \half \tr t^{d_1} ... t^{d_s} t^{d_s} ... t^{d_1} 
\sim (N/2)^{s+1}$. 
In order to find the numerical factor out front 
(which does not scale with $N$ but depends on $p$ and $q$), 
we should determine exactly 
which terms have this structure. 

The generators can appear in opposite order in two pairs of 
traces in (\ref{def:FF}) under the following circumstances. 
First, it can happen 
when $k=l$ and the 
traces are merged as 1 with 3 
and 2 with 4. The factors which arise are: 
$[1/p!]^2$ from symmetrizations in the 2-nd and 4-th traces; 
$[1/(p+q)!]^2$ from symmetrizations in the 2-nd and 4-th traces; 
$p!$ because for any ordering in the 1-st trace there is an identical 
one in the 3-d trace; 
$(p+q)!$ for the same reason for the 2-nd and 4-th trace; 
$k! (p-k)!$ since any permutation of just $t^a$-s or just $t^b$-s 
in the 1-st trace can be ``undone'' 
by $\sigma$-s and $\rho$-s in the 3-d trace; 
and similarly $k! (p+q-k)!$ for the 2-nd and 4-th trace; 
$p (p+q)$ because of trace cyclicity.  
There is also an overall factor from the definition (\ref{def:OO}). 
Second, if $q=0$, 
we can 
merge traces the other way: 1 with 4 and 2 with 3; in this case 
$k=p-l$ and all other factors are the same. 
Thus, the leading contributions%
\footnote{
	The error we are committing is of order $\OO(N^{-2})$. 
	}
add up to 
\begin{eqnarray}
\label{o-o:leading}
\langle {\OO}_{[p,q,p]}(x) \bar{\OO}_{[p,q,p]}(y) \rangle |_{\rm free} 
\sim \left( \half N 
G(x,y) \right)^{(2p+q)} 
\hspace{-3.5em} 
\hspace{-16em}&&
\nonumber\\&&~~~~\times
\sum_{k=0}^p 
(-)^{k} 
\left[ {p! \over k! (p-k)!} \right]^2
{ k! (p-k)! k! (p+q-k)! \over (p-1)! (p+q-1)! } 
\left[ (-)^k + \delta_{q,0} (-)^{p-k} \right]
\nonumber\\&&=
\left( \half N G(x,y) \right)^{(2p+q)} 
\left[ 1 + \delta_{q,0} (-)^{p} \right] 
{ p (p+q) (p+q+1) \over (q+1) } 
\end{eqnarray}
The reproduces the leading order correlators 
in the low dimensional cases considered in Sections 
\ref{2-0-2}, \ref{2-1-2}, and \ref{6 and higher}.
Also note that if $q=0$ and $p$ 
is odd, both operators 
$\OO_{[p,q,p]}$ and $\KK_{[p,q,p]}$ 
vanish identically, in agreement with (\ref{o-o:leading}).

Now consider the 
corrections to this result. 
Diagrams contributing 
to two point functions of scalar composite operators 
at $\OO(g^2)$ 
fall into two categories, see Figure \ref{fig:contributing diagrams}. 
On the one hand, there are 
Feynman graphs involving a gauge boson exchange 
(these are proportional to $A$ or $B$). 
On the other hand, we also have 
gauge independent ones (proportional to $\tilde B$) 
coming entirely from the 
$z z \bar z \bar z$-vertex. 
These two types of corrections have different combinatorial 
(index) structure, and we shall handle them separately.

\subsubsection{Gauge dependent contributions: Combinatorial Argument}
\label{section:OO-combinatorial}

In Section \ref{section:gauge-dependent}, we argued that 
two point functions of gauge-invariant operators can not 
contain pieces proportional to the gauge dependent functions
$A$ and $B$. Here we show this explicitly for operators
${\OO}_{[p,q,p]}$ and ${\KK}_{[p,q,p]}$. This is the only 
part of Section \ref{section:large N} which is exact in $N$,
and is not just a large $N$ approximation.

The simplest order $g^2$ contribution 
to $\langle {\OO}_{[p,q,p]} \bar{\OO}_{[p,q,p]} \rangle$ 
comes from corrections to the scalar propagator 
(diagrams of type (a) and (b) in 
Figure \ref{fig:contributing diagrams}). 
It has the same index structure as the 
free field result, and so is the 
same up to a factor $(p+q) N A$ for (a)- 
and $p N A$ for (b)-type diagrams. 
These factors simply count the number of $z_1$-s
and $z_2$-s.

Next consider the other diagrams where the 
correction comes from blocks with the same flavor 
in the four legs, ones of type (d) and (e). 
Each term in the $k$, $l$ sum in (\ref{def:RR}) 
receives corrections of the form 
\begin{eqnarray}
\label{e:basic}
&&
(\half)(\half)(-1)(2) B 
\sum_{\sigma,\rho} \sum_{i \ne j = 1}^{p+q} 
\left[ \tr t^{b_{k+1}} ... t^{b_p} t^{a_1} ... t^{a_k} \right] 
\left[ \tr t^{a_{k+1}} ... t^{a_{p+q}} t^{b_1} ... t^{b_k} \right] 
\nonumber \\ && \quad 
\left[ t^{b_{\rho(l+1)}} ... t^{b_{\rho(p)}} 
t^{a_{\sigma(1)}} ... 
\left[ t^{a_{\sigma(i)}} , t^c \right] 
... 
\left[ t^{a_{\sigma(j)}} , t^c \right] 
... t^{a_{\sigma(p+q)}} 
t^{b_{\rho(1)}} ... t^{b_{\rho(l)}} \right] 
\end{eqnarray}
where all traces have to be symmetrized, 
and the second line also contains two traces.%
\footnote{
	We will omit the $[G(x,y)]^{(2p+q)}$ factor 
	which is common to all diagrams. 
	} 
The 
prefactor multiplying the sum 
comes about in the following manner: 
a factor of (\half), since the sum is on $i \ne j$ rather than 
on $i < j$; similarly the other (\half) arises because using
$\{ \sigma(i) , \sigma(j) \}$ and $\{ \sigma(j) , \sigma(i) \}$ 
counts the same pair twice; a factor of 2 has to be taken into 
account as the two cross-symmetric pairs give the same contribution;
and finally $(-1)$ is there from two factors of $i$ which are 
needed to convert $f$-s to commutators. 

The structure in the sum (\ref{e:basic}) 
consists of four kinds of terms. 
The two commutators can be both in the third trace, 
or both in the fourth trace, 
or one in either trace. 
When both commutators are 
in the same trace, we can play the same game as for \half-BPS 
operators. 
Fix $i$ and do the sum on $j$ first; this assembles 
$
( ... [t^{a_{\sigma(1)}} , t^c] ... ) + ... + 
( ... [t^{a_{\sigma(l)}} , t^c] ... ) 
= ( ... [t^{a_{\sigma(1)}} ... t^{a_{\sigma(l)}} , t^c] ... )
$
for example. 
Then, use trace cyclicity in the form 
$\tr [A,B]C = \tr A[B,C]$
to move one of the traces over 
to the other commutator and $t^{b_\rho}$-s; 
here we pick up a minus sign which cancels the $(-1)$ in 
(\ref{e:basic}). 
As $[[ t^a , t^c],t^c] = N t^a$, the first bit is easy --- 
just like in the \half-BPS case, it is proportional to 
$(+\half B N)$ times the free-field combinatorial factor; 
when we do the sum on $i$ we also get a factor of $(p+q)$ here. 
As $B+2A=0$ (by \NN=4 SUSY), this part cancels 
the diagrams of type (a). 
The leftovers, together with the terms with 
commutators in different traces, add up to 
\begin{eqnarray}
\label{leftover} 
&&
\half B 
\left[ \tr t^{b_{k+1}} ... t^{b_p} t^{a_1} ... t^{a_k} \right] 
\left[ \tr t^{a_{k+1}} ... t^{a_{p+q}} t^{b_1} ... t^{b_k} \right] 
\sum_{\sigma,\rho} 
\nonumber \\ && \quad 
\left[ \tr t^{b_{\rho(l+1)}} ... t^{b_{\rho(p)}} 
t^{a_{\sigma(1)}} ... t^{a_{\sigma(l)}} \right] 
\left[ \tr 
\left[ t^{a_{\sigma(l+1)}} ... t^{a_{\sigma(p+q)}} , t^c \right] 
\left[ t^{b_{\rho(1)}} ... t^{b_{\rho(l)}}  , t^c \right] 
\right] 
\nonumber \\ && 
+ 
\left[ \tr 
\left[ t^{b_{\rho(l+1)}} ... t^{b_{\rho(p)}}  , t^c \right] 
\left[ t^{a_{\sigma(1)}} ... t^{a_{\sigma(l)}} , t^c \right] 
\right] 
\left[ \tr 
t^{a_{\sigma(l+1)}} ... t^{a_{\sigma(p+q)}} 
t^{b_{\rho(1)}} ... t^{b_{\rho(l)}} 
\right] 
\nonumber \\ && 
- 2
\left[ \tr 
t^{b_{\rho(l+1)}} ... t^{b_{\rho(p)}} 
\left[ t^{a_{\sigma(1)}} ... t^{a_{\sigma(l)}} , t^c \right] 
\right] 
\left[ \tr 
\left[ t^{a_{\sigma(l+1)}} ... t^{a_{\sigma(p+q)}} , t^c \right] 
t^{b_{\rho(1)}} ... t^{b_{\rho(l)}} 
\right] 
\nonumber\\
\end{eqnarray}

Similarly, 
there are diagrams of type (e) in Figure \ref{fig:contributing diagrams}, 
where all of the flavors are ``2'' in all four legs. 
Here, we have a term proportional to the free-field result (the only
difference being an overall factor of $p$ rather than $p+q$), 
which cancels contributions from diagrams of type (b). 
The leftover term 
is the same as what we have just computed. 
This removes the factor of \half~ from 
(\ref{leftover}). 

Finally consider the diagrams of type (f). 
Now we have both 
flavors ``1'' and ``2'' in the four-scalar blocks, while the 
index structure is the same as that of (d) and (e). 
The discussion goes through as above, 
but with a few small modifications. 
First, 
the prefactor is just $(-1) B$ as now the indices $i$ and $j$
run over different flavors (so there is no 
``$i \ne j$ overcounting,'' 
no ``\{$\sigma(i), \sigma(j)\}$ overcounting,''
and no factor of 2 from crossing symmetry). 
Second, 
we do not pick up a minus sign when transferring commutators
under the traces (before, both commutators with $t^c$ were on, say, 
$t^a$-s, whereas now one is on $t^a$ and one on $t^b$). 
Therefore, the result of 
adding the diagrams of type (f) is to precisely cancel 
the whole leftover structure of (twice that given in) equation 
(\ref{leftover}). 

Thus we have explicitly reproduced the general result of 
Section \ref{section:gauge-dependent},
but with a lot more work: $A$ and $B$ contributions to two-point
functions of arbitrary gauge-invariant scalar composite 
operators combine as $2 A + B$, which vanishes by 
\NN=4 supersymmetry.

\subsubsection{Contributions proportional to $\tilde B$}
\label{section:OO-F-term}

So far we have established that adding all 
gauge dependent Feynman graphs, i.e. 
diagrams of types (a), (b), (d), (e), and (f) 
shown in Figure \ref{fig:contributing diagrams}, 
gives a vanishing $\OO(g^2)$ contribution to the two-point 
functions (\ref{def:RR}), once we impose 
\NN=4 SUSY. 
Just as in the cases of low dimensional operators considered 
in Sections \ref{section:simplest} and \ref{6 and higher}, the 
$\OO(g^2)$ corrections to the two-point function 
of $[p,q,p]$ operators 
come exclusively from diagrams of type (c), 
and are proportional to $\tilde B$. 
To find the combinatorial factor multiplying $\tilde B$, 
we need to perform a calculation similar to the one 
in Section \ref{section:OO-combinatorial}.

Here, ``contracted with $f$-s'' are $z$-s with $z$-s and 
$\bar z$-s with $\bar z$-s (unlike in diagrams of types 
(d), (e) and (f), where it was one $z$ and one $\bar z$), 
and it is more convenient to label the generators 
slightly differently. For example, the free-field 
result (\ref{def:FF}) can be rewritten as 
\begin{eqnarray}
\label{def:FF-other}
\FF_{k,l}^{p+q,p} &=& 
\sum_{\sigma , \rho } 
\left[ \str 
t^{a_{\sigma(k+1)}} ... t^{a_{\sigma(p+q)}} 
t^{b_{1}} ... t^{b_{k}} 
\right] 
\left[ \str 
t^{a_{\sigma(1)}} ... t^{a_{\sigma(k)}} 
t^{b_{k+1}} ... t^{b_{p}} 
\right] 
\nonumber \\ && \quad 
\left[ \str 
t^{a_{l+1}} ... t^{a_{p+q}} 
t^{b_{\rho(1)}} ... t^{b_{\rho(l)}} 
\right] 
\left[ \str 
t^{a_{1}} ... t^{a_{l}} 
t^{b_{\rho(l+1)}} ... t^{b_{\rho(p)}} 
\right] 
\end{eqnarray}
For the Born level combinatorial factor 
it makes no difference, 
but in calculating the order $g^2$ diagrams of type (c) 
we ``$f$-contract'' 
$i$-th $z_1$ with $\rho(j)$-th $z_2$, and 
$\sigma(i)$-th $\bar z_1$ with $j$-th $\bar z_2$. 
This will exhaust all pairs without overcounting 
(because $i$ and $j$ are again on different flavors), 
so the prefactor will be just $(-1) \tilde B(x,y)$. Apart from 
this prefactor (and a factor of $[G(x,y)]^{(2p+q)}$), 
the (c)-type correction reads 
(sums on $\sigma$ and $\rho$ implied) 
\begin{eqnarray}
\label{h2 contribution}
&&
\!\!\!\!\!\!\!
\sum
\left[ \tr 
t^{a_{\sigma(k+1)}} ... t^{a_{\sigma(p+q)}} 
t^{b_{1}} ... t^{b_{j-1}} [t^{a_{i}}, t^c] t^{b_{j+1}} ... t^{b_{k}} 
\right] 
\left[ \tr 
t^{a_{\sigma(1)}} ... t^{a_{\sigma(k)}} 
t^{b_{k+1}} ... t^{b_{p}} 
\right] 
\nonumber \\ && \quad 
\left[ \tr 
t^{a_{l+1}} ... t^{a_{p+q}} 
t^{b_{\rho(1)}} ... t^{b_{\rho(l)}} 
\right] 
\left[ \tr 
t^{a_{1}} ... t^{a_{i-1}} [t^{b_{j}}, t^c] t^{a_{i+1}} ... t^{a_{l}} 
t^{b_{\rho(l+1)}} ... t^{b_{\rho(p)}} 
\right] 
\nonumber\\&+&
\!\!\!\!\!\!\!
\sum
\left[ \tr 
t^{a_{\sigma(k+1)}} ... t^{a_{\sigma(p+q)}} 
t^{b_{1}} ... t^{b_{j-1}} [t^{a_{i}}, t^c] t^{b_{j+1}} ... t^{b_{k}} 
\right] 
\left[ \tr 
t^{a_{\sigma(1)}} ... t^{a_{\sigma(k)}} 
t^{b_{k+1}} ... t^{b_{p}} 
\right] 
\nonumber \\ && \quad 
\left[ \tr 
t^{a_{l+1}} ... t^{a_{i-1}} [t^{b_{j}}, t^c] t^{a_{i+1}} ... t^{a_{p+q}} 
t^{b_{\rho(1)}} ... t^{b_{\rho(l)}} 
\right] 
\left[ \tr 
t^{a_{1}} ... t^{a_{l}} 
t^{b_{\rho(l+1)}} ... t^{b_{\rho(p)}} 
\right] 
\nonumber\\&+&
\!\!\!\!\!\!\!
\sum
\left[ \tr 
t^{a_{\sigma(k+1)}} ... t^{a_{\sigma(p+q)}} 
t^{b_{1}} ... t^{b_{k}} 
\right] 
\left[ \tr 
t^{a_{\sigma(1)}} ... t^{a_{\sigma(k)}} 
t^{b_{k+1}} ... t^{b_{j-1}} [t^{a_{i}}, t^c] t^{b_{j+1}} ... t^{b_{p}} 
\right] 
\nonumber \\ && \quad 
\left[ \tr 
t^{a_{l+1}} ... t^{a_{p+q}} 
t^{b_{\rho(1)}} ... t^{b_{\rho(l)}} 
\right] 
\left[ \tr 
t^{a_{1}} ... t^{a_{i-1}} [t^{b_{j}}, t^c] t^{a_{i+1}} ... t^{a_{l}} 
t^{b_{\rho(l+1)}} ... t^{b_{\rho(p)}} 
\right] 
\nonumber\\&+&
\!\!\!\!\!\!\!
\sum
\left[ \tr 
t^{a_{\sigma(k+1)}} ... t^{a_{\sigma(p+q)}} 
t^{b_{1}} ... t^{b_{k}} 
\right] 
\left[ \tr 
t^{a_{\sigma(1)}} ... t^{a_{\sigma(k)}} 
t^{b_{k+1}} ... t^{b_{j-1}} [t^{a_{i}}, t^c] t^{b_{j+1}} ... t^{b_{p}} 
\right] 
\nonumber \\ && \quad 
\left[ \tr 
t^{a_{l+1}} ... t^{a_{i-1}} [t^{b_{j}}, t^c] t^{a_{i+1}} ... t^{a_{p+q}} 
t^{b_{\rho(1)}} ... t^{b_{\rho(l)}} 
\right] 
\left[ \tr 
t^{a_{1}} ... t^{a_{l}} 
t^{b_{\rho(l+1)}} ... t^{b_{\rho(p)}} 
\right] 
\nonumber\\
\end{eqnarray}
where all traces have to be symmetrized again; 
the sums on $\{i,j\}$ are: 
in the first line, from $\{1,1\}$ to $\{l,k\}$,  
in the second line, from $\{l+1,1\}$ to $\{p+q,k\}$, 
in the third line, from $\{1,k+1\}$ to $\{l,p\}$, 
and in the last line, from $\{l+1,k+1\}$ to $\{p+q,p\}$.  

In the large $N$ limit, such terms can scale as 
$\half (N/2)^{(2p+q-1)}$, at best. 
(Because of the commutators with 
$t^c$, we have to merge traces three times: 
they don't eat each other up in pairs as they did before). 
After including the factor of $\tilde B$, 
together with (\ref{o-o:leading}) this means that 
\begin{eqnarray}
\langle {\OO}_{[p,q,p]} (x) \bar{\OO}_{[p,q,p]} (y) \rangle 
&\!\!=\!\!& 
\left( {N\over2} G(x,y) \right)^{(2p+q)} 
\\&&\hspace{-4em} \times 
\left( 
\left[ 1 + \delta_{q,0} (-)^{p} \right] 
{ p (p+q) (p+q+1) \over (q+1) } 
+ 
\tilde B(x,y) N \times \OO(N^{-2}) 
\right) 
\nonumber
\end{eqnarray}
to order $g^2$. 
We might be tempted to stop here. 
By observing that since to working precision, 
the two point function of ${\OO}_{[p,q,p]}$ with itself 
does not get $\OO(g^2)$ corrections, 
we could try to conclude it is chiral, and 
in particular has a protected $\Delta = 2 p + q$. 
However, as the explicit examples of 
Sections \ref{section:simplest} and \ref{6 and higher} 
show, ${\OO}_{[p,q,p]}$ may not be a pure operator, 
in which case it doesn't make sense to talk 
about its scaling dimension.

\subsection{General correlators 
$\langle {\KK}_{[p,q,p]} \bar{\OO}_{[p,q,p]} \rangle$}

The analysis exactly parallels that of the previous section. 
Again, 
\begin{eqnarray}
\label{def:PP}
\langle {\KK}_{[p,q,p]}(x) \bar{\OO}_{[p,q,p]}(y) \rangle 
&=&
\sum_{k,l=0}^p 
{(-1)^k p! \over k! (p-k)!} ~ {(-1)^l p! \over l! (p-l)!} ~
\PP_{k,l}^{p+q,p}
\end{eqnarray}
with
\begin{eqnarray}
\PP_{k,l}^{p+q,p} &=& 
\langle 
\tr \left[ 
\left( (z_1)^{(p+q-l)} (z_2)^{l} \right)_s 
\left( (z_1)^{l} (z_2)^{(p-l)} \right)_s 
\right] (x) 
\nonumber\\&&~\,
\left[ 
\str (z_1)^{(p+q-k)} (z_2)^{k} 
\right] (y) 
\left[ 
\str (z_1)^{k} (z_2)^{(p-k)} 
\right] (y) 
\rangle 
\end{eqnarray}

The leading large $N$ contributions to the free correlators 
now come from terms which contain the combinatorial factor 
similar to 
\begin{eqnarray}
\label{k-o:leading}
&&\hspace{-4em}
(\tr t^{a_1} ... t^{a_k} t^{b_{k+1}} ... t^{b_p}
t^{a_{k+1}} ... t^{a_{p+q}} t^{b_{1}} ... t^{b_k}) 
\nonumber\\&&\hspace{-4em} \quad \times
(\tr t^{b_{p}} ... t^{b_{k+1}} t^{a_k} ... t^{a_1})
(\tr t^{b_{k}} ... t^{b_1} t^{a_{p+q}} ... t^{a_{k+1}})
\nonumber\\&&\hspace{4em} \sim
(\half)^2 N (N/2)^{(2p+q-2)} = \half (N/2)^{(2p+q-1)}
\end{eqnarray}
(the two halves and $-2$ in the exponent 
are because we have to merge traces twice, 
and the factor of $N=\tr \bone$ is there as usual). 
To get the other numerical factors we have to carefully analyze
which terms in the sums and symmetrizations scale with $N$ this way. 
So far, we do not need them.

As before, individual terms in the $k$, $l$ sum get corrections 
similar to (\ref{e:basic}): 
\begin{eqnarray}
&&
(\half)(\half)(-1)(2) B 
\sum_{\sigma,\rho} \sum_{i \ne j = 1}^{p+q} 
 \tr 
\left[ 
\left( t^{b_{k+1}} ... t^{b_p} t^{a_1} ... t^{a_k} \right) 
\left( t^{a_{k+1}} ... t^{a_{p+q}} t^{b_1} ... t^{b_k} \right)
\right]  
\nonumber \\ && \quad 
\left[ t^{b_{\rho(l+1)}} ... t^{b_{\rho(p)}} 
t^{a_{\sigma(1)}} ... 
\left[ t^{a_{\sigma(i)}} , t^c \right] 
... 
\left[ t^{a_{\sigma(j)}} , t^c \right] 
... t^{a_{\sigma(p+q)}} 
t^{b_{\rho(1)}} ... t^{b_{\rho(l)}} \right] 
\end{eqnarray}
with proper symmetrizations (and omitted factor of $[G(x,y)]^{(2p+q)}$). 
The only difference is that now 
there are three traces (rather than four). 
The discussion of gauge dependent diagrams 
(all types other than (c), see Figure \ref{fig:contributing diagrams}) 
goes through verbatim, since we were only manipulating 
the second set of traces. 
As before, when we impose \NN=4 SUSY they cancel, 
and order $g^2$ corrections to the two-point functions (\ref{def:PP}) 
are due to diagrams of type (c) only.  

Diagrams of type (c) are only slightly different
from those contributing to the 
$\langle \OO \bar{\OO} \rangle$ correlator; they add up 
(apart from the $(-1) \tilde B(x,y) [G(x,y)]^{(2p+q)}$ prefactor)
to 
\begin{eqnarray}
\label{h2 contribution for OK}
&&
\!\!\!\!\!\!\!
\sum
\tr \!\! \left[ 
\left( 
t^{a_{\sigma(k+1)}} ... t^{a_{\sigma(p+q)}} 
t^{b_{1}} ... t^{b_{j-1}} [t^{a_{i}}, t^c] t^{b_{j+1}} ... t^{b_{k}} 
\right) 
\left( 
t^{a_{\sigma(1)}} ... t^{a_{\sigma(k)}} 
t^{b_{k+1}} ... t^{b_{p}} 
\right) 
\right] 
\nonumber \\ && \quad 
\left[ \tr 
t^{a_{l+1}} ... t^{a_{p+q}} 
t^{b_{\rho(1)}} ... t^{b_{\rho(l)}} 
\right] 
\left[ \tr 
t^{a_{1}} ... t^{a_{i-1}} [t^{b_{j}}, t^c] t^{a_{i+1}} ... t^{a_{l}} 
t^{b_{\rho(l+1)}} ... t^{b_{\rho(p)}} 
\right] 
\nonumber\\&+&
\!\!\!\!\!\!\!
\sum
\tr \!\! \left[ 
\left( 
t^{a_{\sigma(k+1)}} ... t^{a_{\sigma(p+q)}} 
t^{b_{1}} ... t^{b_{j-1}} [t^{a_{i}}, t^c] t^{b_{j+1}} ... t^{b_{k}} 
\right) 
\left( 
t^{a_{\sigma(1)}} ... t^{a_{\sigma(k)}} 
t^{b_{k+1}} ... t^{b_{p}} 
\right) 
\right] 
\nonumber \\ && \quad 
\left[ \tr 
t^{a_{l+1}} ... t^{a_{i-1}} [t^{b_{j}}, t^c] t^{a_{i+1}} ... t^{a_{p+q}} 
t^{b_{\rho(1)}} ... t^{b_{\rho(l)}} 
\right] 
\left[ \tr 
t^{a_{1}} ... t^{a_{l}} 
t^{b_{\rho(l+1)}} ... t^{b_{\rho(p)}} 
\right] 
\nonumber\\&+&
\!\!\!\!\!\!\!
\sum
\tr \!\! \left[ 
\left( 
t^{a_{\sigma(k+1)}} ... t^{a_{\sigma(p+q)}} 
t^{b_{1}} ... t^{b_{k}} 
\right) 
\left( 
t^{a_{\sigma(1)}} ... t^{a_{\sigma(k)}} 
t^{b_{k+1}} ... t^{b_{j-1}} [t^{a_{i}}, t^c] t^{b_{j+1}} ... t^{b_{p}} 
\right) 
\right] 
\nonumber \\ && \quad 
\left[ \tr 
t^{a_{l+1}} ... t^{a_{p+q}} 
t^{b_{\rho(1)}} ... t^{b_{\rho(l)}} 
\right] 
\left[ \tr 
t^{a_{1}} ... t^{a_{i-1}} [t^{b_{j}}, t^c] t^{a_{i+1}} ... t^{a_{l}} 
t^{b_{\rho(l+1)}} ... t^{b_{\rho(p)}} 
\right] 
\nonumber\\&+&
\!\!\!\!\!\!\!
\sum
\tr \!\! \left[ 
\left( 
t^{a_{\sigma(k+1)}} ... t^{a_{\sigma(p+q)}} 
t^{b_{1}} ... t^{b_{k}} 
\right) 
\left( 
t^{a_{\sigma(1)}} ... t^{a_{\sigma(k)}} 
t^{b_{k+1}} ... t^{b_{j-1}} [t^{a_{i}}, t^c] t^{b_{j+1}} ... t^{b_{p}} 
\right) 
\right] 
\nonumber \\ && \quad 
\left[ \tr 
t^{a_{l+1}} ... t^{a_{i-1}} [t^{b_{j}}, t^c] t^{a_{i+1}} ... t^{a_{p+q}} 
t^{b_{\rho(1)}} ... t^{b_{\rho(l)}} 
\right] 
\left[ \tr 
t^{a_{1}} ... t^{a_{l}} 
t^{b_{\rho(l+1)}} ... t^{b_{\rho(p)}} 
\right] 
\nonumber\\
\end{eqnarray}
with proper symmetrizations; 
the sums on $\{i,j\}$ are: 
in the first line, from $\{1,1\}$ to $\{l,k\}$,  
in the second line, from $\{l+1,1\}$ to $\{p+q,k\}$, 
in the third line, from $\{1,k+1\}$ to $\{l,p\}$, 
and in the last line, from $\{l+1,k+1\}$ to $\{p+q,p\}$.  

We can get correlators 
$\langle {\OO}_{[p,q,p]}(x) \bar{\KK}_{[p,q,p]}(y) \rangle$
by just complex conjugating the sum 
(\ref{h2 contribution for OK})
times the same prefactor; 
both the free propagator $G(x,y)$ and the function $\tilde B(x,y)$
are real and depend only on $(x-y)^2$, so exchanging the arguments 
$x \lra y$ and conjugating doesn't change anything.

Large $N$ dependence of (\ref{h2 contribution for OK})
can be again estimated by merging traces. This time, we have to 
merge traces only twice (there are three traces total), 
so it scales a power of $N$ higher than a similar 
$\langle {\OO}(x) \bar{\OO}(y) \rangle |_{g^2}$
correction (where traces had to be merged three times). 
Hence, 
$\langle {\OO}_{[p,q,p]}(x) \bar{\KK}_{[p,q,p]}(y) \rangle |_{g^2} 
\sim (N/2)^{(2p+q)}$.

\subsection{General correlators 
$\langle {\KK}_{[p,q,p]} \bar{\KK}_{[p,q,p]} \rangle$}
\label{section:KK}

The analysis is similar as for 
$\langle {\KK}_{[p,q,p]}(x) \bar{\OO}_{[p,q,p]}(y) \rangle$. 
The only surviving contribution to 
$\langle {\KK}_{[p,q,p]} \bar{\KK}_{[p,q,p]} \rangle|_{g^2}$
is due to diagrams of type (c) again, and 
equals 
\begin{eqnarray}
\label{h2 contribution for KK}
&&
\!\!\!\!\!\!\!
\sum
\tr \!\! \left[ 
\left( 
t^{a_{\sigma(k+1)}} ... t^{a_{\sigma(p+q)}} 
t^{b_{1}} ... t^{b_{j-1}} [t^{a_{i}}, t^c] t^{b_{j+1}} ... t^{b_{k}} 
\right) 
\left( 
t^{a_{\sigma(1)}} ... t^{a_{\sigma(k)}} 
t^{b_{k+1}} ... t^{b_{p}} 
\right) 
\right] 
\nonumber \\ && \quad 
\tr \!\! \left[ 
\left( 
t^{a_{l+1}} ... t^{a_{p+q}} 
t^{b_{\rho(1)}} ... t^{b_{\rho(l)}} 
\right) 
\left( 
t^{a_{1}} ... t^{a_{i-1}} [t^{b_{j}}, t^c] t^{a_{i+1}} ... t^{a_{l}} 
t^{b_{\rho(l+1)}} ... t^{b_{\rho(p)}} 
\right) 
\right] 
\nonumber\\&+&
\!\!\!\!\!\!\!
\sum
\tr \!\! \left[ 
\left( 
t^{a_{\sigma(k+1)}} ... t^{a_{\sigma(p+q)}} 
t^{b_{1}} ... t^{b_{j-1}} [t^{a_{i}}, t^c] t^{b_{j+1}} ... t^{b_{k}} 
\right) 
\left( 
t^{a_{\sigma(1)}} ... t^{a_{\sigma(k)}} 
t^{b_{k+1}} ... t^{b_{p}} 
\right) 
\right] 
\nonumber \\ && \quad 
\tr \!\! \left[ 
\left( 
t^{a_{l+1}} ... t^{a_{i-1}} [t^{b_{j}}, t^c] t^{a_{i+1}} ... t^{a_{p+q}} 
t^{b_{\rho(1)}} ... t^{b_{\rho(l)}} 
\right) 
\left( 
t^{a_{1}} ... t^{a_{l}} 
t^{b_{\rho(l+1)}} ... t^{b_{\rho(p)}} 
\right) 
\right] 
\nonumber\\&+&
\!\!\!\!\!\!\!
\sum
\tr \!\! \left[ 
\left( 
t^{a_{\sigma(k+1)}} ... t^{a_{\sigma(p+q)}} 
t^{b_{1}} ... t^{b_{k}} 
\right) 
\left( 
t^{a_{\sigma(1)}} ... t^{a_{\sigma(k)}} 
t^{b_{k+1}} ... t^{b_{j-1}} [t^{a_{i}}, t^c] t^{b_{j+1}} ... t^{b_{p}} 
\right) 
\right] 
\nonumber \\ && \quad 
\tr \!\! \left[ 
\left( 
t^{a_{l+1}} ... t^{a_{p+q}} 
t^{b_{\rho(1)}} ... t^{b_{\rho(l)}} 
\right) 
\left( 
t^{a_{1}} ... t^{a_{i-1}} [t^{b_{j}}, t^c] t^{a_{i+1}} ... t^{a_{l}} 
t^{b_{\rho(l+1)}} ... t^{b_{\rho(p)}} 
\right) 
\right] 
\nonumber\\&+&
\!\!\!\!\!\!\!
\sum
\tr \!\! \left[ 
\left( 
t^{a_{\sigma(k+1)}} ... t^{a_{\sigma(p+q)}} 
t^{b_{1}} ... t^{b_{k}} 
\right) 
\left( 
t^{a_{\sigma(1)}} ... t^{a_{\sigma(k)}} 
t^{b_{k+1}} ... t^{b_{j-1}} [t^{a_{i}}, t^c] t^{b_{j+1}} ... t^{b_{p}} 
\right) 
\right] 
\nonumber \\ && \quad 
\tr \!\! \left[ 
\left( 
t^{a_{l+1}} ... t^{a_{i-1}} [t^{b_{j}}, t^c] t^{a_{i+1}} ... t^{a_{p+q}} 
t^{b_{\rho(1)}} ... t^{b_{\rho(l)}} 
\right) 
\left( 
t^{a_{1}} ... t^{a_{l}} 
t^{b_{\rho(l+1)}} ... t^{b_{\rho(p)}} 
\right) 
\right] 
\nonumber\\
\end{eqnarray}
up to a factor of $(-1) \tilde B(x,y) [G(x,y)]^{(2p+q)}$, 
and proper symmetrizations. 
The sums on $\{i,j\}$ are as before: 
in the first line, from $\{1,1\}$ to $\{l,k\}$,  
in the second line, from $\{l+1,1\}$ to $\{p+q,k\}$, 
in the third line, from $\{1,k+1\}$ to $\{l,p\}$, 
and in the last line, from $\{l+1,k+1\}$ to $\{p+q,p\}$).

At Born level, 
$\langle {\KK}_{[p,q,p]} \bar{\KK}_{[p,q,p]} \rangle|_{\rm free} 
\sim (\half N)^{(2p+q)}$ at large $N$:
we have to merge traces once, and there are 
$2p+q$ generators involved.

\subsection{Quarter BPS operators}
\label{section:consistency}

Given the leading large $N$ dependence of a 
$\langle \OO \bar{\KK} \rangle |_{\rm free}$
correlator, it's easy to determine the 
leading large $N$ dependence of the corresponding 
$\langle \KK \bar{\KK} \rangle |_{\rm free}$. 
Indeed, suppose that a particular term e.g.
\begin{eqnarray} 
\label{term:o-k}
(\tr t^a ... t^a t^b ... t^b)(\tr t^a ... t^a t^b ... t^b)
\tr (t^a ... t^a t^b ... t^b)(t^a ... t^a t^b ... t^b)
\sim\nonumber\\
(2/N)
(\tr t^a ... t^a t^b ... t^b)(\tr t^c t^c t^a ... t^a t^b ... t^b)
\tr (t^a ... t^a t^b ... t^b)(t^a ... t^a t^b ... t^b)
\end{eqnarray}
contributes to $\langle \OO \bar{\KK} \rangle$. 
We can insert contents of the first and second 
traces into the third trace 
(using $2 (\tr A t^r) (\tr B t^r) \sim \tr A B$)
to reduce this expression to a single trace. 
On the other hand, the term in the $\langle \KK \bar{\KK} \rangle$
with the same order of generators can be written as 
\begin{eqnarray} 
\label{term:k-k}
\tr (t^a ... t^a t^b ... t^b)(t^a ... t^a t^b ... t^b)
\tr (t^a ... t^a t^b ... t^b)(t^a ... t^a t^b ... t^b)
\sim\nonumber\\
2
(\tr t^a ... t^a t^b ... t^b t^c)(\tr t^c t^a ... t^a t^b ... t^b)
\tr (t^a ... t^a t^b ... t^b)(t^a ... t^a t^b ... t^b)
\end{eqnarray}
and we can insert the first and second traces into the third trace 
again in the same locations. 
This term gives a leading contribution provided all 
generators collapse 
after consecutively applying $2 t^r t^r \sim N \bone$
without having to commute generators past one another. 
In this case, the term in (\ref{term:o-k}) also gives a leading 
contribution. 
The only other terms 
which 
have the same large $N$ behavior are the ones 
that differ from it by cyclic permutations within 
the first and second traces. 
This gives an overall factor of 
$p(p+q)$. Comparing (\ref{term:o-k}) and (\ref{term:k-k}) then 
shows that to leading order in $N$, 
the difference is a factor of 
$\beta \equiv p(p+q)/N$. 

For large $N$, the analysis of order $g^2$ corrections 
is analogous to the case of free field contributions. 
The structure of terms 
in (\ref{h2 contribution for OK}) and (\ref{h2 contribution for KK})
is similar, 
and leading contributions come from terms with 
the same order of generators (modulo cyclic permutations 
for $\langle \OO \bar{\KK} \rangle$ corrections); 
the difference is the multiplicative factor 
$\beta$, the same for all such terms. 

Bringing this together with the results of 
Sections \ref{section:OO}-\ref{section:KK}, we can write down the 
large $N$ leading order 
two point functions as 
\begin{eqnarray}
\label{pqp:largeN}
\left( \matrix{
\langle \KK \bar{\KK} \rangle & 
\langle \KK \bar{\OO} \rangle \cr 
\langle \OO \bar{\KK} \rangle & 
\langle \OO \bar{\OO} \rangle 
} \right) 
= 
\alpha
(G N)^{2p+q}
\left\{
\left( \matrix{
1 & \beta \cr 
\beta & * 
} \right)
+ 
\tilde \alpha 
(\tilde B N) 
\left( \matrix{
1 & \beta \cr 
\beta & \OO(N^{-2}) 
} \right)
\right\}
\end{eqnarray}
where $\alpha$, $*$, 
and $\tilde \alpha$ are some constants of 
order $\OO(N^0)$, 
and $\beta = p(p+q)/N$. 
As before, $G \equiv G(x,y)$; 
$\tilde B \equiv \tilde B(x,y)$; 
and 
$\langle \OO \bar{\OO} \rangle \equiv 
\langle \OO_{[p,q,p]}(x) \bar{\OO}_{[p,q,p]}(y) \rangle$, etc. 
Each (order one) coefficient 
above is valid 
to $\OO(N^{-2})$, and of course (\ref{pqp:largeN}) is 
perturbative in the coupling constant --- 
we have neglected $\OO(g^4)$ terms. 

Diagonalizing the matrix of corrections 
with respect to the matrix of free correlators 
as in Section \ref{6 and higher}, we find that 
\begin{eqnarray}
\label{K-hat:large N}
\tilde{\YY}_{[p,q,p]} &=& 
\KK_{[p,q,p]}   + \OO(N^{-2}) 
\\
\label{O-hat:large N}
\YY_{[p,q,p]} &=& 
\OO_{[p,q,p]} - {p(p+q)\over N} \KK_{[p,q,p]}  + \OO(N^{-2}) 
\end{eqnarray}
are pure operators, 
and as such have well defined scaling dimension. 
At this order, 
the scaling dimension of $\tilde{\YY}_{[p,q,p]}$ receives 
an $\OO(g^2N)$ correction proportional to the coefficient $\tilde \alpha$ 
in (\ref{pqp:largeN}), while 
the scaling dimension $\Delta_{\tiny \YY} = 2 p + q$ 
of $\YY_{[p,q,p]}$ 
is protected. 
Finally, the normalization of $\YY_{[p,q,p]}$ does not 
get any $g^2$ corrections, and is given by the 
Born level expression 
\begin{eqnarray}
\label{normalization:general}
\langle \YY_{[p,q,p]}(x) \bar{\YY}_{[p,q,p]}(y) \rangle 
&\!\!=\!\!& 
\left[ 1 + \delta_{q,0} (-)^{p} \right]
{ p (p+q) (p+q+1) \over (q+1) } 
\left[ {N \over 8 \pi (x-y)^2} \right]^{2p+q}
\nonumber\\
&& \quad 
\times \left( 1 + \, \OO(N^{-2}; g^4) \right) 
\end{eqnarray}
This is found from formulae (\ref{o-o:leading}), 
(\ref{pqp:largeN}), and (\ref{O-hat:large N}). 
The exact expressions of 
Sections \ref{section:simplest} and \ref{6 and higher} 
agree with (\ref{O-hat:large N}) and (\ref{normalization:general})
in the large $N$ limit. 
We conclude that to working precision, 
$\YY_{[p,q,p]}$ is a \quarter-BPS chiral primary 
operator. 

\section{Conclusion}

In this paper we studied local, polynomial, gauge invariant 
scalar composite operators
in $[p,q,p]$ representations of $SU(4)$ in 
\NN=4 SYM in four dimensions. 
We found that certain such operators 
have protected 
two-point functions at order $g^2$, with each other 
as well as with other operators. 
We presented ample evidence that these $\OO(g^2)$ protected operators 
are \quarter-BPS chiral primaries 
in the fully interacting theory. 

These operators are 
not just the double trace operators from 
the classification of \cite{AFSZ}, 
but mixtures of all 
gauge invariant local composite operators 
made of the same scalars:
single trace operators, 
other double trace operators, 
triple trace operators, etc. 
As our exact in $N$, explicit 
construction of \quarter-BPS 
primaries of low scaling dimension 
($\Delta = 2p+q < 8$) shows, 
the $N$ dependence of the coefficients 
in these linear combinations is quite complicated. 

Apart from 
operators
of low dimensions for arbitrary $N$, 
we considered the large $N$ behavior 
of two point functions for all $[p,q,p]$. 
A leading and subleading analysis 
reveals that for every $[p,q,p]$, 
there is a \quarter-BPS operator 
which is a 
certain linear combinations of 
double and single trace operators. 
We give closed form expressions 
for the operators involved in this 
linear combination and the coefficients with 
which they enter, 
all valid to next-to-leading order in $N$.

\section{Acknowledgments}

I would like to thank Sergio Ferrara for useful discussions, 
as well as pointing out references \cite{BKRS}. 
I am especially grateful to Eric D'Hoker, who 
suggested the problem and without whose 
involvement this paper would not be finished.

\pagebreak

{\LARGE \bf \appendixname}
\appendix

\section{\NN = 4 SUSY in various forms}
\label{n=4 susy:section}

In the \NN=1 component notation, the classical Lagrangian 
(see \cite{Sohnius}, p. 158) takes the form
(in geometric notation, i.e. with ${1\over g^2}$ 
multiplying the whole action)
\begin{eqnarray}
\LL &=& \mbox{$1\over g^2$ } \tr 
\left\{ 
- \mbox{$1\over 4$} F_{\mu\nu} F^{\mu\nu} 
+ \ihalf \bar \lambda \gamma^\mu D_\mu \lambda 
+ \half D^2
\right. \\ \nonumber &&\quad~~ \left. 
+ \half D_\mu A_j D^\mu A_j 
+ \half D_\mu B_j D^\mu B_j 
+ \ihalf \bar \psi_j \gamma^\mu D_\mu \psi_j
+ \half F_j F_j 
+ \half G_j G_j 
\right. \\ \nonumber &&\quad~~ \left. 
- i [ A_j , B_j ] D 
- i \bar \psi_j [ \lambda , A_j ] 
- i \bar \psi_j \gamma_5 [ \lambda , B_j ] 
\right. \\ \nonumber &&\quad~~ \left. 
- \ihalf \e_{jkl} 
\left( \right.\!\!  
\bar \psi_j [\psi_k , A_l ] - \bar \psi_j \gamma_5 [\psi_k , B_l ] 
\right. \\ \nonumber &&\quad\quad\quad\quad\quad~ \left. 
+ [ A_j , A_k ] F_l - [ B_j , B_k ] F_l 
+ 2 [ A_j , B_k ] G_l 
\left.\!\! \right) 
\right\}
\end{eqnarray}
(with Lorentz signature); 
there should be no confusion between the auxiliary field $D$ of 
the vector multiplet and the covariant derivative 
$D_\mu = \partial_\mu + i A_\mu$. 

This Lagrangian can also be rewritten in a manifestly 
$SU(4)$-invariant form. 
Combining the three chiral spinors and 
the gaugino into 
\begin{eqnarray}
\lambda_4 = \lambda ; \quad\quad
\lambda_j = \psi_j , \quad j = 1,2,3 
\end{eqnarray}
and making $4\times4$ antisymmetric matrices of scalars and pseudoscalars by
\begin{eqnarray}
A_{j k} = - \e_{jkl} A_l ; &\quad& A_{4 j} = - A_{j 4} = A_j ; \\
B_{j k} = \phantom{-} \e_{jkl} B_l ; &\quad& B_{4 j} = - B_{j 4} = B_j 
\end{eqnarray}
the Lagrangian becomes (sums on indices $j,k,l$ now run from 1 to 4)
\begin{eqnarray}
\LL &=& \mbox{$1\over g^2$ } \tr 
\left\{ 
- \mbox{$1\over 4$} F_{\mu\nu} F^{\mu\nu} 
+ \ihalf \bar \lambda_j \gamma^\mu D_\mu \lambda_j 
+ \half D_\mu A_{jk} D^\mu A_{jk} 
+ \half D_\mu B_{jk} D^\mu B_{jk} 
\right.\nonumber  \\ \nonumber &&\quad~~ \left. 
+ \ihalf \bar \lambda_j [ \lambda_k , A_{jk} ] 
+ \ihalf \bar \lambda_j \gamma_5 [ \lambda_k , B_{jk} ] 
+ \mbox{$1\over32$} [ A_{jk} , B_{lm} ] [ A_{jk} , B_{lm} ] 
\right. \\ &&\quad~~ \left. 
+ \mbox{$1\over64$} [ A_{jk} , A_{lm} ] [ A_{jk} , A_{lm} ] 
+ \mbox{$1\over64$} [ B_{jk} , B_{lm} ] [ B_{jk} , B_{lm} ] 
\right\}
\end{eqnarray}
after integrating out the auxiliary fields $D$, $F_j$, and $G_j$. 
The $A_{jk}$ and $B_{jk}$ are self-dual and antiself-dual tensors of
$O(4)$: 
\begin{equation}
A_{jk} = \half \e_{jklm} A_{lm} ; 
\quad 
B_{jk} = - \half \e_{jklm} B_{lm} 
\end{equation}
Alternatively, the fields $A_i$ and $B_i$ form a {\bf 6} of the $R$-symmetry 
group $SU(4) \sim SO(6)$: we can group them as 
$\phi^i = A_i$, $\phi^{i+3} = B_i$, $i=1,2,3$.

This form of the Lagrangian is only manifestly $O(4)$ symmetric, however. 
If we define a complex matrix of scalars 
\begin{equation}
M_{jk} \equiv \half \left( A_{lm} + i B_{jk} \right) 
\end{equation}
subject to a reality condition 
\begin{equation}
\bar M^{jk} \equiv (M_{jk})^\dagger = \half \e^{jklm} M_{lm} 
\end{equation}
the \NN=4 Lagrangian 
\begin{eqnarray}
\LL &=& \mbox{$1\over g^2$ } \tr 
\left\{ 
- \mbox{$1\over 4$} F_{\mu\nu} F^{\mu\nu} 
+ i \lambda_j \sigma^\mu D_\mu \bar \lambda^j 
+ \half D_\mu M_{jk} \bar D^\mu \bar M^{jk} 
\right. \\ \nonumber &&\quad~~ \left. 
+ i \lambda_j [ \lambda_k , \bar M^{jk} ] 
+ i \bar \lambda^j [ \bar \lambda^k , M_{jk} ] 
+ \mbox{$1\over 4$} [ M_{jk} , M_{lm} ] [ \bar M^{jk} , \bar M^{lm} ] 
\right\}
\end{eqnarray}
is then manifestly $SU(4)$ covariant, as are the SUSY transformation laws 
\begin{eqnarray}
\delta A_\mu &=& 
i \zeta_j \sigma_\mu \bar \lambda^j - 
i \lambda_j \sigma_\mu \bar \zeta^j \\
\delta M_{jk} &=& 
\zeta_j \lambda_k - \zeta_k \lambda_j + 
\e_{jklm} \bar \zeta^l \bar \lambda^m \\
\delta \lambda_j &=& 
- \ihalf \sigma^{\mu\nu} F_{\mu\nu} \zeta_j 
+ 2 i \sigma^\mu D_\mu M_{jk} \bar \zeta^k 
+ 2 i [ M_{jk}, \bar M^{kl} ] \zeta_l 
\end{eqnarray}
(notice that now $\lambda_j$ and $\bar \lambda^j$ are Weyl spinors; there
should be no confusion: when spinors are multiplied by $2\times2$ 
$\sigma$ matrices they are Weyl, and when the $4\times4$ $\gamma$ matrices
are used, they are Dirac).

We can also rewrite this Lagrangian in terms of three unconstrained
(unlike the $M_{jk}$) complex scalar fields 
\begin{equation}
     z_j = \mbox{$1\over\sqrt{2}$} \left( A_j + i B_j \right) , \quad
\bar z^j = \mbox{$1\over\sqrt{2}$} \left( A_j - i B_j \right) 
\end{equation}
and the original fermions $\psi_j$ and $\lambda$:
\begin{eqnarray}
\label{L su3 form}
\LL &=& \mbox{$1\over g^2$ } \tr 
\left\{ 
- \mbox{$1\over 4$} F_{\mu\nu} F^{\mu\nu} 
+ i \lambda \sigma^\mu D_\mu \bar \lambda 
+ i \psi_j \sigma^\mu D_\mu \bar \psi^j 
+ D_\mu z_j \bar D^\mu \bar z^j 
\right. \\ \nonumber &&\quad~~ \left. 
+ \mbox{$i\over\sqrt2$} \lambda [ \psi_j , \bar z^j ] 
- \mbox{$i\over\sqrt2$} \psi_j [ \lambda , \bar z^j ] 
- \mbox{$i\over\sqrt2$} \e^{jkl} \psi_j [ \psi_k , z_l ] 
\right. \\ \nonumber &&\quad~~ \left. 
+ \mbox{$i\over\sqrt2$} \bar \lambda [ \bar \psi^j , z_j ] 
- \mbox{$i\over\sqrt2$} \bar \psi^j [ \bar \lambda , z_j ] 
- \mbox{$i\over\sqrt2$} \e_{jkl} \bar \psi^j [ \bar \psi^k , \bar z^l ] 
\right. \\ \nonumber &&\quad~~ \left. 
+ [ z_j , z_k ] [ \bar z^j , \bar z^k ] 
- \half [ z_j , \bar z^j ] [ z_k , \bar z^k ] 
\right\}
\end{eqnarray}
and the SUSY transformations now are 
\begin{eqnarray}
\label{SUSY-su(3) form:begin}
\delta A_\mu &=& 
i \zeta_j \sigma_\mu \bar \psi^j - 
i \psi_j \sigma_\mu \bar \zeta^j + 
i \zeta \sigma_\mu \bar \lambda - 
i \lambda \sigma_\mu \bar \zeta \\
\delta z_j &=& 
\sqrt{2} \left(
\zeta \psi_j - \zeta_j \lambda - 
\e_{jkl} \bar \zeta^k \bar \psi^l 
\right) \\
\delta \lambda &=& 
- \ihalf \sigma^{\mu\nu} F_{\mu\nu} \zeta 
+ i \sqrt{2} \sigma^\mu D_\mu z_j \bar \zeta^j 
+ i \e^{jkl} [ z_j, z_k ] \zeta_l 
- i [ z_j, \bar z^j ] \zeta \\ 
\delta \psi_j &=& 
- \ihalf \sigma^{\mu\nu} F_{\mu\nu} \zeta_j 
+ i \sqrt{2} \e_{jkl} \sigma^\mu \bar D_\mu \bar z^k \bar \zeta^l 
- i \sqrt{2} \sigma^\mu D_\mu z_j \bar \zeta 
\nonumber \\ && \quad~~
+ i \left( 
[ z_k , \bar z^k ] \zeta_j 
- 2 [ z_j , \bar z^k ] \zeta_k 
- \e_{jkl} [ \bar z^k , \bar z^l ] \zeta 
\right)
\\
\noalign{\noindent and their conjugates}
\delta \bar z^j &=& 
\sqrt{2} \left(
\bar \zeta \bar \psi^j - \bar \zeta^j \bar \lambda - 
\e^{jkl} \zeta_k \psi_l 
\right) \\
\delta \bar \lambda &=& 
+ \ihalf \bar \sigma^{\mu\nu} F_{\mu\nu} \bar \zeta 
- i \sqrt{2} \bar \sigma^\mu \bar D_\mu \bar z^j \zeta_j 
- i \e_{jkl} [ \bar z^j, \bar z^k ] \bar \zeta^l 
+ i [ \bar z^j, z_j ] \bar \zeta \\ 
\delta \bar \psi^j &=& 
+ \ihalf \bar \sigma^{\mu\nu} F_{\mu\nu} \bar \zeta^j 
- i \sqrt{2} \e^{jkl} \bar \sigma^\mu D_\mu z_k \zeta_l 
+ i \sqrt{2} \bar \sigma^\mu \bar D_\mu \bar z^j \zeta 
\nonumber \\ && \quad~~
- i \left( 
[ \bar z^k , z_k ] \bar \zeta^j 
- 2 [ \bar z^j , z_k ] \bar \zeta^k 
- \e^{jkl} [ z_k , z_l ] \bar \zeta 
\right)
\label{SUSY-su(3) form:end}
\end{eqnarray}
This way of writing the Lagrangian and SUSY transformations 
hides the full $SU(4)$ $R$-symmetry of the 
theory; now, only the $SU(3) \times U(1)$ subgroup of it is manifest.

\section{Miscellaneous identities for $SU(N)$}
\label{suN-identities}

We can use the following property of generators of 
$SU(N)$ (for $N\ge3$) in the fundamental representation:
\begin{equation}
\{ t^a, t^b \} = {1\over N} \delta^{ab} + d^{abc} t^c
\end{equation}
Together with $[ t^a, t^b ] = i f^{abc} t^c$ (valid in any
representation), we find 
\begin{equation}
t^a t^b = {1\over 2 N} \delta^{ab} \bone + 
{1\over 2} \left( d^{abc} + i f^{abc} \right) t^c
\end{equation}

Let 
\begin{equation}
g^{a_1 ... a_k} \equiv \tr t^{a_1} ... t^{a_k}
\end{equation}
Then with the standard normalization 
$\tr t^a t^b = {1\over 2} \delta^{ab}$ 
for $SU(N)$ generators in the fundamental, we can 
in principle recursively determine the 
trace of any string of generators in terms of 
$\delta^{ab}$, $d^{abc}$, and $f^{abc}$: 
\begin{eqnarray}
&&
g^{a} = \tr t^{a} = 0 ,~
g^{a b} = \half \delta^{ab} ,~
g^{a b c} =  \quarter \left( d^{abc} + i f^{abc} \right) ,\quad\mbox{and}\\
&&
g^{a_1 ... a_k} = \mbox{$1\over 2 N$} \, \delta^{a_1 a_2} g^{a_3 ... a_k} 
+ 2\, g^{a_1 a_2 c} g^{c a_3 ... a_k} 
\end{eqnarray}
(for completeness we can define $g^{(0)} = \tr \bone = N$).

Now we can set up a recursion relation for
\begin{eqnarray}
\label{eq:recursion p}
P_k \equiv g^{a_1 ... a_k} g^{a_1 ... a_k} \\
\noalign{\noindent and}
\label{eq:recursion p-tilde}
\tilde P_k \equiv g^{a_1 ... a_k} g^{a_k ... a_1} 
\end{eqnarray}
(with sums on repeated $a_1 , ... ,  a_k$ implied). 
Using 
$t^a t^a = {N^2 - 1 \over 2 N} \bone$, we find 
\begin{eqnarray}
P_k = {N^2 - 1 \over 4 N^2} P_{k-2} 
+ {4\over N^2 - 1} P_3 P_{k-1} \\
\noalign{\noindent and similarly}
\tilde P_k = {N^2 - 1 \over 4 N^2} \tilde P_{k-2} 
+ {4\over N^2 - 1} \tilde P_3 \tilde P_{k-1} 
\end{eqnarray}
The values of $P_2$, $\tilde P_2$, 
$P_3$ and $\tilde P_3$ have to be computed explicitly;
they are 
\begin{equation}
\label{p3 and p3-tilde}
P_2 = \tilde P_2 = {N^2 - 1 \over 4} , 
\quad 
P_3 = - {N^2 - 1 \over 4 N} , 
\quad \mbox{and}\quad 
\tilde P_3 = {(N^2 - 1)(N^2 - 2) \over 8 N} 
\end{equation}
For large $N$, the leading behavior is given by 
\begin{eqnarray}
\label{leading order traces}
&&
g^{a_1 ... a_k} \sim 
\left( 2^{k-3} \right) 
g^{a_1 a_2 c_3} g^{c_3 a_3 c_4} ... 
g^{c_{k-2} a_{k-2} c_{k-1} } g^{c_{k-1} a_{k-1} a_k} 
\\ \noalign{\noindent and} 
&&
P_{2 k + 1} \sim - {N k \over 4^k} , \quad 
P_{2 k} \sim {N^2 \over 4^k} ; \quad 
\tilde P_{k} \sim {N^k \over 2^k} 
\end{eqnarray}
Dependence on $N$ is%
\footnote{
	Note that the way the recursion formulae 
	(\ref{eq:recursion p-tilde}) and (\ref{eq:recursion p-tilde}) 
	work out 
	together with the initial values (\ref{p3 and p3-tilde}),
	leading order 
	large $N$ results are accurate to order $\OO(N^{-2})$ 
	and not to $\OO(N^{-1})$, as one could have thought naively. 
	}
very different for $P_k$ and $\tilde P_k$; 
in fact, taking generators in reverse order in the second trace
(such as in $\tilde P_k$) grows the fastest with $k$, and taking
them in the same order (as in $P_k$), the slowest.

Here are a few more identities we may have a need for in calculating
two-point functions. First, the normalizations of $SU(N)$ generators 
in an arbitrary representation is defined in terms of a constant $C(r)$ as 
\begin{equation}
\tr\! {}_r \; T^a_r T^b_r = C(r) \delta^{ab}
\end{equation}
and there is a quadratic Casimir, 
\begin{equation}
T^c_r T^c_r = C_2(r) \; \bone 
\end{equation}
In particular, the adjoint and fundamental 
representations will be of interested, and for these 
$C_2(\mbox{adj}) = N$, $C_2(\mbox{fund}) = (\mbox{$N^2 - 1 \over 2N$})$, 
$C(\mbox{adj}) = N$, $C(\mbox{fund}) = \half$. 
Then, for example, 
\begin{eqnarray}
T^a T^a T^b T^b &=& \left[C_2(r) \right]^2 \; \bone \\
T^a T^b T^b T^a &=& \left[C_2(r) \right]^2 \; \bone \\
\left[T^a , T^b \right] \left[T^a , T^b \right] &=& 
- N C_2(r) \; \bone \\
T^a T^b T^a T^b &=& 
C_2(r) \left( C_2(r) - \mbox{$N\over2$} \right) \; \bone  
\end{eqnarray}
(we have omitted the label ``$r$'' on the generators, e.g. $T^a = T^a_r$).  
Longer expressions are just a little more complicated but not by much: 
\begin{eqnarray}
\tr 
T^b \left[ T^a , T^b \right] T^c \left[ T^a , T^c \right] 
&=& \quarter N^2 (N^2 - 1) C(r) \\
\tr 
T^a T^b T^c T^a T^c T^b 
&=& (N^2 - 1) C(r) (C_2(r) - \mbox{$N\over2$})^2 \\
\tr 
T^a T^b T^c T^a T^b T^c 
&=& (N^2 - 1) C(r) (C_2(r) - \mbox{$N\over2$}) (C_2(r) - N) 
\quad\quad
\end{eqnarray}
In particular, the last expression vanishes in the adjoint representation.

Using the fact that $U(N)_C = Gl(N,C)$, 
any $N \times N$ matrix $A$ can be decomposed 
into generators (in the fundamental) 
of $SU(N)$ plus the unit matrix: 
\begin{equation}
\label{u(n):completeness}
A = 
\left( 2 \tr A t^c \right) t^c 
+ 
\left( \mbox{$1\over N$} \tr A \right) \bone 
\end{equation}
Then, for example, 
we can write down the ``trace merging formula''
\begin{equation}
\label{merging traces}
2 \left( \tr A t^c \right) \left( \tr B t^c \right) 
= 
\tr A B - 
\mbox{$1\over N$} \left( \tr A \right) \left( \tr B \right) 
\end{equation}
and we can arrive at an even simpler recursion relations for 
$\tilde P_k$: 
\begin{eqnarray}
\tilde P_{k+1} &=& 
\left( \tr t^{a_1} ... t^{a_k} t^c \right) 
\left( \tr t^{a_k} ... t^{a_1} t^c \right) 
\nonumber\\&=& 
\half \left( \tr t^{a_1} ... t^{a_k} t^{a_k} ... t^{a_1} \right) 
- 
\mbox{$1\over 2N$} 
\left( \tr t^{a_1} ... t^{a_k} \right) 
\left( \tr t^{a_k} ... t^{a_1} \right) 
\nonumber\\&=& 
\mbox{$N\over 2$} (\mbox{$N^2 - 1 \over 2N$})^k
- \mbox{$1\over 2N$} \tilde P_k 
\end{eqnarray}
with $\tilde P_1 = 0$. 
(Naturally, this gives the same values for $\tilde P_k$ as before.)

Another useful relation satisfied by the generators of $SU(N)$ 
in the fundamental, is 
\begin{equation}
\label{double-line}
(t^a)_{ij} (t^a)_{kl} = \half \left( 
\delta_{il} \delta_{jk} 
- \mbox{$1\over N$}
\delta_{ij} \delta_{kl} 
\right) 
.
\end{equation}
Using this identity, one can easily reproduce 
the trace merging formula (\ref{merging traces}), 
as well as the expressions (\ref{p3 and p3-tilde}) 
for $P_2$, $\tilde P_2$, $P_3$ and $\tilde P_3$.

\section{Details for operators of dimension $\Delta \ge 6$}

\subsection{States with weight [2,2,2]}
\label{app:2:2:2}

Instead of computing the symmetry factors by hand, 
we fed {\it Mathematica} an algorithm for calculating 
the contractions for the free correlators as well as 
the $\OO(g^2)$ corrections; the results are 
\begin{eqnarray}
{192 \over 5 N^4 (N^2 - 1)}
\, {\bf F} 
\hspace{-7.5em} && \nonumber\\ 
&=& 
\small
\left( 
\matrix{ 1 - {\frac{4}{{N^2}}} + {\frac{40}{3 {N^4}}} & {\frac{8 
      \left( 3 {N^4} - 22 {N^2} + 60 \right) }{3 {N^5}}} & {\frac{4 
\left( N^2-4 \right)  \left( {N^2}-5 \right) }{{N^5}}} & {\frac{2 
      \left( 7 {N^4} - 18 {N^2} + 40 \right) }{{N^5}}} & {\frac{20 \left({N^2} -2 \right) }
    {{N^4}}} \cr {\frac{8 \left( 3 {N^4} - 22 {N^2} + 60 \right) }{3 {N^5}}} & 8 - {\frac{152}{3 {N^2}} + {\frac{224}{{N^4}} - 
   {\frac{384}{{N^6}}} } } & {\frac{-8 
      \left( N^2-4 \right)  \left( {N^2}-6 \right) }{{N^6}}} & {
     \frac{4 \left( {N^6} + 7 {N^4} + 8 {N^2} -48 \right) }{{N^6}}} & {\frac{16 
      \left( 2 {N^4} - 7 {N^2} + 6 \right) }{{N^5}}} \cr {\frac{4 
\left( N^2-4 \right)  \left({N^2} -5 \right) }{{N^5}}} & {\frac{-8 
      \left( N^2-4 \right)  \left({N^2} -6 \right) }{{N^6}}} & {
     \frac{3 {{\left( {N^2}-4 \right) }^2} \left( {N^2}-2 \right) }{{N^6}}} & {\frac{6 
      \left( {N^4}-16 \right) }{{N^6}}} & {\frac{-12 \left( {N^2}-4 \right) }
    {{N^5}}} \cr {\frac{2 \left( 7 {N^4} - 18 {N^2} + 40 \right) }{{N^5}}} & {\frac{4 
      \left( {N^6} + 7 {N^4} + 8 {N^2} -48 \right) }{{N^6}}} & {\frac{6 
      \left( {N^4}-16 \right) }{{N^6}}} & {\frac{12 \left( 1 + {N^2} \right)  
      \left( {N^4} + 6 {N^2} -8 \right) }{{N^6}}} & {\frac{12 
      \left( 3 {N^4} - 3 {N^2} + 4 \right) }{{N^5}}} \cr {\frac{20 \left( {N^2}-2 \right) }
    {{N^4}}} & {\frac{16 \left( 2 {N^4} - 7 {N^2} + 6 \right) }{{N^5}}} & {\frac{-12 
      \left( {N^2}-4 \right) }{{N^5}}} & {\frac{12 \left( 3 {N^4} - 3 {N^2} + 4 \right) }
    {{N^5}}} & {\frac{12 \left( N^2 - 2 \right)  \left( N^2 +1 \right) }{{N^4}}} \cr  }
\right) 
\hspace{-11em} 
\nonumber\\
\end	{eqnarray}
for the matrix of free combinatorial factors, and 
\begin{eqnarray}
{96 \over 25 N^4 (N^2 - 1) }
\, {\bf G} 
\hspace{-8em} && \nonumber\\ 
&=& 
\left( 
\matrix{ 1 + {\frac{4}{{N^2}}} + {\frac{8}{{N^4}}} & {\frac{8 
      \left( {N^4} - 2 {N^2} + 12 \right) }{{N^5}}} & {\frac{3 {{\left( {N^2}-4 \right) }^2}}
    {{N^5}}} & {\frac{12 \left( 2 {N^4} + {N^2} + 4 \right) }{{N^5}}} & {\frac{12 
      \left( 3 {N^2} -2 \right) }{{N^4}}} \cr {\frac{8 \left( {N^4} - 2 {N^2} + 12 \right) }
    {{N^5}}} & {\frac{64 \left( {N^4} - 6 {N^2} + 18 \right) }{{N^6}}} & {\frac{24 
      \left( N^2 -4 \right)  \left( {N^2} -6\right) }{{N^6}}} & {
     \frac{48 \left( 3 {N^4} - 5 {N^2} + 12 \right) }{{N^6}}} & {\frac{96 
      \left( 2 {N^2} -3 \right) }{{N^5}}} \cr {\frac{3 {{\left( {N^2} -4 \right) }^2}}
    {{N^5}}} & {\frac{24 \left( N^2 -4 \right)  
      \left( {N^2} -6 \right) }{{N^6}}} & {\frac{18 {{\left( {N^2}-4 \right) }^2}}
    {{N^6}}} & {\frac{18 {{\left( {N^2}-4 \right) }^2}}{{N^6}}} & {\frac{36 
      \left( {N^2} -4\right) }{{N^5}}} \cr {\frac{12 \left(2 {N^4} + {N^2} + 4 \right) }
    {{N^5}}} & {\frac{48 \left( 3 {N^4} - 5 {N^2} + 12 \right) }{{N^6}}} & {\frac{18 
      {{\left( {N^2} -4 \right) }^2}}{{N^6}}} & {\frac{18 
      \left( {N^6} + 17 {N^4} - 8 {N^2} + 16 \right) }{{N^6}}} & {\frac{36 
      \left( {N^4} + 7 {N^2} -4 \right) }{{N^5}}} \cr {\frac{12 \left( 3 {N^2} -2 \right) }
    {{N^4}}} & {\frac{96 \left( 2 {N^2} -3 \right) }{{N^5}}} & {\frac{36 
      \left( {N^2} -4 \right) }{{N^5}}} & {\frac{36 \left( {N^4} + 7 {N^2} -4 \right) }
    {{N^5}}} & {\frac{72 \left( {N^2} + 1 \right) }{{N^4}}} \cr  }
\right) 
\hspace{-10em} 
\nonumber\\
\end	{eqnarray}
for the matrix of corrections proportional to $\tilde B(x,y) N$; 
for the notation see Section \ref{6 and higher}.

\subsection{States with weight [2,3,2]}
\label{app:2:3:2}

In the same fashion, for the operators 
defined in Section \ref{section:2:3:2} 
(also of scaling dimension $\Delta = 7 + \OO(g^2)$), we find 
\begin{eqnarray}
{\frac{64}{3 \left(N^2-1 \right) \left( N^2-4 \right) {N^3}}}
\, {\bf F} 
\hspace{-11em} && \nonumber\\ 
&=& 
\left( 
\matrix{ {\frac{\left( 2 {N^2} + 15 \right) }{2 {N^2}}} & {\frac{10 
      \left( {N^4} - 3 {N^2} + 30 \right) }{{N^5}}} & {\frac{5  
      \left( 7 {N^4} - 3 {N^2} + 30 \right) }{{N^5}}} & {\frac{15 
      \left( {N^4} - 8 {N^2} + 30 \right) }{{N^5}}} \cr {\frac{10 
      \left( {N^4} - 3 {N^2} + 30 \right) }{{N^5}}} & {\frac{10 
      \left( {N^6} + 12 {N^2} -60 \right) }{{N^6}}} & {\frac{5 
      \left( {N^6} + 45 {N^4} - 78 {N^2} -60 \right) }{{N^6}}} & {\frac{30 
      \left( {N^2} -6 \right)  \left( {N^2} +5 \right) }{{N^6}}} \cr {\frac{5 
      \left( 7 {N^4} - 3 {N^2} + 30 \right) }{{N^5}}} & {\frac{5 
      \left( {N^6} + 45 {N^4} - 78 {N^2} -60 \right) }{{N^6}}} & {\frac{10 
      \left( 4 {N^6} + 45 {N^4} - 42 {N^2} -15 \right) }{{N^6}}} & {
     \frac{30 \left( 8 {N^4} - 23 {N^2} -15 \right) }{{N^6}}} \cr {\frac{15 
      \left( {N^4} - 8 {N^2} + 30 \right) }{{N^5}}} & {\frac{30 
      \left( {N^2} -6 \right)  \left( {N^2} +5 \right) }{{N^6}}} & {
     \frac{30 \left( 8 {N^4} - 23 {N^2} -15 \right) }{{N^6}}} & {\frac{15 
      \left( {N^2} -3 \right)  \left( {N^4} - 6 {N^2} + 30 \right) }{{N^6}}}
   \cr {\frac{15 \left( 5 {N^4} - 16 {N^2} + 60 \right) }
    {2 {N^5}}} & {\frac{30 \left( 2N^2+3 \right)  \left( 3N^2-10 \right)  
      }{{N^6}}} & {\frac{30 \left( 18 {N^4} - 28 {N^2} -15 \right)  
      }{{N^6}}} & {\frac{15 \left( {N^6} - 4 {N^4} + 18 {N^2} -90 \right)  
      }{{N^6}}} \cr {\frac{30 
      \left( 9 N^2 -25 \right) }{{N^4}}} & {\frac{30 
      \left( {N^2} -2 \right)  \left( 11 {N^2} -25 \right) }{{N^5}}} &
   {\frac{30 \left( 13 {N^4} + 14 {N^2} + 25 \right) }{{N^5}}} & {\frac{30 
      \left( 2 {N^4} - 17 {N^2} + 75 \right) }{{N^5}}} 
    \cr -{\frac{15 \left( 3 {N^2} -10 \right) }{{N^4}}} & -{\frac{30 
      \left( {N^2} -2 \right)  \left( 4 {N^2} -5 \right) }{{N^5}}} & {
     \frac{-30 \left( 2 {N^4} + {N^2} + 5 \right) }{{N^5}}} & {\frac{30 
      \left( 2 {N^4} - 2 {N^2} -15 \right) }{{N^5}}} \cr  }
\right.
\hspace{-9em} 
\nonumber\\&&\phantom{blah}
\nonumber\\&&\hspace{9em}
\left. 
\matrix{ {\frac{15 
      \left( 5 {N^4} - 16 {N^2} + 60 \right) }{2 {N^5}}} & {\frac{30 
      \left( 9 N^2 -25 \right) }{{N^4}}} & -{\frac{15 
      \left( 3 {N^2} -10 \right) }{{N^4}}} \cr {
     \frac{30 \left( 2 {N^2} +3 \right)  \left( 3 {N^2} -10 \right) }
    {{N^6}}} & {\frac{30 \left( N^2 -2 \right)  \left( 11 N^2-25 \right)  
      }{{N^5}}} & -{\frac{30 \left( {N^2} -2 \right)  
      \left( 4 {N^2} -5 \right) }{{N^5}}} \cr {\frac{30 
      \left( 18 {N^4} - 28 {N^2} -15 \right) }{{N^6}}} & {\frac{30 
      \left( 13 {N^4} + 14 {N^2} + 25 \right) }{{N^5}}} & -{\frac{30 
      \left( 2 {N^4} + {N^2} + 5 \right) }{{N^5}}} \cr 
    {\frac{15 \left( {N^6} - 4 {N^4} + 18 {N^2} -90 \right) }
    {{N^6}}} & {\frac{30 \left( 2 {N^4} - 17 {N^2} + 75 \right) }
    {{N^5}}} & {\frac{30 \left( 2 {N^4} - 2 {N^2} -15 \right) }
    {{N^5}}} \cr {\frac{15 
      \left( 7 {N^6} + 32 {N^4} - 24 {N^2} -180 \right) }{2 {N^6}}} & 
    {\frac{30 \left( 7 {N^4} + 8 {N^2} + 75 \right) }{{N^5}}} & -{\frac{15 
      \left( {N^4} + 14 {N^2} + 30 \right) }{{N^5}}} \cr {\frac{30 
      \left( 7 {N^4} + 8 {N^2} + 75 \right) }{{N^5}}} & {\frac{30 
      \left( {N^2} +5 \right)  \left( 11 {N^2} -25 \right) }{{N^4}}} & 
    -{\frac{30 \left( {N^2}+5\right)  \left( 4 {N^2} -5 \right) }{{N^4}}}
    \cr -{\frac{15 
      \left( {N^4} + 14 {N^2} + 30 \right) }{{N^5}}} & -{\frac{30 
      \left( {N^2} +5 \right)  \left( 4 {N^2} -5 \right) }{{N^4}}} & {
     \frac{15 \left( 7 {N^4} + 15 {N^2} -10 \right) }{{N^4}}} \cr  }
\right)
\hspace{-9em} 
\nonumber\\
\end	{eqnarray}
for the matrix of free combinatorial factors, and 
\begin{eqnarray}
{\frac{16}{27 \left(N^2-1 \right) \left( N^2-4 \right) {N^3} }}
  {\bf G} 
\hspace{-11em} && \nonumber\\ 
&=& 
\left( 
\matrix{ {\frac{{N^2}+29}{{N^2}}} & {\frac{10 \left( {N^2}+ 12\right) }{{N^3}}} & {\frac{10 
      \left( 5 {N^2} +21 \right) }{{N^3}}} & {\frac{10 \left( {N^2} +3 \right) }
    {{N^3}}} \cr {\frac{10 \left( {N^2} +12 \right) }{{N^3}}} & {\frac{100 
      \left( {N^4} +24 \right) }{{N^6}}} & {\frac{400 \left( {N^4} +3\right) }{{N^6}}} & {
     \frac{100 \left( {N^4} - 6 {N^2} + 36 \right) }{{N^6}}} \cr {\frac{10 
      \left( 5 {N^2} +21 \right) }{{N^3}}} & {\frac{400 \left( {N^4} +3 \right) }
    {{N^6}}} & {\frac{50 \left( {N^6} + 19 {N^4} + 12 \right) }{{N^6}}} & {\frac{300 
      \left( {N^4} - {N^2} + 6 \right) }{{N^6}}} \cr {\frac{10 \left( {N^2} +3 \right) }
    {{N^3}}} & {\frac{100 \left( {N^4} - 6 {N^2} + 36 \right) }{{N^6}}} & {\frac{300 
      \left( {N^4} - {N^2} + 6 \right) }{{N^6}}} & {\frac{200 
      \left( {N^4} - 9 {N^2} + 27 \right) }{{N^6}}} \cr {\frac{60 
      \left( {N^2} +3 \right) }{{N^3}}} & {\frac{150 \left( 3 {N^4} - 4 {N^2} + 24 \right) }
    {{N^6}}} & {\frac{300 \left( 4 {N^4} - {N^2} + 6 \right) }{{N^6}}} & {\frac{300 
      \left( {N^4} - 6 {N^2} + 18 \right) }{{N^6}}} \cr {\frac{420}{{N^2}}} & {\frac{3000 
      \left( {N^2} -2 \right) }{{N^5}}} & {\frac{300 \left( {N^4} + 17 {N^2} -10 \right) }
    {{N^5}}} & {\frac{600 \left( 4 {N^2} -15 \right) }{{N^5}}} \cr -{\frac{90}{{N^2}}} & {
     \frac{-600 \left( {N^2} -2 \right) }{{N^5}}} & -{\frac{600 \left( 2 {N^2} -1 \right) }
    {{N^5}}} & -{\frac{300 \left( {N^2} -6 \right) }{{N^5}}} \cr  }
\right. 
\hspace{-9em} 
\nonumber\\&&\phantom{blah}
\nonumber\\&&\hspace{5em}
\left. 
\matrix{ {\frac{60 \left( {N^2} +3 \right) }{{N^3}}} & {\frac{420}{{N^2}}} & -{\frac{90}
    {{N^2}}} \cr {\frac{150 
      \left( 3 {N^4} - 4 {N^2} + 24 \right) }{{N^6}}} & {\frac{3000 \left( {N^2} -2 \right) }
    {{N^5}}} & -{\frac{600 \left( {N^2} -2 \right) }{{N^5}}} \cr {\frac{300 
      \left( 4 {N^4} - {N^2} + 6 \right) }{{N^6}}} & {\frac{300 
      \left( {N^4} + 17 {N^2} -10 \right) }{{N^5}}} & -{\frac{600 
      \left( 2 {N^2} -1 \right) }{{N^5}}} \cr {\frac{300 
      \left( {N^4} - 6 {N^2} + 18 \right) }{{N^6}}} & {\frac{600 \left( 4 {N^2} -15 \right) }
    {{N^5}}} & -{\frac{300 \left( {N^2} -6 \right) }{{N^5}}} \cr {\frac{75 
      \left( {N^6} + 17 {N^4} - 24 {N^2} + 72 \right) }{{N^6}}} & {\frac{300 
      \left( {N^4} + 20 {N^2} -30 \right) }{{N^5}}} & -{\frac{150 
      \left( {N^4} + 8 {N^2} -12 \right) }{{N^5}}} \cr {\frac{300 
      \left( {N^4} + 20 {N^2} -30 \right) }{{N^5}}} & {\frac{3000 \left( {N^2} +5 \right) }
    {{N^4}}} & -{\frac{600 \left( {N^2} +5 \right) }{{N^4}}} \cr -{\frac{150 
      \left( {N^4} + 8 {N^2} -12 \right) }{{N^5}}} & -{\frac{600 \left( {N^2} +5 \right) }
    {{N^4}}} & {\frac{300 \left( {N^2} +2 \right) }{{N^4}}} \cr  }
\right) 
\hspace{-9em} 
\nonumber\\&&\phantom{blah}
\nonumber\\
\end	{eqnarray}
for the matrix of corrections proportional to $\tilde B(x,y) N$.



\end{document}